\documentstyle[aps,multicol,epsfig]{revtex}

\renewcommand{\narrowtext}{\begin{multicols}{2} \global\columnwidth20.5pc}
\renewcommand{\widetext}{\end{multicols} \global\columnwidth42.5pc}

\def\top#1{\vskip #1\begin{picture}(290,80)(80,500)\thinlines \put(65,500){\line( 1, 0){255}}\put(320,500){\line( 0, 1){5}}\end{picture}}
\def\bottom#1{\vskip #1\begin{picture}(290,80)(80,500)\thinlines \put(330,500){\line( 1, 0){255}}\put(330,500){\line( 0, -1){5}}\end{picture}}

\newcommand{\bq}{\begin{equation}}
\newcommand{\eq}{\end{equation}}
\newcommand{\bqa}{\begin{eqnarray}}
\newcommand{\eqa}{\end{eqnarray}}
\newcommand{\nn}{\nonumber \\}

\begin{document}
\draft 
\title{Holon-pair boson theory based on the U(1) and SU(2) slave-boson approaches to the t-J Hamiltonian}
\author{Sung-Sik Lee and Sung-Ho Suck Salk$^1$}
\address{Department of Physics, Pohang University of Science and Technology,\\
Pohang, Kyoungbuk, Korea 790-784\\
$^1$ Korea Institute of Advanced Studies, Seoul 130-012, Korea\\}

\date{\today}
\maketitle

\begin{abstract}
To supplement our recent brief report on the theory of holon-pair boson approach to the t-J Hamiltonian [S.-S. Lee and Sung-Ho Suck Salk, Phys. Rev. B {\bf 64}, 052501(2001)], in this paper we present a full exposure to the theory, detailed physical implications and predicted various physical properties of high $T_c$ cuprates.
We discuss the significance of coupling (interplay) between the spin and charge degrees of freedom in the Heisenberg interaction term of the t-J Hamiltonian.
We discuss its importance in causing the arch-shaped superconducting transition temperature $T_c$ and the pseudogap (spin gap) temperature $T^*$ tangential to $T_c$ in the overdoped region in the observed phase diagram of high $T_c$ cuprates. 
A universal parabolic scaling behavior of $T^*/T_c$ (or $T_c/T^*$) with hole doping concentration is predicted in agreement with observations, indicating that there exists correlation between the pseudogap (spin gap) phase and the superconducting phase through antiferromagnetic fluctuations.
Our proposed holon-pair boson theory is shown to be self-consistent in that it not only yields the arch (dome) shape structure of $T_c$ but also reproduces numerous other physical properties such as superfluid weight, bose condensation energy, spectral function, optical conductivity and spin susceptibility, including their temperature and doping dependence.
\end{abstract}
\begin{multicols}{2}

\newpage
\section{Introduction}
\vspace{-0.3cm}
Since the discovery of high $T_c$ cuprates\cite{BEDNORZ}, the pseudogap phase and /superconducting phase have been revealed in various measurements\cite{TIMUSK} such as 
the angle resolved photoemission spectroscopy (ARPES)\cite{DING,SHEN,DING2,SHEN2}, 
tunneling spectroscopy\cite{TAO,ODA,YEH},
nuclear magnetic resonance (NMR)\cite{WALSTEDT,YASUOKA,ISHIDA,JULIEN}, 
magnetic susceptibility\cite{ODA},
inelastic neutron scattering\cite{KEIMER},
optical conductivity\cite{ORENSTEIN,HOMES,UCHIDA},
Raman scattering\cite{KENDZIORA,KANG}, 
resistivity\cite{BUCHER}
and
specific heat\cite{MOMONO_IDO,LORAM_PSEUDO,TALLON,MOMONO_PG}.
The arch shaped bose condensation (superconducting transition) temperature $T_c$ and the tangential appearance of the pseudogap temperature $T^*$ to $T_c$ in the overdoped region are commonly observed in the phase diagram of both high $T_c$ cuprates $La_{2-x} Sr_x Cu O_4$ and $Bi_2 Sr_2 Ca Cu_2 O_{8+\delta}$\cite{ODA,DING,WALSTEDT,YASUOKA,ISHIDA,JULIEN,KENDZIORA,KANG,MOMONO_IDO}.
Most recently, from the Nernst effect measurements in the high $T_c$ cuprates, $Bi_2 Sr_2 Ca Cu_2 O_{8+\delta}$ Wang et al. showed that in agreement with the observed trend of the ARPES gap\cite{DING2} the Cooper pairing potential monotonically decreases with increasing hole concentration\cite{WANG}.
The pseudogap also decreases with increasing hole concentration.
Paying attention to the ratio of the observed spin gap (pseudogap) temperature $T^*$ to the superconducting temperature $T_c$, Nakano et al.\cite{ODA} observed that there is a good correlation between $T^*$ and $T_c$, and that there exists a universal scaling behavior with hole doping, that is, the sample independence in the ratio $T^*/T_c$ (or $T_c/T^*$) for the entire range of hole doping concentration for both $La_{2-x} Sr_x Cu O_4$ and $Bi_2 Sr_2 Ca Cu_2 O_{8+\delta}$.
This hints that superconductivity is affected by (correlated with) spin fluctuations or spin singlet pair excitations.
The generic features of peak-dip-hump structures have been observed in ARPES (angle resolved photoemission spectroscopy)\cite{DING2,SHEN2}, STM (scanning tunneling measurement)\cite{TAO,ODA,YEH} and optical conductivity measurements\cite{ORENSTEIN}, suggesting a possibility of common origin. 

From the very beginning of high $T_c$ superconductivity, Anderson\cite{ANDERSON} has stressed that the t-J Hamiltonian is an appropriate model for the study of high $T_c$ superconductivity.
Resonating-valence-bond (RVB) mean field approach to the t-J Hamiltonian\cite{BASKARAN} was introduced to unsatisfactorily predict a trend of monotonically decreasing superconducting transition temperature with increasing hole doping.
Later by applying the U(1) slave-boson theory of the t-J Hamiltonian\cite{KOTLIAR,FUKUYAMA,NAGAOSA,UBBENS,ICHINOSE}, decreasing pseudogap temperature with hole doping concentration was predicted and identified to be the spin gap temperature.
Superconductivity was proposed to arise as a result of bose condensation of composite particles made of the spinon singlet pair and single holons.
Later, Wen and Lee\cite{WEN} proposed an extended SU(2) slave-boson theory to impose symmetry not only for the spinon but for the holon for hole doped systems.
In theories which pay attention to the bose condensation of single holons, the predicted bose condensation temperature $T_c$ showed a linear increase with hole concentration $x$.
In these theories $T_c$ is scaled by the kinetic energy of doped holes (of the order of hopping energy $tx$) and the pseudogap temperature $T^*$ by the Heisenberg coupling $J$.
Accordingly $T_c$ was shown to increase linearly with increasing $x$.
These theories fails to predict correlation between $T_c$ and $T^*$ and thus the experimentally observed universal behavior of $T^*/T_c$ can not be explained.
Gimm et al.\cite{GIMM} proposed a theory of holon-pair bose condensation by paying attention to the $-\frac{J}{4}n_i n_j$ term in the Heisenberg interaction term of $J({\bf S}_{i} \cdot {\bf S}_{j} - \frac{1}{4}n_{i}n_{j})$.
As a result, the bose condensation (superconducting transition) temperature scales with the Heisenberg coupling $J$ but failed to predict the arch (dome) shaped superconducting transition temperature in the observed phase diagram.

To remedy this failure, lately we proposed an improved version of holon-pair boson theory\cite{LEE} which takes into account coupling between the charge (holon) and spin (spinon) degrees of freedom in the slave-boson representation of the Heisenberg interaction term in the t-J Hamiltonian.
Theories which neglect the charge contribution in treating the interaction term are expected to fail in providing superconductivity.
This is because the Cooper pairs represent pairing of charged particles ($-2e$ or $+2e$).
In such incomplete theories only the spin (spinon) paring is possible. 
On the other hand, in our theory both the charge and spin pairing are possible since the Heisenberg interaction term is represented by coupling between the charge and spin contributions.
Such coupling is proven to cause the arch-shaped superconducting transition temperature\cite{LEE}.
Further, correlation between the spin gap temperature $T^*$ and the superconducting transition temperature $T_c$ is predicted in agreement with observation\cite{ODA}.
The coupling between the spin and charge degrees of freedom results in salient predictions of various physical properties; they are
the arch-shaped superconducting temperature\cite{LEE},
bose condensation energy\cite{LEE_COND},
spectral function\cite{LEE_SPEC},
the peak-dip-hump structure in the optical conductivity\cite{LEE_OPT},
the universal scaling behavior of $T^*/T_c$ with hole density $x$\cite{LEE_SPEC},
the superfluid weight vs. $T_c$ and
the doping and temperature dependence of superfluid weight\cite{LEE_SUPER}.

It is reminded that our approach is pivotally different from other proposed U(1) and SU(2) slave-boson theories\cite{KOTLIAR,FUKUYAMA,NAGAOSA,UBBENS,WEN} in that 
the coupling of the spin and charge degrees of freedom naturally appears in the slave-boson representation of the Heisenberg interaction term and 
that physical spin-charge separation does not appear even in the mean field Hamiltonian.
As a result of the coupling, this theory predicts the spin-gap (pseudogap) temperature $T^*$ tangential to $T_c$ in the overdoped region and a universal behavior of the ratio $T^*/T_c$ with $x$.
Realizing that the pseudogap is identified as the spin gap from NMR measurements\cite{WALSTEDT,YASUOKA,ISHIDA,JULIEN} and that $T_c$ is observed to be correlated with $T^*$\cite{ODA}, we discuss how the spin singlet pair excitations (spin gap) affect the high $T_c$ superconductivity.
Our earlier presentation of this holon-pair boson theory\cite{LEE} is excessively brief.
In this paper a full disclosure to the details of the holon-pair boson theory will be made with emphasis of physical implications, and theoretical predictions, highlighted. 
By means of self-containment we show the self-consistency of our theory by revealing that not only the phase diagram but also various other physical properties are well reproduced in agreement with observations.

\section{Interpretation of the Heisenberg interaction term in the U(1) slave-boson representation}
The t-J Hamiltonian of interest is given by,
\begin{eqnarray}
H & = & -t\sum_{<i,j>}(c_{i\sigma}^{\dagger}c_{j\sigma} + c.c.) 
+ J\sum_{<i,j>}( {\bf S}_{i} \cdot {\bf S}_{j} - \frac{1}{4}n_{i}n_{j}) \nn 
&& -\mu \sum_i c_{i\sigma}^{\dagger} c_{i\sigma} 
\label{tjmodel}
\end{eqnarray}
and the Heisenberg interaction term is rewritten 
\bqa
H_J & = & J \sum_{<i,j>} ({\bf S}_{i} \cdot {\bf S}_{j} - \frac{1}{4}n_{i}n_{j}) \nn
& = & -\frac{J}{2} \sum_{<i,j>} ( c_{i\downarrow}^{\dagger}c_{j\uparrow}^{\dagger}-c_{i\uparrow}^{\dagger}c_{j\downarrow}^{\dagger}) (c_{j\uparrow}c_{i\downarrow}-c_{j\downarrow} c_{i\uparrow}).
\label{Heisen}
\eqa
Here $t$ is the hopping energy and ${\bf S}_{i}$, the electron spin operator at site $i$, ${\bf S}_{i}=\frac{1}{2}c_{i\alpha}^{\dagger} {\bbox \sigma}_{\alpha \beta}c_{i\beta}$ with ${\bbox \sigma}_{\alpha \beta}$, the Pauli spin matrix element. $n_i$ is the electron number operator at site $i$, $n_i=c_{i\sigma}^{\dagger}c_{i\sigma}$.
$\mu$ is the chemical potential.

In the U(1) slave-boson representation\cite{KOTLIAR,FUKUYAMA,NAGAOSA,UBBENS}, with single occupancy constraint at site $i$ the electron annihilation operator $c_{i \sigma}$ is taken as a composite operator of the spinon (neutrally charged fermion) annihilation operator $f_{i \sigma}$ and the holon (positively charged boson) creation operator $b_i^\dagger$, and thus, $c_{i \sigma} = f_{i \sigma} b_i^\dagger$.
Rigorously speaking, it should be noted that the expression $c_{i \sigma} = b_i^\dagger f_{i\sigma}$ is not precise since these operators belong to different Hilbert spaces and thus the equality sign here should be taken only as a symbol for mapping.
Using $c_{i \sigma} = f_{i \sigma} b_i^\dagger$ and introducing the Lagrange multiplier term (the last term in Eq.(\ref{tj_u1_slave})) to enforce single occupancy constraint, the t-J Hamiltonian is rewritten,
\bqa
&& H  =  -t\sum_{<i,j>}\left( (f_{i\sigma}^{\dagger}b_i)(b_j^\dagger f_{j\sigma}) + c.c. \right) + H_J \nn
&& - \mu \sum_i f_{i\sigma}^{\dagger}b_i f_{i\sigma} b_i^\dagger \nn
&& - i \sum_i \lambda_i ( b_i^\dagger b_i + f_{i\sigma}^\dagger f_{i\sigma} -1 ) 
\label{tj_u1_slave} 
\eqa
with the Heisenberg interaction term,
\bq
H_J  =   -\frac{J}{2} \sum_{<i,j>} b_i b_j b_j^{\dagger}b_i^{\dagger} (f_{i\downarrow}^{\dagger}f_{j\uparrow }^{\dagger}-f_{i\uparrow}^ {\dagger}f_{j\downarrow}^{\dagger})(f_{j\uparrow}f_{i\downarrow}-f_{j\downarrow} f_{i\uparrow}).
\label{tj_J}
\eq
The first term represents hopping of a spinon from site $j$ to site $i$ and of a holon (positively charged boson) from site $i$ to site $j$.
In the slave-boson representation a charged fermion (electron or hole) is taken as a composite particle of a `spinon' and a `holon'. 
They can conveniently serve as book-keeping labels to discern physical properties or objects involved with the charge or spin degree of freedom (e.g., spin gap phase, spin singlet pairs, hole pairs, ...).
With the single occupancy constraint, electron is allowed to hop from a singly occupied copper site $i$ only to a vacant copper site $j$. 
A site of single occupancy in the $CuO_2$ plane of high $T_c$ cuprates physically represents an electrically neutral site (net charge $0$) with an electron of spin $1/2$ and the vacant site, a site of positive charge $+ e$ with net spin $0$. 
In the slave-boson representation, hopping of an electron (a composite of spinon and holon) from a singly occupied copper site (neutral site) $j$ to an empty site (positively charged site with $+e$) $i$ results in 
the annihilation of a spinon (a fermion of charge $0$ and spin $1/2$) and the creation of a positively charged holon (a boson of charge $+ e$ and spin $0$) at site $j$ while at the copper site $i$ a composite of a spinon (fermion of charge $0$ and spin $1/2$) and a negatively charged holon is created. 
It is of note that as a result of electron hopping the newly occupied copper site $i$ in the $CuO_2$ plane can, also, be labeled as `spinon' since this is an electrically neutral (charge $0$) site with an electron of spin $1/2$ and the vacant site $j$, `holon' since this is a positively charged site with a single charge $+e$ and the net spin $0$ as mentioned above.
Thus in practical sense, there is no distinction between the two different cases above.
At times, we will call the singly occupied site as `spinon' and the vacant (empty) site as `holon' as long as there is no confusion.
This is because any site occupied by a spinon is identified as an electrically neutral site occupied by a single electron with spin $1/2$ and the site with a positive holon is a positively charged vacant site with spin $0$.
Thus physical spin-charge separation is not allowed.

The Heisenberg interaction term, Eq.(\ref{tj_J}) shows coupling between the charge and spin degrees of freedom.
Physics involved with the charge degree of freedom is manifested by the four holon (boson) operator $b_i b_j b_j^\dagger b_i^\dagger$ in the Heisenberg interaction term.
Judging from the intersite charge coupling term $\frac{J}{4} n_i n_j$ present in the Heisenberg interaction term 
$H_J =  J\sum_{<i,j>}( {\bf S}_{i} \cdot {\bf S}_{j} - \frac{1}{4}n_{i}n_{j})$,
it is obvious that this charge contribution can not be neglected in its slave-boson representation.
It is to be noted that the Hubbard Hamiltonian contains repulsive interaction $U$ between charged particles and is mapped into the t-J Hamiltonian $H_{t-J}$ in the large $U$ limit. 
The Coulomb repulsion, $U n_{i\uparrow} n_{i\downarrow} = \frac{U}{4}(n_{i\uparrow} + n_{i\downarrow})^2 - \frac{U}{4}(n_{i\uparrow}-n_{i\downarrow})^2$ obviously manifests the presence of both the charge (the first term) and spin (the second term) degrees of freedom. 
Thus, under mapping the charge part of contribution naturally appears in the Heisenberg interaction term.

Let us now take another look at the importance of the charge contribution.
In general, uncertainty principle between the number density (amplitude ) and the phase of a boson order parameter applies. 
As an example, arbitrarily large fluctuations of the number density fix the phase, or arbitrarily large phase fluctuations fix the number density of the boson. 
The conventional BCS superconductors of long coherence length meet the former classification, and thus the phase fluctuations of the Cooper pair order parameter are minimal.
For charged bosons, e.g., the Cooper pairs, the number density fluctuations refer to charge density fluctuations.
For short coherence length superconductors such as the high $T_c$ cuprate systems of present interest, local charge density fluctuations exist and cause large phase fluctuations.
Thus, both the charge and phase fluctuations need to be taken into account to fully exploit the quantum fluctuations .

Let us now consider the importance of the charge and spin fluctuations.
In generally, coupling between physical quantities $A$ and $B$ is decomposed into terms involving fluctuations of $A$, i.e., $(A-<A>)$ and $B$, i.e., $(B-<B>)$, separately uncorrelated mean field contribution of $<A>$ and $<B>$ and correlation between fluctuations of $A$ and $B$, that is, $(A-<A>)$ and  $(B-<B>)$ respectively;
$AB = (A-<A>)<B> + (B-<B>)<A> + <A><B> + (A-<A>)(B-<B>)$.
Setting $A=b_i b_j b_j^\dagger b_i^\dagger$ for charge (holon) contribution and 
$B=(f_{i \downarrow}^\dagger f_{j \uparrow}^\dagger - f_{i \uparrow}^\dagger f_{j \downarrow}^\dagger ) ( f_{j \uparrow} f_{i \downarrow} - f_{j \downarrow} f_{i \uparrow} )$ for spin (spinon) contribution,
the Heisenberg coupling term, Eq(\ref{tj_J}) can be decomposed into terms involving coupling between 
the charge and spin fluctuations separately, the mean field contributions and coupling (correlation) between fluctuations (charge and spin fluctuations).
Using such decomposition of the Heisenberg interaction term for Eq.(\ref{tj_u1_slave}),  we write the partition function,
\widetext
\top{-2.8cm}
\bq
Z  =  \int Db Df_{\uparrow} Df_{\downarrow} D\lambda e^{-S[b,f,\lambda]},
\label{u1_partition0}
\eq
where 
\bqa
&& S[b,f,\lambda] 
= \int_0^\beta d \tau \Bigl[
\sum_i b_i^\dagger \partial_\tau b_i + \sum_{i} f_{i\sigma}^\dagger \partial_\tau f_{i\sigma} + H_{t-J}^{U(1)} \Bigr]
\label{u1_action0}
\eqa
with $\beta=\frac{1}{k_BT}$, the inverse temperature and $H_{t-J}^{U(1)}$, the U(1) symmetry preserved Hamiltonian,
\bqa
H_{t-J}^{U(1)} & = & -t\sum_{<i,j>}(f_{i\sigma}^{\dagger}f_{j\sigma}b_{j}^{\dagger}b_{i} + c.c.) \nn
&& -\frac{J}{2}  \sum_{<i,j>} \Bigl[  \Bigl< 
 (f_{i\downarrow}^{\dagger}f_{j\uparrow}^{\dagger}-f_{i\uparrow}^{\dagger}f_{j\downarrow}^{\dagger})
 (f_{j\uparrow}f_{i\downarrow}-f_{j\downarrow} f_{i\uparrow}) 
 \Bigr> 
 b_i b_j b_j^{\dagger}b_i^{\dagger}  \nn
&& +  \Bigl< b_i b_j b_j^{\dagger}b_i^{\dagger} \Bigr> 
(f_{i\downarrow}^{\dagger}f_{j\uparrow}^{\dagger}-f_{i\uparrow}^ {\dagger}f_{j\downarrow}^{\dagger})
(f_{j\uparrow}f_{i\downarrow}-f_{j\downarrow} f_{i\uparrow}) \nn
& & -   \Bigl< b_i b_j b_j^{\dagger}b_i^{\dagger} \Bigr> 
\Bigl< 
(f_{i\downarrow}^{\dagger}f_{j\uparrow}^{\dagger}-f_{i\uparrow}^ {\dagger}f_{j\downarrow}^{\dagger})
(f_{j\uparrow}f_{i\downarrow}-f_{j\downarrow} f_{i\uparrow}) 
\Bigr> \nn
&& + \Bigl( b_i b_j b_j^{\dagger}b_i^{\dagger} - \Bigl< b_i b_j b_j^{\dagger}b_i^{\dagger}\Bigr> \Bigr)  
\Bigl( (f_{i\downarrow}^{\dagger}f_{j\uparrow}^{\dagger}-f_{i\uparrow}^ {\dagger}f_{j\downarrow}^{\dagger} ) 
( f_{j\uparrow}f_{i\downarrow}-f_{j\downarrow} f_{i\uparrow}) 
- \Bigl<(f_{i\downarrow}^{\dagger}f_{j\uparrow}^{\dagger}-f_{i\uparrow}^ {\dagger}f_{j\downarrow}^{\dagger})
(f_{j\uparrow}f_{i\downarrow}-f_{j\downarrow} f_{i\uparrow})\Bigr> \Bigr) \Bigr] \nn
&& - \mu \sum_{i} f_{i\sigma}^{\dagger} f_{i\sigma} ( 1 + b_i^\dagger b_i )
- i\sum_{i} \lambda_{i}(f_{i\sigma}^{\dagger}f_{i\sigma}+b_{i}^{\dagger}b_{i} -1).
\label{u1_tjmodel0}
\end{eqnarray}
\bottom{-2.8cm}
\narrowtext
\noindent

\section{U(1) mean field Hamiltonian}

Noting that $[b_i, b_j^{\dagger}] = \delta_{ij}$ for boson, the intersite charge (holon) interaction term (the second term) in Eq.(\ref{u1_tjmodel0}) is rewritten,
\bqa
&& -\frac{J}{2} 
\Bigl< (f_{i\downarrow}^{\dagger}f_{j\uparrow}^{\dagger}-f_{i\uparrow}^ {\dagger}f_{j\downarrow}^{\dagger})(f_{j\uparrow}f_{i\downarrow}-f_{j\downarrow} f_{i\uparrow}) \Bigr> 
b_i b_j b_j^{\dagger}b_i^{\dagger} \nn
&& = -\frac{J}{2} <|\Delta_{ij}^f|>^2 \left( 1 + b_i^\dagger b_i  +  b_j^\dagger b_j + b_i^\dagger b_j^\dagger b_j  b_i \right),
\label{u1_mf_holon_result}
\eqa
with $\Delta^f_{ij} = f_{j\uparrow}f_{i\downarrow}-f_{j\downarrow} f_{i\uparrow}$, the spinon pairing field.
The third term in Eq.(\ref{u1_tjmodel0}) represents the intersite spin (spinon) interaction and is rewritten,
\bqa
&& -\frac{J}{2} \Bigl< b_i b_j b_j^{\dagger}b_i^{\dagger}\Bigr> (f_{i\downarrow}^{\dagger}f_{j\uparrow}^{\dagger}-f_{i\uparrow}^ {\dagger}f_{j\downarrow}^{\dagger})(f_{j\uparrow}f_{i\downarrow}-f_{j\downarrow} f_{i\uparrow}) \nn
& = & -\frac{J_p}{2} (f_{i\downarrow}^{\dagger}f_{j\uparrow}^{\dagger}-f_{i\uparrow}^ {\dagger}f_{j\downarrow}^{\dagger})(f_{j\uparrow}f_{i\downarrow}-f_{j\downarrow} f_{i\uparrow}),
\label{u1_mf_spinon_result}
\eqa
where $J_p = J(  1 + <b_i^\dagger b_i> + <b_j^\dagger b_j> + <b_i^\dagger b_i b_j^\dagger b_j >  )$ or $J_p = J(1-x)^2$ with $x$, the uniform hole doping concentration\cite{1_DELTA}. 
The fourth term in Eq.(\ref{u1_tjmodel0}) is written,
\bqa
\lefteqn{ \frac{J}{2} \Bigl< b_i b_j b_j^{\dagger}b_i^{\dagger} \Bigr> \Bigl< (f_{i\downarrow}^{\dagger}f_{j\uparrow}^{\dagger}-f_{i\uparrow}^{\dagger}f_{j\downarrow}^{\dagger})(f_{j\uparrow}f_{i\downarrow}-f_{j\downarrow} f_{i\uparrow}) \Bigr> } \nn
& = & \frac{J}{2}  \left( 1 + <b_i^\dagger b_i> + <b_j^\dagger b_j> + <b_i^\dagger b_i b_j^\dagger b_j > \right) <|\Delta^f_{ij}|^2>. \nn
\label{u1_mf_const}
\eqa

The intersite spinon interaction term in Eq.(\ref{u1_mf_spinon_result}) is decomposed into the direct, exchange and pairing channels\cite{UBBENS},
\bqa
&& -\frac{J_p}{2} (f_{i\downarrow}^{\dagger}f_{j\uparrow}^{\dagger}-f_{i\uparrow}^ {\dagger}f_{j\downarrow}^{\dagger})(f_{j\uparrow}f_{i\downarrow}-f_{j\downarrow} f_{i\uparrow}) \nn
& = & \frac{J_p}{4} \Bigl[ \sum_{k=1}^3 (f_{i\alpha}^{\dagger} \sigma_{\alpha \beta} ^k f_{i\beta})( f_{j\gamma }^{\dagger} \sigma_{\gamma \delta} ^k f_{j\delta} ) - ( f_{i\alpha}^\dagger \sigma^0_{\alpha \beta} f_{i\beta} ) ( f_{i\gamma}^\dagger \sigma^0_{\gamma \delta} f_{j\delta} ) \Bigr] \nn
& = & v_D + v_E + v_P 
\label{eq:mf_1} 
\eqa
with $\sigma^0 = I$, the identity matrix and $\sigma^{1,2,3}$, the Pauli spin matrices,
where $v_D$, $v_E$ and $v_P$ are the spinon interaction terms of the direct, exchange and pairing channels respectively\cite{LEE},
\bqa
v_D & = & -\frac{J_p}{8} \sum_{k=0}^{3} ( f_i^{\dagger} \sigma^k f_i ) ( f_j^{\dagger} \sigma^k f_j ),  \label{vD} \\
v_E & =  & -\frac{J_p}{4} \Bigl( (f_{i\sigma}^{\dagger}f_{j\sigma})(f_{j\sigma}^{\dagger}f_{i\sigma}) - n_i \Bigr), \label{vE} \\
v_P & = &  -\frac{J_p}{2} (f_{i\downarrow}^{\dagger}f_{j\uparrow}^{\dagger}-f_{i\uparrow}^{\dagger}f_{j\downarrow}^{\dagger}) (f_{j\uparrow}f_{i\downarrow}-f_{j\downarrow}f_{i\uparrow}). \label{vP}
\eqa
Here $\sigma^0$ is the unit matrix and $\sigma^{1,2,3}$, the Pauli spin matrices.

Combining Eq.(\ref{u1_mf_holon_result}) and Eq.(\ref{u1_mf_const}), we have
\bqa
&& -\frac{J}{2}<|\Delta_{ij}^f|^2> \left( 1 + b_i^\dagger b_i  +  b_j^\dagger b_j + b_i^\dagger b_j^\dagger b_j  b_i \right) \nn
&& + \frac{J}{2} <|\Delta_{ij}^f|^2> \left( 1 + <b_i^\dagger b_i> + <b_j^\dagger b_j> + <b_i^\dagger b_i b_j^\dagger b_j > \right) \nn
&& = -\frac{J}{2} <|\Delta_{ij}^f|^2>  b_i^\dagger b_j^\dagger b_j  b_i
+ \frac{J}{2} <|\Delta^f_{ij}|^2> <b_i^\dagger b_i b_j^\dagger b_j >  \nn
&& -\frac{J}{2} <|\Delta_{ij}^f|^2> \left[ \left(b_i^\dagger b_i -<b_i^\dagger b_i>\right) + \left(b_j^\dagger b_j-<b_j^\dagger b_j> \right) \right].
\label{u1_mf_holon_const}
\eqa

Collecting the decomposed terms Eq.(\ref{u1_mf_holon_result}) through Eq.(\ref{u1_mf_const}) in association with Eqs.(\ref{eq:mf_1}) through Eq.(\ref{u1_mf_holon_const}), we write 
\bqa
&&H_J= 
-\frac{J}{2} \sum_{<i,j>} |\Delta_{ij}^f|^2  b_i^\dagger b_j^\dagger b_j  b_i \nn
&& -J_p \sum_{<i,j>} \Bigl[ 
\frac{1}{2} (f_{i\downarrow}^{\dagger}f_{j\uparrow}^{\dagger}-f_{i\uparrow}^{\dagger}f_{j\downarrow}^{\dagger}) (f_{j\uparrow}f_{i\downarrow}-f_{j\downarrow}f_{i\uparrow})  \nn
&& +\frac{1}{4} \Bigl( (f_{i\sigma}^{\dagger}f_{j\sigma})(f_{j\sigma}^{\dagger}f_{i\sigma}) - n_i \Bigr) \nn
&& + \frac{1}{8} \sum_{k=0}^{3} ( f_i^{\dagger} \sigma^k f_i ) ( f_j^{\dagger} \sigma^k f_j ) 
\Bigr] \nn
&& + \frac{J}{2} \sum_{<i,j>} |\Delta^f_{ij}|^2 <b_i^\dagger b_i> <b_j^\dagger b_j >  \nn
&& -\frac{J}{2} \sum_{<i,j>} |\Delta_{ij}^f|^2 \left[ \left(b_i^\dagger b_i -<b_i^\dagger b_i>\right) + \left(b_j^\dagger b_j-<b_j^\dagger b_j> \right) \right],
\label{u1_tjmodel}
\end{eqnarray}
where we considered $<|\Delta_{ij}^f|^2> = |\Delta_{ij}^f|^2$ and ignored the fifth term in Eq.(\ref{u1_tjmodel0}).

\widetext
\top{-2.8cm}

Hubbard Stratonovich transformation for the holon pairing term (the second term of Eq.(\ref{u1_tjmodel})) leads to
\bqa
\displaystyle
e^{\sum_{<i,j>} \frac{J}{2}|\Delta^{f}_{ij}|^2 b_{i}^\dagger b_{j}^\dagger b_{i} b_{j}}  \propto 
\int \prod_{<i,j>} d\Delta_{ij}^{b*} d\Delta_{ij}^{b} 
e^{ -\sum_{<i,j>} \frac{J}{2}|\Delta^{f}_{ij}|^2 
\Bigl[ |\Delta_{ij}^{b}|^{2} - \Delta_{ij}^{b*} (b_{i}b_{j}) - \Delta_{ij}^{b} (b_{j}^\dagger b_{i}^\dagger ) \Bigr] }, 
\label{u1_holon_Hubbard_Stratonovich}
\eqa
and the saddle point approximation yields,
\bqa
H^b_P = \sum_{<i,j>} \frac{J}{2}|\Delta^{f}_{ij}|^2 \Bigl[ |\Delta_{ij}^{0b}|^{2} 
- \Delta_{ij}^{0b*} (b_{j}b_{i}) - \Delta_{ij}^{0b} (b_{j}^\dagger b_{i}^\dagger ) \Bigr],
\label{eff_Hb}
\eqa
where $\Delta^{0b}_{ij} = <b_i b_j>$ is the saddle point for the holon pairing order parameter $\Delta^b_{ij}$.
Since confusion is not likely to occur, we will use the notation $\Delta^{b}_{ij}$ in place of $\Delta^{0b}_{ij}$ for the saddle point.
As are shown in Eqs.(\ref{vD}) through (\ref{vP}) the spinon interaction term is decomposed into the direct, exchange and pairing channels respectively.
Proper Hubbard-Stratonovich transformations corresponding to these channels and saddle point approximation leads to the effective Hamiltonian,
\bqa
H_{eff} & = & 
\frac{J_p}{4} \sum_{<i,j>} \Bigl[ |\chi_{ij}|^2 - \chi_{ij}^* ( f_{i\sigma}^{\dagger}f_{j\sigma} + \frac{4t}{J_p} b_i^\dagger b_j )  - c.c. \Bigr]  \nn
&& + \frac{J}{2} \sum_{<i,j>} |\Delta^f_{ij}|^2 \Bigl[ |\Delta_{ij}^{b}|^{2} - \Delta_{ij}^{b*} (b_{j}b_{i}) - c.c \Bigr] \nn
&& + \frac{J_p}{2} \sum_{<i,j>} \Bigl[ |\Delta^f_{ij}|^2 - \Delta^{f}_{ij} (f_{i\downarrow}^\dagger f_{j\uparrow}^\dagger - f_{i\uparrow}^\dagger  f_{j\downarrow}^\dagger )  - c.c. \Bigr] \nn
&& + \frac{J_p}{2} \sum_{<i,j>} \sum_{l=0}^{3} \Bigl[ (\rho^l_{j})^2 -  \rho^l_{j} ( f_i^{\dagger} \sigma^l f_i )  \Bigr]  + \frac{J_p}{2} \sum_{i} (f_{i\sigma}^\dagger f_{i\sigma}) \nn
&& + \frac{4t^2}{J_p} \sum_{<i,j>} ( b_i^\dagger b_j )(b_j^\dagger b_i ) \nn
&& + \frac{J}{2} \sum_{<i,j>} |\Delta^f_{ij}|^2 <b_i^\dagger b_i> <b_j^\dagger b_j >  \nn
&&  -\frac{J}{2} \sum_{<i,j>} |\Delta_{ij}^f|^2 \left[ \left(b_i^\dagger b_i - <b_i^\dagger b_i>\right) + \left(b_j^\dagger b_j - <b_j^\dagger b_j> \right) \right] \nn
&&  - \mu \sum_{i} f_{i\sigma}^{\dagger}f_{i\sigma}( 1+b_i^\dagger b_i) - i\sum_{i} \lambda_{i}(f_{i\sigma}^{\dagger}f_{i\sigma} + b_{i}^{\dagger}b_{i} -1),
\label{u1_eff_hamil1}
\eqa
\bottom{-2.8cm}
\narrowtext
\noindent
where $\Delta_{ij}^{b} = <b_{i}b_{j}>$, $\chi_{ij}= < f_{i\sigma}^{\dagger}f_{j\sigma} + \frac{4t}{J_p} b_{i}^{\dagger}b_{j}>$, $\Delta^f_{ij} = < f_{j\uparrow}f_{i\downarrow}-f_{j\downarrow}f_{i\uparrow} >$ and $\rho_{i}^{k}=<\frac{1}{2}f_{i}^\dagger \sigma^k f_i>$ are proper saddle points.

We note that $\rho^l_{i} = \frac{1}{2} < f_i^{\dagger} \sigma^l f_i > = < S^l_i > = 0 $ for $l=1,2,3$, $\rho^0_{i} = \frac{1}{2} < f_{i \sigma}^\dagger f_{i \sigma} > = \frac{1}{2}$ for $l=0$ for the contribution of the direct spinon interaction term (the fourth term). 
The expression $(b_{j}^{\dagger}b_{i})(b_{i}^{\dagger}b_{j})$ in the fifth term of Eq.(\ref{u1_eff_hamil1}) represents the exchange interaction channel. 
The exchange channel will be ignored owing to a large cost in energy, $U \approx \frac{4t^2}{J}$\cite{UBBENS}\cite{WEN}.
The resulting effective Hamiltonian is
\bq
H^{MF} = H^{\Delta,\chi} + H^b + H^f,
\label{mean_H_u1}
\eq
where $H^{\Delta,\chi}$ represents the the saddle point energy involved with the spinon pairing order parameter $\Delta^f$, the holon pairing order parameter $\Delta^b$ and the hopping order parameter $\chi$,
\bqa
H^{\Delta,\chi} & = & 
 J \sum_{<i,j>} \Bigl[ \frac{1}{2} |\Delta^f_{ij}|^2 |\Delta_{ij}^{b}|^{2}  + \frac{1}{2} |\Delta^f_{ij}|^2 x^2 \Bigr] \nn
&& + \frac{J_p}{2} \sum_{<i,j>} \Bigl[ |\Delta_{ij}^{f}|^{2} + \frac{1}{2} |\chi_{ij}|^{2} + \frac{1}{4} \Bigr], 
\label{u1_Delta_chi}
\eqa
$H^b$ is the holon Hamiltonian,
\bqa
H^b  & = & -t \sum_{<i,j>} \Bigl[ \chi_{ij}^{*}(b_{i}^{\dagger}b_{j}) + c.c.  \Bigr] \nn
     & - & \sum_{<i,j>} \frac{J}{2}|\Delta^f_{ij}|^2 \Bigl[ \Delta_{ij}^{b*} (b_{i}b_{j}) + c.c. \Bigr] \nn
    & - & \sum_{i} \mu^b_i ( b_{i}^{\dagger}b_{i} -x ),
\label{u1_holon_sector}
\eqa
where $\mu^b_i = i\lambda_i + \frac{J}{2}\sum_{j=i\pm \hat x, i \pm \hat y} |\Delta_{ij}^f|^2$
and $H^f$, the spinon Hamiltonian,
\bqa
H^f & = & 
-  \frac{J_p}{4} \sum_{<i,j>} \Bigl[ \chi_{ij}^{*} (f_{i\sigma}^{\dagger}f_{j\sigma}) + c.c. \Bigr] \nn
& - & \frac{J_p}{2} \sum_{<i,j>} \Bigl[ \Delta_{ij}^{f*} (f_{j\uparrow}f_{i\downarrow}-f_{j\downarrow}f_{i\uparrow}) + c.c. \Bigr]   \nn
& - & \sum_{i} \mu^f_i \left( f_{i\sigma}^{\dagger} f_{i\sigma} -(1-x) \right),
\label{u1_spinon_sector}
\eqa
where $\mu^f_i = \mu (1-x)  + i\lambda_i$.

As can be seen from Eqs.(\ref{u1_Delta_chi}) through (\ref{u1_spinon_sector}), Eq.(\ref{mean_H_u1}) reveals the importance of coupling between the spin and charge degrees of freedom, that is, coupling between the spinon pairing and holon pairing. 
Thus no spin-charge separation appears in the mean-field Hamiltonian above contrary to other mean field theories\cite{KOTLIAR,FUKUYAMA,NAGAOSA,UBBENS,WEN} which pay attention to the single-holon bose condensation.
As can be seen from the second term in Eq.(\ref{u1_holon_sector}) which represents holon pairing contribution
it is expected that, owing to the coupling effect, bose condensation (or superconducting phase transition) will occur only in the presence of the non-vanishing spin singlet pairing order, $\Delta^f$ owing to the coupling effects mentioned above.
Indeed, in high $T_c$ cuprates superconductivity is not observed above the pseudogap (spin gap) temperatures $T^*$ where the spin singlet pairing order disappears.

\section{U(1) free energy}
The diagonalized Hamiltonian for Eq.(\ref{mean_H_u1}) above is obtained to be (see Appendix A for detailed derivations),
\bqa
&& H^{MF}_{U(1)}  = 
N J \Bigl[ \Delta_f^2 \Delta_b^{2}  + \Delta_f^2 x^2 \Bigr] 
+ N J_p \Bigl[ \Delta_{f}^{2} + \frac{1}{2} \chi^{2} +\frac{1}{4} \Bigr]  \nn
& & + \sum_{k,s}^{'} E_{ks}^f ( \alpha_{ks\uparrow}^{\dagger}  \alpha_{ks\uparrow} - \alpha_{ks\downarrow}^{\dagger} \alpha_{ks\downarrow} ) - N x \mu^f \nn
 & & + \sum_{k,s=\pm 1}^{'}  E_{ks}^b  \beta_{ks}^{\dagger}  \beta_{ks} + \sum_{k,s=\pm 1}^{'} \frac{1}{2}(E_{ks}^b + \mu^b) + \mu^b N x, \nn
\label{u1_diagonalized_hamiltonian}
\eqa
where $E_{ks}^{f}$ is the quasispinon excitation energy, 
\begin{equation}
E_{ks}^{f}  =  \sqrt{(\epsilon_{ks}^{f}-\mu^f)^{2} + ( \Delta^f_0 )^{2}}
\label{u1_spinon_energy_app}
\end{equation}
with the spinon pairing energy (gap), $\Delta^f_0 = J_p \xi_{k}(\tau^{f}) \Delta_f$,
and $E_{ks}^{b}$ is the quasiholon excitation energy,
\begin{equation}
E_{ks}^{b}  =  \sqrt{(\epsilon_{ks}^{b} - \mu^b)^{2} - ( \Delta^b_0 )^{2}},
\label{u1_holon_energy_app}
\end{equation}
where the holon pairing energy, $\Delta^b_0 =  J\Delta_f^2 \xi_{k}(\tau^{b}) \Delta_b$ and
with $\phi=\theta$, $\tau^f$ or $\tau^b$,
\begin{eqnarray}
\xi_{k}(\phi) & = & \sqrt{ \gamma_{k}^{2} \cos^{2} \phi + \varphi_{k}^{2} \sin^{2} \phi }, \label{u1_xi_app},  \\
\epsilon_{ks}^{f} & = & \frac{J_p}{2}s\chi \xi_{k}(\theta), \\
\epsilon_{ks}^{b}  & = & 2ts \chi \xi_{k}(\theta),
\end{eqnarray}
with $ \gamma_{k} = (\cos k_{x} + \cos k_{y})$ and 
$\varphi_{k} = ( \cos k_{x} - \cos k_{y})$.
$\sum^{'}$ denotes the summation over momentum $k$ in the half reduced Brillouin zone, and $s= +1$ and $-1$ represent the upper and lower energy bands of quasiparticles respectively. 
Here $\alpha_{ks\uparrow}( \alpha_{ks\uparrow}^{\dagger})$  and $\alpha_{ks\downarrow}( \alpha_{ks\downarrow}^{\dagger})$ are the annihilation(creation) operators of spinon quasiparticles of spin up and spin down respectively, and $\beta_{ks}(\beta_{ks}^{\dagger})$, the annihilation(creation) operators of holon quasiparticles of spin $0$.
$\epsilon_{ks}^{f}$ and $\epsilon_{ks}^{b}$ are the kinetic energies for spinons and holons respectively.
The minus sign ($-\Delta^2$) in the expression of the holon quasiparticle energy $\sqrt{ (\epsilon-\mu)^2 - \Delta^2}$ arises as a consequence of the Bose Einstein statistics\cite{NOZIERE}.
From the diagonalized Hamiltonian Eq.(\ref{u1_diagonalized_hamiltonian}), we calculate the total free energy.

Rewriting Eq.(\ref{u1_diagonalized_hamiltonian}) as
\bqa
H^{MF}_{U(1)} & = & \sum_{k, s= \pm 1}^{'} \Bigl[ E_{ks}^{f}(\alpha_{ks\uparrow}^{\dagger}\alpha_{ks\uparrow} - \alpha_{ks\downarrow}\alpha_{ks\downarrow}^{\dagger}) +  E_{ks}^{b} \beta_{ks} ^{\dagger} \beta_{ks}\Bigr] \nn
&& +  H_{c}
\label{eq:diagonalized_hamiltonian_2}
\eqa
with 
\bqa
&&H_{c} = 
NJ\Delta_f^2 \Bigl( \Delta_b^{2} + x^2 \Bigr) 
+ NJ_p \Bigl( \Delta_f^{2} + \frac{\chi^{2}}{2} +\frac{1}{4} \Bigr) \nn
&&+ \sum_{k, s=\pm 1}^{'}  \frac{ E_{ks}^{b} +\mu^b }{2} - N x \mu^{f} + N x \mu^b,
\label{u1_Hc}
\eqa
the partition function is derived to be,
\bqa
Z &=& \exp (-\beta H_c ) \prod_{k,s= \pm 1}^{'} 
( 2 \cosh \frac{\beta E_{ks}^f}{2} )^2 (1-e^{-\beta E_{ks}^b})^{-1}.
\label{eq:partition_function}
\eqa
Using the above expression, the total free energy is given by
\bqa
&& F_{U(1)}  =   
NJ \Delta_f^2 \Bigl( \Delta_b^{2} + x^2 \Bigr) 
+ NJ_p \Bigl( \Delta_f^{2} + \frac{\chi^{2}}{2} +\frac{1}{4} \Bigr)  \nn
&& - 2k_{B}T \sum_{k,s=\pm 1}^{'} \ln ( \cosh (\beta E_{ks}^{f}/2) ) - N x \mu^{f} - 2N k_{B}T \ln2 \nn
&& + k_{B}T \sum_{k,s=\pm 1}^{'} \ln (1 - e^{-\beta E_{ks}^{b}}) + \sum_{k,s=\pm 1}^{'} \frac{ E_{ks}^{b} + \mu^b }{2} + N x \mu^b. \nn
\label{u1_free_energy}
\eqa


The set of uniform phase ($\theta =0$) for the hopping order parameter, d-wave symmetry ($\tau^f=\pi/2$) for the spinon pairing order parameter and s-wave symmetry ($\tau^b=0$) for the holon pairing order parameter is found to yield a stable saddle point energy for both the underdoped and overdoped regions.
There is another set of order parameters which yield the same energy as the above one; $2\pi$-flux phase ($\theta =\pi/2$) for the hopping order parameter, s-wave symmetry ($\tau^f=0$) for the spinon pairing order parameter and d-wave symmetry ($\tau^b=\pi/2$) for the holon pairing order parameter. 
In both cases, the d-wave symmetry of the electron or hole (not holon) pairs occurs as a composite of the d-wave (s-wave) symmetry of spinon pairs and s-wave (d-wave) symmetry of holon pairs.
Only at very low doping near half filling, the flux phase\cite{UBBENS} becomes more stable.
Thus, the phase of the order parameters of present interest are $\theta =0$, $\tau^f=\pi/2$ and $\tau^b=0$.
Then the d-wave symmetry of the electron or hole (not holon) pairs is a composite of the d-wave symmetry of spinon pairs and s-wave symmetry of holon pairs.
Minimizing the free energy with respect to the amplitudes of the order parameters $\chi$, $\Delta_b$ and $\Delta_f$, we obtain the self-consistent equations for the order parameters,
\bqa
\frac{\partial F_{U(1)}}{\partial \chi}  & = &  
N J_p \chi  - \sum_{ks}^{'} \left( \tanh \frac{\beta E^f_{ks}}{2} \right) \left( \frac{\partial E^f_{ks}}{\partial \chi} \right) \nn
&& + \sum_{ks}^{'} \left( \frac{1}{e^{\beta E_{ks}^{b}}-1} + \frac{1}{2} \right) \left( \frac{\partial E^b_{ks}}{\partial \chi} \right) =0, \label{u1_self1} \\
\frac{\partial F_{U(1)}}{\partial \Delta_b} & = & 
2 N J \Delta_f^2 \Delta_b \nn
&& + \sum_{ks}^{'} \left( \frac{1}{e^{\beta E_{ks}^{b}}-1} + \frac{1}{2} \right) \left( \frac{\partial E^b_{ks}}{\partial \Delta_b} \right) =0, \label{u1_self2} \\
\frac{\partial F_{U(1)}}{\partial \Delta_f} & = & 
2 N J_p \Delta_f + 2N \Delta_f (\Delta_b^2 + x^2) \nn
&& - \sum_{ks}^{'} \left( \tanh \frac{\beta E^f_{ks}}{2} \right) \left( \frac{\partial E^f_{ks}}{\partial \Delta_f} \right)  \nn
&& + \sum_{ks}^{'} \left( \frac{1}{e^{\beta E_{ks}^{b}}-1} + \frac{1}{2} \right) \left( \frac{\partial E^b_{ks}}{\partial \Delta_f} \right) =0. \label{u1_self3} 
\eqa
For fixed numbers of spinon and holon at a given hole concentration, we obtain, for the chemical potentials, $\mu^f$ and $\mu^b$,
\bqa
\frac{\partial F_{U(1)}}{\partial \mu^{f}} & = & \sum_{k,s=\pm 1}^{'} \left( \tanh \frac{\beta E^f_{ks}}{2} \right) \left( \frac{ \epsilon_{ks}^{f} - \mu^f } { E_{ks}^f} \right) - Nx = 0, \label{u1_d_muf_F} \\
\frac{\partial F_{U(1)}}{\partial \mu^{b}} & = & -\sum_{k,s=\pm 1}^{'} \Bigl[ \frac{1}{e^{\beta E_{ks}^{b}}-1} \frac{\epsilon_{ks}^{b} - \mu^b}{E_{ks}^{b}} \nn
&& + \frac{\epsilon_{ks}^{b} - \mu^b -E_{ks}^{b}}{2E_{ks}^{b}} \Bigr] + Nx  = 0. 
\label{u1_d_mub_F}
\eqa
Using the five self-consistent equations of Eqs.(\ref{u1_self1}) through (\ref{u1_d_mub_F}), we determine $\chi$, $\Delta_b$, $\Delta_f$, $\mu^f$ and $\mu^b$ at each doping and temperature.
Both the pseudogap temperature $T^*$ and the superconducting transition (bose condensation) temperature $T_c$ are determined to be the temperatures at which the spin gap $\Delta^f_0$ and the holon pairing energy (gap) $\Delta^b_0$ respectively begin to open.

\section{SU(2) action from the U(1) action}
The t-J Hamiltonian is manifestly invariant under the local SU(2) transformation 
$g_i = e^{i {\bbox \sigma} \cdot {\bbox \theta}}$ 
for both the spinon and holon spinors with
$
\left( \begin{array}{c} f_{i1} \\ f_{i2}^\dagger \end{array} \right) 
= g_i
\left( \begin{array}{c} f_{i\uparrow} \\ f_{i\downarrow}^\dagger \end{array} \right)
$,
$
\left( \begin{array}{c} f_{i2} \\ -f_{i1}^\dagger \end{array} \right) 
= g_i
\left( \begin{array}{c} f_{i\downarrow} \\ -f_{i\uparrow}^\dagger \end{array} \right)
$
and
$
\left( \begin{array}{c} b_{i1} \\ b_{i2} \end{array} \right) 
= g_i
\left( \begin{array}{c} b_{i} \\ 0 \end{array} \right)
$, 
satisfying
$
c_{i \uparrow}  = b_i^\dagger f_{i\uparrow}  
=  ( b_i^\dagger, 0 ) 
\left( \begin{array}{c} f_{i\uparrow} \\ f_{i\downarrow}^\dagger \end{array} \right) 
=  ( b_i^\dagger, 0 ) 
g_i^\dagger g_i
\left( \begin{array}{c} f_{i\uparrow} \\ f_{i\downarrow}^\dagger \end{array} \right)
=  ( b_{i1}^\dagger, b_{i2}^\dagger ) 
\left( \begin{array}{c} f_{i1} \\ f_{i2}^\dagger \end{array} \right)
$
and
$
c_{i \downarrow}  =  b_i^\dagger f_{i\downarrow} 
= ( b_i^\dagger, 0 ) 
\left( \begin{array}{c} f_{i\downarrow} \\ -f_{i\uparrow}^\dagger \end{array} \right) 
= ( b_i^\dagger, 0 ) 
g_i^\dagger g_i
\left( \begin{array}{c} f_{i\downarrow} \\ -f_{i\uparrow}^\dagger \end{array} \right) 
 =  ( b_{i1}^\dagger, b_{i2}^\dagger ) 
\left( \begin{array}{c} f_{i2} \\ -f_{i1}^\dagger \end{array} \right)
$\cite{WEN}.
We introduce additional Lagrange multiplier terms involved with the constraints 
$f_{i \uparrow}^\dagger f_{i \downarrow}^\dagger=0$ and $f_{i \downarrow} f_{i \uparrow}=0$ to write
\widetext
\top{-2.8cm}
\bq
-  i \sum_i \lambda_i  (f_{i\sigma}^{\dagger}f_{i\sigma}+b_{i}^{\dagger}b_{i} -1)
- i \sum_i \lambda_i^{'} f_{i \uparrow}^\dagger f_{i \downarrow}^\dagger
-  i \sum_i \lambda_i^{''} f_{i \downarrow} f_{i \uparrow}
\label{u1_tjmodel_app1}
\eq
in order to allow for SU(2) symmetry. 
Thus writing spinors 
$ \psi_{i1}^{0} = \left( \begin{array}{r}
   f_{i \uparrow} \\
   f_{i \downarrow}^\dagger
\end{array} \right)$ and
$ \psi_{i2}^{0} = \left( \begin{array}{r}
   f_{i \downarrow} \\
   -f_{i \uparrow}^\dagger
\end{array} \right)$ for spinon,
$ h_{i}^{0} = \left( \begin{array}{r}
   b_{i} \\
   0
\end{array} \right)$ for holon
and the three-component Lagrangian multiplier field ${\bf a}_i^{0}$ with $a_i^{0(1)} = \frac{i\lambda^{'} + i\lambda^{''}}{2}$, $a_i^{0(2)} = \frac{-\lambda^{'} + \lambda^{''}}{2}$, $a_i^{0(3)} = i \lambda_i$, 
the U(1) action in Eq.(\ref{u1_action0}) can be rewritten as
\begin{eqnarray}
S_{U(1)}[b,f,a] & = & 
   \int_0^\beta d \tau \Bigl[
\sum_i h_i^{0\dagger} \partial_\tau h^0_i
+ \frac{1}{2} \sum_{i,\alpha} \psi_{i\alpha}^{0\dagger} \partial_\tau \psi_{i\alpha}^0
- t\sum_{<i,j>} \left( ( \psi_{i\alpha}^{0\dagger} h^0_i) ( h_{j}^{0\dagger} \psi_{j\alpha}^0 ) + c.c. \right) \nn
&& -\frac{J}{2} \sum_{<i,j>} h_{i\alpha}^0  h_{j\beta}^0 h_{j\beta}^{0\dagger} h_{i\alpha}^{0\dagger} 
 (f_{i\downarrow}^{\dagger}f_{j\uparrow}^{\dagger}-f_{i\uparrow}^ {\dagger}f_{j\downarrow}^{\dagger})(f_{j\uparrow}f_{i\downarrow}-f_{j\downarrow} f_{i\uparrow}) \nn
&& - \mu \sum_i f_{i\sigma}^{\dagger} f_{i\sigma} ( b_i^\dagger b_i + 1) 
 - \sum_{i} {\bf a}_{i}^0 \cdot  ( \frac{1}{2} \psi_{i\alpha}^{0\dagger} \bbox{\sigma} \psi_{i\alpha}^0 + h_{i}^{0\dagger} \bbox{\sigma}  h_{i}^0 )
 \Bigr].
\label{u1_tjmodel_app}
\end{eqnarray}
Here the fourth term is the Heisenberg interaction term,
$H_J = -\frac{J}{2} \sum_{<i,j>} 
b_{i} b_{j} b_{j}^{\dagger} b_{i}^{\dagger}  
 (f_{i\downarrow}^{\dagger}f_{j\uparrow}^{\dagger}-f_{i\uparrow}^ {\dagger}f_{j\downarrow}^{\dagger})(f_{j\uparrow}f_{i\downarrow}-f_{j\downarrow} f_{i\uparrow})$.

We rewrite the spinon part of the Heisenberg interaction,
\bqa
&& -\frac{J}{2} (f_{i\downarrow}^{\dagger}f_{j\uparrow}^{\dagger}-f_{i\uparrow}^ {\dagger}f_{j\downarrow}^{\dagger})(f_{j\uparrow}f_{i\downarrow}-f_{j\downarrow} f_{i\uparrow}) \nn
& = & \frac{J}{4} \Bigl[  \sum_{k=1}^{3} (f_{i\alpha}^{\dagger} \sigma_{\alpha \beta}^k f_{i\beta}) 
	       (f_{j\gamma}^{\dagger} \sigma_{\gamma \delta}^k f_{j\delta})
-  (f_{i\alpha}^{\dagger}f_{i\alpha})(f_{j\beta}^{\dagger}f_{j\beta}) \Bigr] \nn
& = & 
\frac{J}{4} \Bigl[  
      \frac{1}{4}
      \left(tr \Psi_i^{0\dagger} \Psi_i^0 (\sigma^k)^T \right)
      \left(tr \Psi_j^{0\dagger} \Psi_j^0 (\sigma^k)^T \right)
      - (f_{i\alpha}^{\dagger}f_{i\alpha})(f_{j\beta}^{\dagger}f_{j\beta})
  \Bigr],
  \label{c6}
\eqa
where $\Psi_i^{0} \equiv \left( \begin{array}{cc} 
f_{i \uparrow} & f_{i \downarrow} \\
f_{i \downarrow}^\dagger & -f_{i \uparrow}^\dagger 
\end{array} \right)$
and 
$(f_{i\alpha}^{\dagger} \sigma_{\alpha \beta}^k f_{i\beta}) = \frac{1}{2} tr \left( \Psi_i^{0\dagger} \Psi_i^0 (\sigma^k)^T \right)$\cite{AFFLECK}. 
Here $(\sigma^k)^T$ denotes the transpose of the Pauli matrices for $k=1,2,3$.

Realizing
$h_i = g_i h_i^0$ 
and
$\Psi_i = 
\left( \begin{array}{cc} 
f_{i 1} & f_{i 2} \\
f_{i 2}^\dagger & -f_{i 1}^\dagger 
\end{array} \right)
=
g_i 
\left( \begin{array}{cc} 
f_{i \uparrow} & f_{i \downarrow} \\
f_{i \downarrow}^\dagger & -f_{i \uparrow}^\dagger 
\end{array} \right)$, 
and using Eq.(\ref{c6}), the SU(2) symmetric Heisenberg interaction term is given by
\bqa
H^{SU(2)}_J
& = &
 \frac{J}{4} \sum_{<i,j>} 
 (1 + h_i^{\dagger} h_i ) ( 1 + h_j^{\dagger} h_j ) 
 \left[
     \frac{1}{4} 
      \left(tr \Psi_i^{\dagger}  \Psi_i (\sigma^k)^T \right)
      \left(tr \Psi_j^{\dagger}  \Psi_j (\sigma^k)^T \right)
 - (f_{i\alpha}^{\dagger}f_{i\alpha})(f_{j\beta}^{\dagger}f_{j\beta})
\right] \nn
& = &
-\frac{J}{2} \sum_{<i,j>} 
 ( 1 + h_i^{\dagger} h_i ) ( 1 + h_j^{\dagger} h_j )
(f_{i2}^{\dagger}f_{j1}^{\dagger}-f_{i1}^ {\dagger}f_{j2}^{\dagger})
(f_{j1}f_{i2}-f_{j2} f_{i1}).
 \label{c8}
\eqa

Taking decomposition of the Heisenberg interaction term above into terms involving charge and spin fluctuations separately, uncorrelated mean field contributions and correlated fluctuations, i.e., correlations between charge and spin fluctuations as in the U(1) case, the SU(2) action is rewritten,
\bqa
&& S[b_{\alpha},f_{\alpha},\lambda_i]  =  \int_0^\beta d \tau \Bigl[
\sum_{i,\alpha=1,2} ( b_{i\alpha}^\dagger \partial_\tau b_{i\alpha} + f_{i\alpha}^\dagger \partial_\tau f_{i\alpha} ) 
+ H^{SU(2)}_{t-J} \Bigr],
\eqa
where
\bqa
H^{SU(2)}_{t-J} &=& - \frac{t}{2} \sum_{<i,j>}  \Bigl[ (f_{i \alpha}^{\dagger}f_{j \alpha})(b_{j1}^{\dagger}b_{i1}-b_{i2}^{\dagger}b_{j2}) + c.c. \nn 
&& + (f_{i2}f_{j1}-f_{i1}f_{j2}) (b_{j1}^{\dagger}b_{i2} + b_{i1}^{\dagger}b_{j2}) + c.c. \Bigr] \nn 
&& -\frac{J}{2} \sum_{<i,j>} \Bigl[
\Bigl< (f_{i2}^{\dagger}f_{j1}^{\dagger}-f_{i1}^ {\dagger}f_{j2}^{\dagger})(f_{j1}f_{i2}-f_{j2} f_{i1}) \Bigr> ( 1 + h_i^\dagger h_i) ( 1 +  h_j^{\dagger}h_j ) \nn
&& + \Bigl< ( 1 + h_i^\dagger h_i) ( 1 +  h_j^{\dagger}h_j ) \Bigr> (f_{i2}^{\dagger}f_{j1}^{\dagger}-f_{i1}^ {\dagger}f_{j2}^{\dagger})(f_{j1}f_{i2}-f_{j2} f_{i1})  \nn
 & & -  \Bigl< ( 1 + h_i^\dagger h_i) ( 1 +  h_j^{\dagger}h_j ) \Bigr> \Bigl< (f_{i2}^{\dagger}f_{j1}^{\dagger}-f_{i1}^ {\dagger}f_{j2}^{\dagger})(f_{j1}f_{i2}-f_{j2} f_{i1}) \Bigr> \nn
  & & + \Bigl( ( 1 + h_i^\dagger h_i) ( 1 +  h_j^{\dagger}h_j ) - \Bigl< ( 1 + h_i^\dagger h_i) ( 1 +  h_j^{\dagger}h_j ) \Bigr> \Bigr) \times \nn
  && \Bigl( (f_{i2}^{\dagger}f_{j1}^{\dagger}-f_{i1}^ {\dagger}f_{j2}^{\dagger} ) ( f_{j1}f_{i2}-f_{j2} f_{i1}) - \Bigl<(f_{i2}^{\dagger}f_{j1}^{\dagger}-f_{i1}^ {\dagger}f_{j2}^{\dagger})(f_{j1}f_{i2}-f_{j2} f_{i1})\Bigr> \Bigr) \Bigr] \nn
&& - \mu \sum_i ( 1 - h_i^{\dagger} h_i ) \nn
&& -  \sum_i  \Bigl(
 i\lambda_{i}^{(1)} ( f_{i1}^{\dagger}f_{i2}^{\dagger} + b_{i1}^{\dagger}b_{i2}) 
+ i \lambda_{i}^{(2)} ( f_{i2}f_{i1} + b_{i2}^\dagger b_{i1} ) \nn
 && + i \lambda_{i}^{(3)} ( f_{i1}^{\dagger}f_{i1} -  f_{i2} f_{i2}^{\dagger} + b_{i1}^{\dagger}b_{i1} - b_{i2}^{\dagger}b_{i2} ) \Bigr).
\label{su2_tjmodel2}
\eqa
\bottom{-2.8cm}
\narrowtext
\noindent

\section{SU(2) mean-field Hamiltonian}
The intersite holon interaction (the third term in Eq.(\ref{su2_tjmodel2})) is rewritten,
\bqa
&& -\frac{J}{2} \Bigl< (f_{i2}^{\dagger}f_{j1}^{\dagger}-f_{i1}^ {\dagger}f_{j2}^{\dagger})(f_{j1}f_{i2}-f_{j2} f_{i1}) \Bigr> ( 1 + h_i^\dagger h_i) ( 1 +  h_j^{\dagger}h_j ) \nn
&& =  -\frac{J}{2} <|\Delta^f_{ij}|^2> ( 1 + h_i^\dagger h_i +  h_j^{\dagger}h_j + h_i^\dagger h_i h_j^{\dagger}h_j ) \nn
&& \approx  -\frac{J}{2} |\Delta^f_{ij}|^2 ( 1 + b_{i\alpha}^\dagger b_{i\alpha} +  b_{j\alpha}^\dagger b_{j\alpha}  + b_{i\alpha}^{\dagger}b_{j\beta}^{\dagger}b_{j\beta}b_{i\alpha}  ) \nn
\label{eq:su2_mf_holon_result}
\eqa
where $\Delta^f_{ij} = \Bigl<f_{j1}f_{i2}-f_{j2} f_{i1} \Bigr>$ is the spinon singlet pairing order parameter.
The intersite spinon interaction (the fourth term in Eq.(\ref{su2_tjmodel2})) is rewritten in terms of decomposed Hatree-Fock-Bogoliubov channels in the same way as in the U(1) case, 
\bq
-\frac{J_p}{2} (f_{i2}^{\dagger} f_{j1}^{\dagger}-f_{i1}^ {\dagger}f_{j2}^{\dagger})(f_{j1}f_{i2}-f_{j2} f_{i1})  = v_D + v_E + v_P,
\label{eq:mf_spinon_spinon} 
\eq
where $v_D$, $v_E$ and  $v_P$ are the interactions corresponding to the direct, exchange and pairing channels respectively,
\bqa
v_D & = & -\frac{J_p}{8} \sum_{l=0}^{3} ( f_i^{\dagger} \sigma^l f_i ) ( f_j^{\dagger} \sigma f_j ), \label{su2_vD} \\
 v_E & = & -\frac{J_p}{4} \Bigl( (f_{i\sigma}^{\dagger}f_{j\sigma})(f_{j\sigma}^{\dagger}f_{i\sigma}) - n_i \Bigr), \label{su2_vE} \\
 v_P & = & -\frac{J_p}{2} (f_{i2}^{\dagger}f_{j1}^{\dagger}-f_{i1}^{\dagger}f_{j2}^{\dagger}) (f_{j1}f_{i2}-f_{j2}f_{i1}). \label{su2_vP}
\eqa
Here $\sigma^0$ is the unit matrix and $\sigma^{1,2,3}$, the Pauli spin matrices.
The fifth term in Eq.(\ref{su2_tjmodel2}) is written,
\bqa
&& \frac{J}{2} \Bigl<  ( 1 + h_i^\dagger h_i) ( 1 +  h_j^{\dagger}h_j ) \Bigr> \times \nn
&& \Bigl< (f_{i2}^{\dagger}f_{j1}^{\dagger}-f_{i1}^ {\dagger}f_{j2}^{\dagger}) (f_{j1}f_{i2}-f_{j2} f_{i1}) \Bigr> \nn
& \approx & \frac{J}{2} \Bigl<  ( 1 + h_i^\dagger h_i) \Bigr> \Bigl< ( 1 +  h_j^{\dagger}h_j) \Bigr> \times \nn
&& \Bigl< (f_{i2}^{\dagger}f_{j1}^{\dagger}-f_{i1}^ {\dagger}f_{j2}^{\dagger}) \Bigr> \Bigl< (f_{j1}f_{i2}-f_{j2} f_{i1}) \Bigr>  \nn
& = & \frac{J}{2} |\Delta^f_{ij}|^2 ( 1 + < h_i^\dagger h_i > + < h_j^{\dagger} h_j > + < h_i^\dagger h_i > < h_j^{\dagger}h_j > ). \nn
\label{eq:su2_mf_const}
\eqa
We introduced $\Bigl< (f_{i2}^{\dagger}f_{j1}^{\dagger}-f_{i1}^ {\dagger}f_{j2}^{\dagger})(f_{j1}f_{i2}-f_{j2} f_{i1}) \Bigr> \approx \Bigl< (f_{i2}^{\dagger}f_{j1}^{\dagger}-f_{i1}^ {\dagger}f_{j2}^{\dagger}) \Bigr> \Bigl< (f_{j1}f_{i2}-f_{j2} f_{i1}) \Bigr> = |\Delta^f_{ij}|^2$ and $ \Bigl< ( 1 + h_i^\dagger h_i) ( 1 +  h_j^{\dagger}h_j ) \Bigr> \approx \Bigl< ( 1 + h_i^\dagger h_i)\Bigr> \Bigl< ( 1 +  h_j^{\dagger}h_j ) \Bigr>$.

To obtain the effective Hamiltonian, we introduce the Hubbard-Stratonovich fields, $\Delta^b_{ij\alpha\beta}$ concerned with the holon pairing channel in Eq.(\ref{eq:su2_mf_holon_result}), and ${\rho}_{i}^{k}$, $\chi_{ij}$ and $\Delta_{ij}$ concerned with the spinon direct, exchange and pairing channels. 
We perform the Hubbard-Stratonovich transformation for the holon pairing interaction,
\widetext
\top{-2.8cm}
\bqa
&& \exp \left( \frac{J}{2} |\Delta^f_{ij}|^2 b_{i\alpha}^{\dagger}b_{j\beta}^{\dagger}b_{j\beta}b_{i\alpha} \right)  \nn
& \propto & \int  \Pi_{\alpha,\beta} d \Delta_{ij\alpha\beta}^{b*} d \Delta_{ij\alpha\beta}^b \exp 
\left(  - \frac{J}{2} |\Delta^f_{ij}|^2 \sum_{\alpha,\beta} ( |\Delta^b_{ij\alpha\beta}|^2 
- \Delta_{ij\alpha\beta}^{b*} b_{i\alpha} b_{j\beta} - \Delta_{ij\alpha\beta}^{b} b_{j\alpha}^\dagger b_{i\beta}^\dagger ) \right). 
\label{hs_su2_p2}
\eqa
After saddle point approximation, we obtain, 
\bq
H^P_{SU(2)} = \frac{J}{2} |\Delta^f_{ij}|^2 \sum_{\alpha,\beta} ( |\Delta^b_{ij\alpha\beta}|^2 
- \Delta_{ij\alpha\beta}^{b*} b_{i\alpha} b_{j\beta} - \Delta_{ij\alpha\beta}^{b} b_{j\alpha}^\dagger b_{i\beta}^\dagger ),
\label{hs_su2_p3}
\eq
with $\Delta_{ij\alpha\beta} = < b_{i\alpha} b_{j\beta} >$, the saddle point of the holon pairing order parameter.

By introducing the Hubbard-Stratonovich fields, 
${\rho}_{i}^{k}$, $\chi_{ij}$ and $\Delta_{ij}$ for the spinon direct, exchange and pairing order shown in Eqs.(\ref{su2_vD}), (\ref{su2_vE}), (\ref{su2_vP}),
we rewrite the effective Hamiltonian for Eq.(\ref{su2_tjmodel2}), 
\begin{eqnarray}
&& H_{SU(2)}^{MF}  = 
\frac{J_p}{4} \sum_{<i,j>} \Bigl[ |\chi_{ij}|^2 - \chi_{ij}^* \{ f_{i \sigma}^{\dagger}f_{ \sigma j}  + \frac{2t}{J_p} (b_{i1}^{\dagger}b_{j1}-b_{j2}^{\dagger}b_{2i }) \}  - c.c. \Bigr] \nn
& + &  \frac{J}{2} \sum_{<i,j>} |\Delta^f_{ij}|^2 \Bigl[ \sum_{\alpha,\beta} ( |\Delta^b_{ij\alpha\beta}|^2 - \Delta_{ij\alpha\beta}^{b*} b_{i\alpha} b_{j\beta} - c.c.) \Bigr] \nn
 & + & \frac{J_p}{2} \sum_{<i,j>} \Bigl[ |\Delta_{ij}|^2 - \Delta_{ij} \{ (f_{i2}^{\dagger}f_{j1}^{\dagger}-f_{i1}^{\dagger}f_{j2}^{\dagger}) - \frac{t}{J_p} (b_{j1}^{ \dagger}b_{i2} + b_{i1}^{\dagger}b_{j2}) \}  - c.c. \Bigr]  \nn
&+&  \frac{J_p}{2} \sum_{<i,j>} \sum_{l=0}^{3} \Bigl( (\rho^l_{ij})^2 - \rho^l_{ij} ( f_i^{\dagger} \sigma^l f_i ) \Bigr) \nn
& + &  \frac{t^2}{J_p}  \sum_{<i,j>} \Bigl[ (b_{i1}^{\dagger}b_{j1}-b_{j2}^{\dagger}b_{i2}) (b_{j1}^{\dagger}b_{i1}-b_{i2}^{\dagger}b_{j2})  + \frac{1}{2} (b_{j1}^{\dagger}b_{i2}+b_{i1}^{\dagger}b_{j2}) (b_{i2}^{\dagger}b_{j1} + b_{j2}^{\dagger}b_{i1}) \Bigr]  \nn
& - & \frac{J}{2} \sum_{<i,j>} |\Delta^f_{ij}|^2 \Bigl[  h_j^{\dagger} h_j + h_{i}^\dagger h_{i} - <h_j^{\dagger} h_j > - <h_{i}^\dagger h_{i} >   \Bigr]  + \frac{J}{2} \sum_{<i,j>} |\Delta^f_{ij}|^2 x^2 \nn
& + & \frac{J_p}{2} \sum_{i} (f_{i \sigma}^\dagger f_{i \sigma})
- \mu \sum_i ( 1 - h_i^{\dagger} h_i ) \nn
& - & \sum_i  \Bigl[
 i \lambda_{i}^{1} ( f_{i1}^{\dagger}f_{i2}^{\dagger} + b_{i1}^{\dagger}b_{i2})
+ i \lambda_{i}^{2} ( f_{i2}f_{i1} + b_{i2}^\dagger b_{i1} ) + i \lambda_{i}^{3} ( f_{i1}^{\dagger}f_{i1} -  f_{i2} f_{i2}^{\dagger} + b_{i1}^{\dagger}b_{i1} - b_{i2}^{\dagger}b_{i2} )], 
\label{su2_mf_hamiltonian1}
\end{eqnarray}
\bottom{-2.8cm}
\narrowtext
\noindent
where 
$\chi_{ij} = < f_{i\sigma}^{\dagger}f_{j\sigma} + \frac{2t}{J_p} (b_{i1}^{\dagger}b_{j1}-b_{j2}^{\dagger}b_{i2}) >$,
$\Delta_{ij} = \Bigl< (f_{i1}f_{j2} - f_{i2}f_{j1}) - \frac{t}{J_p} ( b_{i2}^\dagger b_{j1} + b_{j2}^\dagger b_{i1} ) \Bigr> = \Delta_{ij}^f - \frac{t}{J(1-x)} \chi_{ij;12}^b$, with $\chi_{ij;12}^b = \Bigl< b_{i2}^\dagger b_{j1} + b_{j2}^\dagger b_{i1} \Bigr>$ and $x$, the hole doping rate.

The second term in Eq.(\ref{su2_mf_hamiltonian1}) represents the pairing energy term for Cooper pair as a composite of the holon pair and spinon pair.
The scalar boson field, $\Delta_{ij ;\alpha \beta}^b$ represents the holon pairing order parameter for the nearest neighbor intersite holons, namely the $b_{\alpha}$ and $b_{\beta}$ single holons with the holon index, $\alpha, \beta =$ $1$ or $2$\cite{WEN}.
To simplify the effective Hamiltonian, we rearrange each term in Eq.(\ref{su2_mf_hamiltonian1}).

For the paramagnetic state, we obtain $\rho^l_{i} = \frac{1}{2} ( f_i^{\dagger} \sigma^l f_i ) = < S^l_i > = 0 $ for $l=1,2,3$ and $\rho^0_{i} = \frac{1}{2} < f_{i \sigma}^\dagger f_{i \sigma} > = \frac{1}{2}$  for $l=0$. 
The one-body holon term (the sixth term) in the above Hamiltonian is incorporated into the effective chemical potential term.
By setting $\Delta_{ij} = \Delta_{ij}^f - \frac{t}{J(1-x)} \chi_{ij;12}^b$ where $\Delta_{ij}^f = < f_{i1}f_{j2} - f_{i2}f_{j1}>$ and $\chi_{ij;12}^b = \Bigl< b_{i2}^\dagger b_{j1} + b_{j2}^\dagger b_{i1} \Bigr>$,
we rearrange the third term and the second term in the bracket of the fifth term to obtain the effective Hamiltonian,
\begin{eqnarray}
&& H^{MF}_{SU(2)} = 
\frac{J}{2} \sum_{<i,j>} |\Delta^f_{ij}|^2 \Bigl[ \sum_{\alpha,\beta} |\Delta_{ij; \alpha \beta}^{b}|^{2} + x^2  \Bigr]   \nn
&& +\frac{J_p}{2} \sum_{<i,j>} \Bigl[ |\Delta_{ij}^{f}|^{2} + \frac{1}{2} |\chi_{ij}|^{2} + \frac{1}{4} \Bigr] \nn
& & -\frac{t}{2} \sum_{<i,j>} \Bigl[ \chi_{ij}^* (b_{i1}^{\dagger}b_{j1} - b_{j2}^{\dagger}b_{i2}) -\Delta^f_{ij} (b_{j1}^{\dagger}b_{i2} + b_{i1}^{\dagger}b_{j2})\Bigr] - c.c. \nn
&& -\frac{J_p}{4} \sum_{<i,j>} \Bigl[ \chi_{ij}^* (f_{i \sigma}^{\dagger}f_{j \sigma}) + c.c. \Bigr]  \nn
&& -\frac{J}{2} \sum_{<i,j>,\alpha,\beta} |\Delta^f_{ij}|^2 \Bigl[ \Delta_{ij;\alpha \beta}^{b*} (b_{i \alpha}b_{j \beta}) + c.c. \Bigr]  \nn
& & -\frac{J_p}{2} \sum_{<i,j>} \Bigl[ \Delta_{ij}^{f*} (f_{j1}f_{i2}-f_{j2}f_{i1}) + c.c. \Bigr] \nn
&& -\sum_{i} \mu^b_{i} ( h_{i}^{\dagger}h_{i} - x ) \nn
&& - \sum_i \Bigl[ i\lambda_{i}^{1} ( f_{i1}^{\dagger}f_{i2}^{\dagger} + b_{i1}^{\dagger}b_{i2}^{\dagger}) + i \lambda_{i}^{2} ( f_{i2}f_{i1} + b_{i2}b_{i1} ) \nn
&& + i \lambda_{i}^{3} ( f_{i1}^{\dagger}f_{i1} -  f_{i2} f_{i2}^{\dagger} + b_{i1}^{\dagger}b_{i1} - b_{i2}^{\dagger}b_{i2} ) \Bigr] \nn
&& - \frac{t}{2} \sum_{<i,j>}  \left( \Delta^{f}_{ij} - (f_{j1}f_{i2}- f_{j2}f_{i1}) \right) \chi_{ij;12}^{b*} - c.c. \nn
&& + \frac{t^2}{2J_p} \sum_{<i,j>}  \left| \chi_{ij;12}^b - (b_{i2}^\dagger b_{j1} + b_{j2}^\dagger b_{i1} ) \right|^2  \nn
&& + \frac{t^2}{J_p}\sum_{<i,j>}  (b_{i1}^{\dagger}b_{j1} - b_{j2}^{\dagger}b_{i2}) ( b_{j1}^{\dagger}b_{i1} - b_{i2}^{\dagger}b_{j2}),
\label{su2_mf_hamiltonian3}
\end{eqnarray}
where $\mu^b_i = -\mu + \frac{J}{2} \sum_{j=i\pm \hat x, i \pm \hat y} |\Delta^f_{ij}|^2$.

We neglect correlations between the fluctuations of order parameters.
This is because correlations between the spin (spinon pair) and charge (holon pair) fluctuations are expected to be small as compared to the saddle point contribution of the order parameters (the first  and second terms in the above Hamiltonian) particularly near the pseudogap temperature and the bose condensation temperature.
However, individual fluctuations of the spinon pairing and holon pairing order parameters are not ignored.
We also neglect the fluctuations of order parameters (ninth and tenth terms) in Eq.(\ref{su2_mf_hamiltonian3}).
The ninth term represents the fluctuation of spinon pairing order parameter and we neglect it owing to its vanishment as an expectation value.
The tenth term represents the fluctuations of holon hopping order parameter and is negligible at low temperature as its fluctuations die out in the low temperature regions where pairing order parameters $\Delta^f$ and $\Delta^b$ begin to open.
Owing to the high energy cost involved with the Coulomb repulsion energy the exchange interaction terms (the last positive energy terms) will be ignored\cite{UBBENS}-\cite{WEN}.
We then obtain the mean field Hamiltonian, $H = H^{\Delta,\chi} + H^f + H^b$,
where $H^{\Delta,\chi}$ is the saddle point contribution of order parameters, $\chi$, $\Delta^f$ and $\Delta^b$,
\bqa
H^{\Delta,\chi}_{SU(2)} & = &  
\frac{J}{2} \sum_{<i,j>} |\Delta^f_{ij}|^2 \Bigl[ \sum_{\alpha,\beta} |\Delta_{ij; \alpha \beta}^{b}|^{2} + x^2  \Bigr] \nn
&& + \frac{J_p}{2} \sum_{<i,j>} \Bigl[ |\Delta_{ij}^{f}|^{2} + \frac{1}{2} |\chi_{ij} |^{2} + \frac{1}{4} \Bigr],
\label{su2_mean}
\eqa
$H^b$ is the holon Hamiltonian,
\bqa
&& H^b_{SU(2)}  =   
-\frac{t}{2} \sum_{<i,j>} \Bigl[ \chi_{ij}^* (b_{i1}^{\dagger}b_{j1} - b_{j2}^{\dagger}b_{i2}) \nn
&& -\Delta^f_{ij} (b_{j1}^{\dagger}b_{i2} + b_{i1}^{\dagger}b_{j2})\Bigr] - c.c. \nn
&& - \frac{J}{2} \sum_{<i,j>,\alpha,\beta} |\Delta^f_{ij}|^2 \Bigl[ \Delta_{ij;\alpha \beta}^{b*} (b_{i \alpha}b_{j \beta}) + c.c. \Bigr]  \nn
&& -\sum_{i} \Bigl[ \mu^b_{i} ( h_{i}^{\dagger}h_{i} - x ) + i\lambda_{i}^{1} ( b_{i1}^{\dagger}b_{i2}^{\dagger})  + i \lambda_{i}^{2} ( b_{i2}b_{i1} ) \nn
&& + i \lambda_{i}^{3} ( b_{i1}^{\dagger}b_{i1} - b_{i2}^{\dagger}b_{i2} ) \Bigr],
\label{su2_holon_sector}
\eqa
and $H^f$, the spinon Hamiltonian,
\bqa
&& H^f_{SU(2)}  =  
-  \frac{J_p}{4} \sum_{<i,j>} \Bigl[ \chi_{ij}^* (f_{i \sigma}^{\dagger}f_{j \sigma}) + c.c. \Bigr]  \nn
&& -\frac{J_p}{2} \sum_{<i,j>} \Bigl[ \Delta_{ij}^{f*} (f_{j1}f_{i2}-f_{j2}f_{i1}) + c.c. \Bigr] \nn
&& - \sum_i \Bigl[ i\lambda_{i}^{1} ( f_{i1}^{\dagger}f_{i2}^{\dagger} )  + i \lambda_{i}^{2} ( f_{i2}f_{i1} ) + i \lambda_{i}^{3} ( f_{i1}^{\dagger}f_{i1} - f_{i2} f_{i2}^{\dagger} ) \Bigr], \nn
\label{su2_spinon_sector}
\eqa
where $\chi_{ij}= < f_{i \sigma}^{\dagger}f_{j \sigma} + \frac{2t}{J_p} (b_{i1}^{\dagger}b_{j1} - b_{j2}^\dagger b_{i2} )>$,
$\Delta_{ij;\alpha\beta}^{b} = <b_{i\alpha}b_{j \beta}>$,
$\Delta_{ij}^{f}=< f_{j1}f_{i2}-f_{j2}f_{i1} >$ and
$\mu^b_i = -\mu - \frac{J}{2} \sum_{j=i\pm \hat x, i \pm \hat y} |\Delta^f_{ij}|^2$.

We take the uniform order (site-independent) parameters : $\chi_{ij}=\chi$ for hopping, $ \Delta_{ij}^{f}=\pm \Delta_f$ for d-wave spinon pairing with the sign $+(-)$ for the nearest neighbor link parallel to $\hat x$ ($\hat y$) and $\Delta_{ij;\alpha \beta}^{b}=\Delta^b_{ \alpha \beta}$ for s-wave holon pairing, with $\alpha$ and $\beta=1,2$.
For the case of $\Delta^b_{\alpha \beta}=0$, $\lambda^{(k)}=0$ and $\Delta^f \leq \chi$, the $b_1$-bosons are populated at and near $k=(0,0)$ in the momentum space and the $b_2$-bosons, at and near $k=(\pi,\pi)$.
Pairing of two different($\alpha \neq \beta$) bosons(holons) gives rise to the non-zero center of mass momentum.
On the other hand, the center of mass momentum is zero only for pairing between identical($\alpha = \beta$) bosons.
Thus we write $\Delta^b_{ \alpha \beta} = \Delta_b ( \delta_{\alpha,1}\delta_{\beta,1} \pm \delta_{\alpha,2} \delta_{\beta,2} )$\cite{WEN,LEE} for in-phase ($+$ sign) and out-of-phase ($-$) pairing order only between holons, $b_1$ and $b_2$.
Taking the uniform chemical potential, $\mu^b_{i}=\mu^b$ and the uniform Lagrangian multipliers, $\lambda^l_i = \lambda^l$ for $l=1,2,3$, the mean field Hamiltonian is diagonalized.

The quasiparticle Hamiltonian is obtained to be, as shown in Eq.(\ref{su2_diagonalized_hamiltonian_app}) of Appendix B,
\begin{eqnarray}
H^{MF}_{SU(2)}  & = & 
NJ\Delta_f^2 ( 2\Delta_b^2 + x^2 ) 
+ N J_p \Bigl( \frac{1}{2}\chi^{2} + \Delta_f^{2} + \frac{1}{4} \Bigr) \nn
& + & \sum_{k} E_{k}^{f} (\alpha_{k1}^{\dagger}\alpha_{k1} - \alpha_{k2}\alpha_{k2}^{\dagger}) \nn
& + &  \sum_{k,\alpha=1,2} \Bigl[ E_{k\alpha}^{b} \beta_{k\alpha}^{\dagger}  \beta_{k\alpha}  +  \frac{1}{2}( E_{k\alpha}^b + \mu^b ) \Bigr] + \mu^b N x.
\label{su2_diagonalized_hamiltonian}
\end{eqnarray}
Here the spinon quasiparticle (quasispinon) energy is given by
\bqa
E_{k}^{f}  & = &  \sqrt{(\epsilon_{k}^{f}-a^{(3)} )^{2} + ( \Delta_0^{f} )^2 }
\label{su2_spinon_energy}
\eqa
with the spinon pairing energy 
\bq
\Delta^f_{0} = \sqrt{ \left( J_p \Delta_f \varphi_k  - a^{(1)} \right)^2 + (a^{(2)})^2 }. \nonumber
\eq
The holon quasiparticle (quasiholon) energy is, 
\bqa
E_{k1}^{b}  & =&   \sqrt{ (E_k^b -  \mu^b )^2 - ( \Delta^{b1}_{0} )^{2} }, \nn
E_{k2}^{b}  & =&   \sqrt{ (-E_k^b -  \mu^b )^2 - ( \Delta^{b2}_{0} )^{2} }, 
\label{su2_holon_energy}
\eqa
with the holon pairing energies,
\bqa
\Delta^{b1}_{0} & = & \sqrt{ ( \Delta_b^{'} )^2 + 2 |\mu^b| E_k^b - 2 \sqrt{ (\mu^b E_{k}^{b})^2 - ( \Delta_b^{'} )^{2} ( \Delta_f^{''} - a^{(1)})^2} }, \nn
\Delta^{b2}_{0} & = & \sqrt{ ( \Delta_b^{'} )^2 - 2 |\mu^b| E_k^b + 2 \sqrt{ (\mu^b E_{k}^{b})^2 - ( \Delta_b^{'} )^{2} ( \Delta_f^{''} - a^{(1)})^2} } \nn
\eqa
for the out-of-phase (the phase difference $\pi$ between the $b_1$ holon and $b_2$ holon) holon pair order parameter, $\Delta^b_{ \alpha \beta} = \Delta_b ( \delta_{\alpha,1}\delta_{\beta,1} - \delta_{\alpha,2} \delta_{\beta,2} )$, and the holon pairing energies,
\bqa
\Delta^{b1}_{0} & = & \sqrt{ ( \Delta_b^{'} )^2 + 2 |\mu^b| E_k^b - 2 \sqrt{ (\mu^b E_{k}^{b})^2 - ( \Delta_b^{'} )^{2} ( a^{(2)})^2} }, \nn
\Delta^{b2}_{0} & = & \sqrt{ ( \Delta_b^{'} )^2 - 2 |\mu^b| E_k^b + 2 \sqrt{ (\mu^b E_{k}^{b})^2 - ( \Delta_b^{'} )^{2} ( a^{(2)})^2} }
\eqa
for the in-phase (no phase difference between $b_1$ and $b_2$) holon pairing, $\Delta^b_{ \alpha \beta} = \Delta_b ( \delta_{\alpha,1}\delta_{\beta,1} + \delta_{\alpha,2} \delta_{\beta,2} )$.
Here $\Delta_b^{'} = J \Delta_f^2 \Delta_b \gamma_k$, $\Delta_f^{''} = t  \Delta_f \varphi_k$.
$\epsilon_{k}^{f}$ and $  \epsilon_{k}^{b} $ are 
\begin{eqnarray}
\epsilon_{k}^{f} & = & -\frac{J_p}{2} \chi \gamma_k, \nn
\epsilon_{k}^{b} & = &  -t\chi \gamma_k, \nn 
E_{k}^{b} & = & \sqrt{( \epsilon_{k}^{b} -a^{(3)} )^{2} + ( t\Delta_f \varphi_k - a^{(1)} )^2 + (a^{(2)})^2 },
\end{eqnarray}
with $ \gamma_{k} = (\cos k_{x} + \cos k_{y})$, $\varphi_{k} = ( \cos k_{x} - \cos k_{y})$, $a^{(1)} = i(\lambda^{(1)}+\lambda^{(2)})/2$, $a^{(2)} = (\lambda^{(2)}-\lambda^{(1)})/2$ and  $a^{(3)} = i\lambda^{(3)}$.
$\alpha_{k\alpha}( \alpha_{k\alpha}^{\dagger})$ and $\beta_{k\alpha}(\beta_{k\alpha}^{\dagger})$ are the annihilation(creation) operators of the quasispinons (spinon quasiparticles) and the quasiholons (holon quasiparticles) respectively.

\widetext
\top{-2.8cm}
\section{SU(2) free energy}
We rearrange Eq.(\ref{su2_diagonalized_hamiltonian}) to obtain
\bqa
H^{MF}_{SU(2)} & = & \sum_{k}  E_{k}^{f}(\alpha_{k1}^{\dagger}\alpha_{k1} - \alpha_{k2}\alpha_{k2}^{\dagger}) 
+  \sum_{k, \alpha=1,2}  E_{k\alpha}^{b}  \beta_{k\alpha}^{\dagger}  \beta_{k\alpha} 
+  H_{c},
\label{eq:diagonalized_hamiltonian_su2_2}
\eqa
where $H_{c} = 
NJ \Delta_f^2 \Bigl( 2 \Delta_b^{2} + x^2 \Bigr) 
+ NJ^{'} \Bigl( \Delta_f^{2} + \frac{\chi^{2}}{2} + \frac{1}{4} \Bigr)  
+ \sum_{k} \frac{ E_{k1}^{b} + E_{k2}^{b}}{2} + N \mu^b (1+x)$. 

The partition function is written, 
\bqa
Z & = & tr \exp \Bigl( -\beta H^{MF} \Bigr) \nn
&=& \exp (-\beta H_c ) \prod_{k} 
( 2 \cosh \frac{\beta E_{k}^f}{2} )^2 (1-e^{-\beta E_{k1}^b})^{-1} (1-e^{-\beta E_{k2}^b})^{-1}.
\label{eq:partition_function_su2}
\eqa
From Eq.(\ref{eq:partition_function_su2}), we obtain the total free energy, 
\bqa
F_{SU(2)} & = &  
NJ \Delta_f^2 \Bigl( 2 \Delta_b^{2} + x^2 \Bigr) 
+ NJ^{'} \Bigl( \Delta_f^{2} + \frac{\chi^{2}}{2} + \frac{1}{4} \Bigr)  \nn
&& - 2k_{B}T \sum_{k} \ln ( \cosh (\beta E_{k}^{f}/2) ) - 2N k_{B}T \ln2  \nn
&& + k_{B}T \sum_{k\alpha=1,2} \ln (1 - e^{-\beta E_{k\alpha}^{b}}) + \sum_{k,\alpha=1,2} \frac{ E_{k\alpha}^{b}+\mu^b}{2} + N x \mu^b.
\label{eq:free_energy_su2}
\eqa
\bottom{-2.8cm}
\narrowtext
\noindent

From the diagonalized Hamiltonian Eq.(\ref{su2_diagonalized_hamiltonian}), we readily obtain the total free energy, 
\begin{eqnarray}
&& F_{SU(2)}  =   
NJ\Delta_f^2 ( 2 \Delta_b^{2} + x^2 ) 
+ NJ_p \Bigl( \Delta_f^{2} + \frac{1}{2}\chi^{2} + \frac{1}{4} \Bigr)  \nn
&& - 2k_{B}T \sum_{k} \ln [ \cosh (\beta E_{k}^{f}/2) ] - 2 N k_B T \ln 2 \nn
&& + k_{B}T \sum_{k,\alpha=1,2} ln [1 - e^{-\beta E_{k\alpha}^{b}}] 
 + \sum_{k,\alpha=1,2} \frac{ E_{k\alpha}^{b} + \mu^b }{2}  + N x \mu^b. \nn
\label{su2_free_energy}
\end{eqnarray}

The self-consistent equations for the amplitudes of order parameters ($\chi$, $\Delta_b$ and $\Delta_f$) are
\bqa
\frac{\partial F_{SU(2)}}{\partial \chi}  & = &
N J_p \chi  - \sum_{k} \left( \tanh \frac{\beta E^f_{k}}{2} \right) \left(
 \frac{\partial E^f_{k}}{\partial \chi} \right) \nn
 && + \sum_{k\alpha} \left( \frac{1}{e^{\beta E_{k\alpha}^{b}}-1} + \frac{1}{2} \right)
\left( \frac{\partial E^b_{k\alpha}}{\partial \chi} \right) =0, \label{su2_self1} \\
\frac{\partial F_{SU(2)}}{\partial \Delta_b} & = &
4 N J \Delta_f^2 \Delta_b \nn
&& + \sum_{k, \alpha} 
\left( \frac{1}{e^{\beta E_{k\alpha}^{b}}-1 } + \frac{1}{2} \right) 
\left( \frac{\partial E^b_{k\alpha}}{\partial \Delta_b} \right) =0, \label{su2_self2} \\
\frac{\partial F_{SU(2)}}{\partial \Delta_f} & = &
2 N J_p \Delta_f + 2N \Delta_f (2 \Delta_b^2 + x^2) \nn
&& - \sum_{k} \left( \tanh \frac{\beta E^f_{k}}{2} \right) 
\left( \frac{\partial E^f_{k}}{\partial \Delta_f} \right) \nn
&& + \sum_{k, \alpha} \left( \frac{1}{e^{\beta E_{k\alpha}^{b}}-1} + \frac{1}{2} \right) 
\left( \frac{\partial E^b_{k\alpha}}{\partial \Delta_f} \right) =0. \label{su2_self3} 
\eqa
The self-consistent equations for the holon chemical potential ($\mu^b$) and the Lagrangian multipliers ($a^{(l)}$) are 
\bqa
 \frac{\partial F_{SU(2)}}{\partial \mu^b}  & =   &
 \sum_{k,\alpha} \Bigl[  \frac{1}{e^{\beta E_{k\alpha}^b} -1} 
\frac{ \partial E_{k\alpha}^b} {\partial \mu^b}  \nn
&& + \frac{1}{2}( \frac{\partial E_{k\alpha}^b}{\partial \mu^b} + 1 ) \Bigr] + N x = 0, \label{su2_mu_saddle} \label{su2_mub} \\
\frac{\partial F_{SU(2)}}{ \partial a^{(k)} }  & =  &
-\sum_k \tanh \frac{ \beta E_k^f }{2} \frac{ \partial E_k^f }{ \partial a^{(k)} } \nn
&& + \sum_{k,\alpha} \frac{ e^{\beta E_{k\alpha}^b} + 1 }{ 2(e^{\beta E_{k\alpha}^b}-1) } \frac{ \partial E_{k\alpha}^b }{ \partial a^{(k)} } = 0 , \mbox{ $k=1,2,3$}. \label{constraint_eq}
\eqa
It can be readily proven from Eq.(\ref{constraint_eq}) that $a^{(k)}=0$ satisfies the three constraints above by showing that $F(a^{(1)},a^{(2)},a^{(3)})$ is an even function of $a^{(i)}$, 
\bqa
F(a^{(i)})  & = &  F(0) + \frac{1}{2} \sum_{i,j} \left. \frac{\partial^2 F}{\partial a^{(i)} a^{(j)}} \right|_{a^{(i)}=0} a^{(i)} a^{(j)} \nn
&& + O\left( (a^{(i)})^4 \right).
\eqa

From the remaining four self-consistent equations in Eqs.(\ref{su2_self1}) through (\ref{su2_self3}) and (\ref{su2_mub}), we determine $\chi$, $\Delta_b$, $\Delta_f$ and $\mu^b$ at each temperature and doping.
Both $T^*$ and $T_c$ are then determined to be the temperatures at which the spin gap $\Delta^f_0$ and the holon pairing energy (gap) $\Delta^b_0$ respectively begin to open.

As mentioned before, we consider the two cases of the holon pair order parameter, $\Delta^b_{ \alpha \beta} = \Delta_b ( \delta_{\alpha,1}\delta_{\beta,1} \pm \delta_{\alpha,2} \delta_{\beta,2} )$.
There is no difference in the resulting phase diagram whether we choose $+$ or $-$ for the relative sign between the $b_1$ and $b_2$ pairing.
This is because the quasiparticle energy dispersions become identical upto order $k^2$ near the bottom of the quasiparticle spectrum irrespective of the sign.
At low temperature, the holons (bosons) are largely populated at and near the bottom of the energy.

\section{Predicted results of physical properties for high $T_c$ cuprates}
The holon pair bose condensation is taken care of by neglecting the phase fluctuations and thus by taking phase coherence of the holon pairing order parameter.
In Fig. 1 the predicted phase diagrams from the U(1) theory (dotted lines) and the SU(2) theory (solid lines) are displayed for $J/t =0.3$.
Encouragingly, in agreement with observations\cite{ODA,DING,WALSTEDT,YASUOKA,ISHIDA,JULIEN,KENDZIORA,KANG,MOMONO_IDO} the spin gap temperature (pseudogap temperature) $T^*$ tangentially meets the superconducting transition temperature $T_c$ in the overdoped region.
This is attributed to the effect of coupling between the spin and charge degrees of freedom.
To be more specific, as shown in the pairing energy term, 
$-\frac{J}{2}|\Delta^f_{ij}|^2 b_{i}^\dagger b_{j}^\dagger b_{j} b_{i}$ for the U(1) theory and 
$-\frac{J}{2}|\Delta^f_{ij}|^2 b_{i \alpha}^\dagger b_{j \beta}^\dagger b_{j \beta} b_{i \alpha}$ for the SU(2) theory, 
holon pairing strength depends on the spinon pair (spin singlet pair) order $\Delta^f_{ij}$ which tends to disappear in the overdoped region.
The pairing energy terms here imply that bose condensation arises as a result of condensation of Cooper pairs.
The Cooper pair of d-wave symmetry is a composite of a spinon pair of d-wave symmetry and a holon pair of s-wave symmetry as a consequence of coupling between the spin (spinon) and charge (holon) degrees of freedom.
Thus the bose condensation disappears in the overdoped region where the spinon pairs (spin singlet pairs) no longer exist, as observed by experiments.

As shown in Fig. 1 the predicted pseudogap(spin gap) temperature $T^*_{SU(2)}$ by the SU(2) theory is consistently higher than that of U(1) theory, $T^*_{U(1)}$.
$T^{SU(2)}_c$ at optimal doping is predicted to be lower than $T^{U(1)}_c$. 
The predicted SU(2) optimal doping is shifted to a larger value, showing a closer agreement with observation\cite{ODA} than the U(1) case.
Such improvement is attributed to the amplitude and phase fluctuations of the order parameters $\Delta^f$ which were well taken care of in the SU(2) treatment.

The predicted bose condensation disappears in the overdoping region where the spinon pair order parameter vanishes, in agreement with the observed phase diagram of high $T_c$ cuprates.
Not only the spin gap temperature but also the superconducting temperature is predicted to increase with increasing $J$ as are shown in Figs. 2 and 3.
This is in sharp contrast to the result of the mean field theories involved with the single holon bose condensation\cite{KOTLIAR,FUKUYAMA,NAGAOSA,UBBENS,WEN} in which the bose condensation temperature ($T_c \sim t x \chi$) does not depend on the Heisenberg coupling energy $J$ and increases linearly with hole concentration $x$.
Thus, according to these theories the bose condensation temperature $T_c$ has no correlation with the spin gap temperature $T^*$.
Indeed, the experimentally observed correlation between $T^*$ and $T_c$ favors our proposed holon-pair boson theory in which the Heisenberg interaction term takes care of coupling between the spinon pairing order $\Delta_f$ and the holon pairing order $\Delta_b$.
$T^*$ is determined from $\Delta_f$ and $T_c$ from $\Delta_b$.
Both $\Delta_f$ and $\Delta_b$ depend on $J$.
This is a cause of correlation between $T^*$ and $T_c$ through the $J$ dependence as shown in Figs. 2 and 3. 
However, a $J$ independent universal behavior in the ratio of $T^*/T_c$ for the entire range of hole concentration is predicted, as shown in Fig. 4.
The universal behavior (dotted ($J/t=0.2$), solid ($J.t=0.3$) and dashed ($J/t=0.4$) lines) of showing a hyperbolic scaling of $T^*/T_c$ with hole concentration $x$ is in qualitative agreement with observations (shown by open and solid squares)\cite{ODA,MOMONO_PG}.
This agreement implies that correlation between the pseudogap and superconducting transition temperatures is ``mediated'' by the antiferromagnetic (Heisenberg) coupling $J$.

In Fig. 5 we show the doping dependence of superconducting gap at $T=0K$ for both the U(1) and SU(2) cases.
In both cases the predicted gaps decrease with increasing hole concentration.
This trend is consistent with the ARPES\cite{SHEN,DING2} and with the Nernst effect measurements\cite{WANG}.
That is, a monotonic decrease of the pairing energy with increasing hole concentration in contrast to the arch (dome) shaped doping dependence of superconducting transition temperature is predicted in agreement with observation.
The superconducting gap is given by the spinon pairing energy (spin gap) $\Delta^f_0 = 2 J_p \Delta_f$ at $T=0K$.
This is because the holon quasiparticle excitation energy gap vanishes as will be shown below.

In Fig. 6, we display the doping dependence of the Cooper pair order parameter, $\Delta$ ($=\Delta_f \Delta_b$) at zero temperature for $J/t=0.2$, $0.3$ and $0.4$ based on the U(1) theory. 
Although not shown here, the Cooper pair order parameter of the SU(2) theory shows a similar behavior. 
It is reminded that the Cooper pair order parameter, $\Delta$ is the composite of the spinon pairing order parameter, $\Delta_f$ and the holon pairing order parameter, $\Delta_b$.
The predicted Cooper pair order parameter shows arch shapes in the plane of the Cooper pair order parameter $\Delta$ and hole concentration $x$ as shown in the figure.
This arch shape of the superconducting (Cooper pair) order parameter was also predicted from the numerical study on the projected d-wave state\cite{TRIVEDI}.

In Fig. 7 we display the $J$ dependence of the Cooper pair bose condensation energy $U = E_N(\Delta_b=0) - E_S(\Delta_b)$ as a function of hole concentration at $T=0K$.
The predicted condensation energy for both the U(1) and SU(2) theories shows similar arch shapes as a function of hole concentration, exhibiting consistent increase with $J$ at all doping.
We omitted a plot for the U(1) case.
The arch shaped condensation energy is in agreement with observation\cite{LORAM94,MOMONO}.

In Fig. 8, we display the ratio of the condensation energy to the square of the Cooper pair order parameter $\Delta=\Delta^f \Delta^b$, $\frac{U}{\Delta^2}$.
As shown in the figure, the ratio remains roughly constant for the entire range of hole concentration for each value of $J/t$.
This implies that superconducting condensation energy scales linearly with the square of the Cooper pair order parameter, as is well known from the BCS theory.
It is noted that this ratio increases linearly with the increase of the Heisenberg coupling constant $J$.
Thus, the condensation energy also scales linearly with the coupling energy $J$, satisfying the relation at $T=0K$,
\bq
U = J \Delta_f^2 \Delta_b^2 = J \Delta^2
\eq
for both the U(1) and SU(2) theory.
For the SU(2) theory the Cooper pair order parameter is defined as $\Delta^{SU(2)} = \sqrt{2} \Delta_f \Delta_b$ to take into account the contributions of $b_1$ and $b_2$ holons.

Figs. 9 (a), 9 (b) and 9 (c) display the doping dependence of the superconducting transition temperature $T_c$, bose condensation energy $U$ and superfluid weight, all of which show a common feature of the arch shape, again in agreement with observations\cite{ODA,LORAM94,MOMONO,UEMURA}. 
As shown in Fig. 10, the predicted superfluid weight shows the `boomerang' behavior.
This trend is in complete agreement with the well-known Uemura plot\cite{UEMURA}. 
That is, the predicted boomerang behavior in both $T_c$ and $\frac{n_s(T \rightarrow 0)}{m^*}$ as a function of $x$ is consistent with the measurements\cite{UEMURA,BERNHARD} of muon-spin-relaxation rates $\sigma$, which showed a linear increase of $\sigma \propto n_s/m^* \propto T_c$ in the underdoped region and the reflex behavior for both $T_c$ and $n_s/m^*$ in the overdoped region.
The SU(2) results are qualitatively similar and are not plotted.
Although not shown here, the predicted specific heat shows a non-Fermi liquid behavior in the underdoped region and a Fermi liquid behavior in the overdoped region\cite{LEE_SPECIFIC}.

It is of great interest to see whether there exists a gap in the quasiparticle excitation of holon (boson).
The minimum excitation energy for boson corresponds to the gap, i.e., the energy required for the quasiparticle excitation\cite{NOZIERE}.
As will be discussed below the predicted minimum excitation energy at $T=0K$ for the square lattice of infinite size is found to be $E^b_k = 0^{+}$ at ${\bf k}=(0,0)$ for the U(1) theory and at ${\bf k}=(0,0)$ and $(\pi, \pi)$ for the SU(2) theory.
Excitation energy gap $E^b_0$ referred here should not be confused with the holon pairing energy, $\Delta^b_0$ ($\Delta_0^b$ in Eq.(\ref{u1_holon_energy_app}) or $\Delta_0^{b\alpha}$ in Eq.(\ref{su2_holon_energy}).
We computed the holon quasiparticle excitation energy (gap) $E^b_0$ in the thermodynamic limit $N$.
Fig. 11 shows for $N = 20 \times 20$ the ratio of the computed $E^b_0$ to the effective strength of the holon pairing interaction $J\Delta_f^2$ at zero temperature; 
$J\Delta_f^2$ comes from the pairing energy term $-\frac{J}{2}|\Delta^f_{ij}|^2 b_{i}^\dagger b_{j}^\dagger b_{j} b_{i}$ for the U(1) slave-boson theory and $-\frac{J}{2}|\Delta^f_{ij}|^2 b_{i \alpha}^\dagger b_{j \beta}^\dagger b_{j \beta} b_{i \alpha}$ for the SU(2) theory.
The predicted ratio $\frac{E_0^b}{(J\Delta_f^2)}$ shows nearly a constant value of $1/400$ which is independent of hole concentration for the selected finite square lattice of size, $N = 20 \times 20$.
As seen in Fig. 12 values of $N \frac{\Delta_b^0}{J\Delta_f^2}$ for different values of lattice size $N$, $N=10\times 10$, $16 \times 16$ and $20 \times 20$ are 
independent of the lattice size $N$, satisfying $N \frac{E^b_0}{J\Delta_f^2} = 1$. 
Thus the holon quasiparticle excitation energy at $T=0K$ as a function of lattice size is given by the relation,
\bqa
E^b_0 = \frac{J \Delta_f^2}{N} > 0^{+}.
\eqa
This implies that the holon quasiparticle excitation gap at $T=0K$ approaches $E^b_0 = 0^{+}$ (becomes gapless (massless)) in the thermodynamic limit, $N \rightarrow \infty$.
Thus, the holon-pair bose condensed state is stable against the single-holon bose condensed state.
This is in agreement with the theory of Nozi\`{e}res and Saint James\cite{NOZIERE} who showed that
boson-pair bose condensation is stable against single boson condensation since $E^b_0 \geq 0^{+}$ for the case of boson-pair\cite{NOZIERE}. 

At low temperature the holon quasiparticle excitation energy becomes linear in momentum.
Expanding the holon quasiparticle energy Eq.(\ref{u1_holon_energy_app}) around ${\bf k}=(0,0)$ for the U(1) theory, one obtains the energy dispersion 
\bq
E^b_k = v k,
\eq
with $k=\sqrt{k_x^2 + k_y^2}$, the momentum and $v=\sqrt{2 J \Delta_f^2 \Delta_b (2t\chi + J \Delta_f^2 \Delta_b)}$, the group velocity of the quasiparticle (see derivations in the previous section). 
For the SU(2) theory, the minimum of the quasiparticle energy occurs at ${\bf k}=(0,0)$ and ${\bf k}={\bf Q} \equiv (\pi,\pi)$.
Expanding the quasiparticle energy Eq.(\ref{su2_holon_energy}) around ${\bf k}=(0,0)$ and ${\bf k}={\bf Q}$ respectively, one obtains the energy dispersion relations,
\bqa
E^b_{k,1^{'}} = v k, \nn
E^b_{k,2^{'}} = v |{\bf k}-{\bf Q}|,
\eqa
with $v=\sqrt{ 2 J \Delta_f^2 \Delta_b (t\chi + J \Delta_f^2 \Delta_b)}$ for the case of $\Delta^b_{\alpha,\beta} = \Delta_b ( \delta_{\alpha,1}\delta_{\beta,1} - \delta_{\alpha,2}\delta_{\beta,2})$. 
The above relation holds, also, true in form for the case of $\Delta^b_{\alpha,\beta} = \Delta_b ( \delta_{\alpha,1}\delta_{\beta,1} + \delta_{\alpha,2}\delta_{\beta,2})$. 
The linear dispersion relation with the holon group velocity guarantees the stability of superfluidity\cite{NOZIERE_PINES}.
This is because the quasiparticle group velocity $v$ represents the critical velocity below which the superfluid flow of boson is stable against quasiparticle excitations\cite{NOZIERE_PINES}.
In Fig. 13 we display the doping dependence of the group velocity $v$.
Both the U(1) and SU(2) theories predict an arch shaped group velocity in the plane of $v$ vs. $x$.
The predicted critical hole doping concentration $x_c$ at which the maximum velocity occurs are $0.1$, $0.13$ and $0.15$ for $J/t = 0.2$, $0.3$ and $0.4$ respectively.
Interestingly, these critical values are nearly the same as the critical hole concentrations at which the maximum of the superconducting condensation energy occurs;
$x = 0.1$, $0.12$ and $0.14$ for $J/t = 0.2$, $0.3$ and $0.4$ respectively.
This implies that there exists correlation between the holon (boson) group velocity and the condensation energy.
It is expected that the group velocity should decrease in the heavily overdoped region, since the spinon pair order parameter $\Delta_f$ diminishes in the overdoped region.
Indeed, this is predicted to be the case as shown in Fig. 13.
As the Heisenberg coupling constant increases, the group velocity is shown to increase as is shown in the figure.

The generic features of peak-dip-hump are observed in ARPES, STM and optical conductivity measurements, suggesting a common origin for its cause.
Although not reported here, the predicted sharp quasi-particle peak below $T_c$ in the spectral function is seen to arise as a result of bose condensation below $T_c$ and its reduction above $T_c$ owing to the disappearance of the bose condensation in the pseudogap (spin gap) phase\cite{LEE_SPEC}.
This prediction of the sharp peak as a consequence of the bose condensation is in complete agreement with ARPES measurements\cite{FENG}.
Our earlier study of optical conductivity also showed the peak-dip-hump (Drude peak-dip-mid-infra band) structure\cite{LEE_OPT} below $T_c$, in agreement with observations\cite{ORENSTEIN}.
We find that the hump in both ARPES and optical conductivity is caused by the spin fluctuations (of the shortest possible antiferromagnetic correlation length) or the spin singlet pair excitations.
Although not shown here, the predicted temperature and doping dependence of spectral intensity\cite{LEE_SPEC} and optical conductivity\cite{LEE_OPT} is consistent with observations. 
Further the predicted doping and temperature dependence of spin susceptibility is also in agreement with the INS (inelastic neutron scattering) measurements\cite{KEIMER}.

\section{Summary}
In this paper, by making a full exposure to our earlier holon-pair boson theory\cite{LEE} various physical implications and predicted results are discussed.
We discussed the importance of coupling between the spinon (spin) pairing order and the holon (charge) pairing order.
Superconductivity is seen to result from the bose condensation of the Cooper pairs of the $d$-wave symmetry which is a composite of the $d$-wave symmetry of spinon (spin) pair and the $s$-wave symmetry of holon (charge) pair.
This theory differs from other proposed slave-boson theories in that coupling between the charge and spin degrees of freedom is manifested in the anti-ferromagnetic (Heisenberg) interaction term. 
Further the spin-charge separation no longer appears in the mean field Hamiltonian unlike other mean field theories concerned with single-holon bose condensation. 
Both the pseudogap temperature $T^*$ and the superconducting transition temperature $T_c$ are predicted to scale with the Heisenberg exchange coupling strength $J$.
Interestingly, despite such $J$ dependence on both the $T^*$ and $T_c$, the predicted ratio, $T^*/T_c$ for the entire range of hole doping concentration $x$ is independent of $J$.
Accordingly, the predicted universal behavior of the hyperbolic scaling $T^*/T_c$ with hole concentration $x$ is in agreement with observation\cite{ODA}.
The antiferromagnetic spin fluctuations of the shortest possible correlation length or the spin singlet pair excitations are seen to induce the bose condensation; 
obviously the bose condensation does not occur in the region where the spin singlet pair excitations are absent as shown  in the observed phase diagram for high $T_c$ cuprates. 
The spin-gap temperature or the pseudogap temperature is the temperature where the spin-gap begins to open. 
Both the spin gap temperature and the spin gap are predicted to continuously decrease as the hole doping concentration increases. 
Indeed, we found that the predicted bose condensation (superconducting transition) temperature drops to zero in the overdoped region where the spin gap or spinon pairing order disappears.
This finding is consistent with observations.  
This implies that the superconducting phase transition is attributed to the interplay between the charge and spin degrees of freedom. 
To put it otherwise, coupling between the holon (charge) pair and spinon (spin) pair to form the Cooper pair bosons is essential for causing the bose condensation.
The doping and temperature dependence of various physical properties are in complete agreement with observations.

We have demonstrated that interplay between the charge and spin degrees of freedom are responsible for determining the characteristic features of various observed physical properties; 
the bose condensation energy, superfluid weight, specific heat, spectral functions, optical conductivity and spin susceptibility, not to speak of the arch shape of superconducting transition temperature and the tangential appearance of the pseudogap temperature. 
In short, using the use of the U(1) and SU(2) holon-pair boson theories which fully takes into account coupling between the charge and spin degrees of freedom in the antiferromagnetic (Heisenberg) interaction term, we demonstrated self-consistency in predicting not only the arch shaped bose condensation temperature but also various other physical properties in agreements with observations.

\section{Acknowledgement}
One(SHSS) of us acknowledges the generous supports of Korea Ministry of Education (HakJin Program) and the Institute of Basic Science Research at Pohang University of Science and Technology.

\widetext
\top{-2.8cm}
\renewcommand{\theequation}{A\arabic{equation}}
\setcounter{equation}{0}
\section*{APPENDIX A: Diagonalization of the U(1) holon and spinon Hamiltonians}

We first write the mean-field Hamiltonian in momentum space and diagonalize the resulting kinetic energy term. 
We then express the total Hamiltonian in the basis of the diagonalized kinetic energy.
Finally the total Hamiltonian is diagonalized by using the Bogoliubov-Valatin transformations.
We take the uniform, that is, site-independent hopping order parameter\cite{UBBENS}, $\chi_{ij}=\chi e^{\pm i\theta }$, where the sign $+(-)$ is for the counterclockwise(clockwise) direction around a plaquette and the uniform pairing order parameters, $ \Delta_{ij}^{f}=\Delta_f e^{\pm i\tau^{f}} \mbox{ and } \Delta_{ij}^{b}=\Delta_b e^{\pm i\tau^{b}}$, where the sign $+(-)$ is for the ${\bf ij}$ link parallel to $\hat x$ ($\hat y$) where $\Delta_b$, $\Delta_f$ and $\chi$ are the site-independent amplitudes corresponding to proper order parameters above.
We also take the site-independent chemical potentials, $\mu^f_i = \mu^f$ and $\mu^b_i = \mu^b$\cite{FUKUYAMA,UBBENS,GIMM}.

The saddle point energy term in Eq.(\ref{mean_H_u1}) is, then, 
\bqa
H^{\chi,\Delta} & = & 
J \sum_{<i,j>} \Bigl[ \frac{1}{2} \Delta_f^2 \Delta_b^{2}  + \frac{1}{2} \Delta_f^2 x^2 \Bigr] 
+ \frac{J_p}{2} \sum_{<i,j>} \Bigl[ \Delta_{f}^{2} + \frac{1}{2} \chi^{2} + \frac{1}{4} \Bigr]  \nn
&& = 
N J \Bigl[ \Delta_f^2 \Delta_b^{2}  + \Delta_f^2 x^2 \Bigr]
+ N J_p \Bigl[ \Delta_{f}^{2} + \frac{1}{2} \chi^{2} + \frac{1}{4} \Bigr] 
\label{u1_saddle_point}
\eqa
with $N$, the lattice size.

The kinetic energy term of the holon in Eq.(\ref{u1_holon_sector}) in momentum space is obtained to be
\bqa
H^{b}_{K.E} & = & -t \sum_{<i,j>} ( \chi_{ji}^{*} b_{j}^{\dagger}b_{i} + c.c. ) \nn
&=& -t \frac{1}{N} \sum_{k,k^{'}}\sum_{<i,j>} 
( \chi_{ji}^* e^{-i(r_j \cdot k - r_i \cdot k^{{'}}) } b_{k}^{\dagger} b_{k^{'}} + c.c. ) \nn 
&=&  -2t \sum_{k}^{'}
( \begin{array}{cc} b_{k}^{\dagger} & b_{k+Q}^{\dagger} \end{array})
\left[
\begin{array}{cc} \chi \gamma _k \cos \theta  & i\chi \varphi_k \sin\theta \\
		 -i\chi \varphi_k \sin\theta  & -\chi \gamma_k \cos\theta
\end{array}
\right]
\left( \begin{array}{c} b_{k} \\ b_{k+Q} \end{array} \right),
\label{eq:ke_boson_real_momentum}
\eqa
and the kinetic energy term of the spinon in Eq.(\ref{u1_spinon_sector}) in momentum space,
\bqa
H^{f}_{K.E} & = & -\frac{J_p}{4} \sum_{<i,j>,\sigma} ( \chi_{ji}^{*} f_{j\sigma}^{\dagger}f_{i\sigma} + c.c. ) \nn
&=& -\frac{J_p}{4} \frac{1}{N} \sum_{k,k^{'}}\sum_{<i,j>,\sigma} 
( \chi_{ji}^* e^{-i(r_j \cdot k - r_i \cdot k^{{'}}) } f_{k\sigma}^{\dagger} f_{k^{'}\sigma} + c.c. ) \nn 
&=&  -\frac{J_p}{2} \sum_{k\sigma}^{'} 
( \begin{array}{cc} f_{k\sigma}^{\dagger} & f_{k+Q,\sigma}^{\dagger} \end{array})
\left[
\begin{array}{cc} \chi \gamma _k \cos \theta  & i\chi \varphi_k \sin\theta \\
		 -i\chi \varphi_k \sin\theta  & -\chi \gamma_k \cos\theta
\end{array}
\right]
\left( \begin{array}{c} f_{k\sigma} \\ f_{k+Q\sigma} \end{array} \right) ,
\label{eq:ke_fermion_real_momentum}
\eqa
where $\gamma _k=( \cos k_x + \cos k_y)$ and $ \varphi_k =( \cos k_x - \cos k_y)$.

The holon pairing term in Eq.(\ref{u1_holon_sector}) is
\bqa
H^b_p & = & -J\frac{\Delta_f^2}{2} \sum_{<i,j>} [ \Delta^{b*}_{ji} ( b_{j} b_{i}) + c.c. ] \nn
&=& 
-J\frac{\Delta_f^2}{2} \frac{\Delta_b}{N} \sum_{k,k^{'}}\sum_{j} e^{ir_j \cdot (k+k^{'})} 
[
(e^{i\tau^b+ik_x^{'}} + e^{-i\tau^b + ik_y^{'}})( b_{k} b_{k^{'}}) + c.c. ] \nn
&=& -J\frac{\Delta_f^2}{2} \Delta_b \sum_{k} [ (\cos \tau^b \gamma_k + i \sin \tau^b \varphi_k ) 
(  b_{k} b_{-k  }) + c.c. ],
\label{eq:pairing_boson_real_momentum}
\eqa
and the spinon pairing term in Eq.(\ref{u1_spinon_sector}) in momentum space, 
\bqa
H^f_p & = & -\frac{J_p}{2} \sum_{<i,j>} [ \Delta^{f*}_{ji} (\sigma f_{j\sigma} f_{i-\sigma}) + c.c. ] \nn
&=& 
-\frac{J_p}{2} \frac{\Delta_f}{N} \sum_{k,k^{'}}\sum_{j} e^{ir_j \cdot (k+k^{'})} 
[
(e^{i\tau^f+ik_x^{'}} + e^{-i\tau^f + ik_y^{'}})(\sigma f_{k\sigma} f_{k^{'}-\sigma}) + c.c. ] \nn
&=& -\frac{J_p}{2} \Delta_f \sum_{k} [ (\cos \tau^f \gamma_k + i \sin \tau^f \varphi_k ) 
( \sigma f_{k\sigma} f_{-k -\sigma }) + c.c. ].
\label{eq:pairing_fermion_real_momentum}
\eqa

The chemical potential terms of spinon and holon respectively are 
\bqa
-\mu^f \sum_{i\sigma} \left( f_{i\sigma}^{\dagger}f_{i\sigma} - (1-x) \right) & = & -\mu^f \sum_{k\sigma} f_{k\sigma}^{\dagger}f_{k\sigma} + \mu^f N (1-x), \\ \label{spinon_chemical_momentum}
-\mu^b \sum_i \left( b_{i}^{\dagger}b_{i} -x \right)  & = & -\mu^b \sum_k b_{k}^{\dagger}b_{k} + \mu^b N x. \label{holon_chemical_momentum}
\eqa

Finally, combining Eqs.(\ref{eq:ke_fermion_real_momentum}), (\ref{eq:pairing_fermion_real_momentum}) and (\ref{spinon_chemical_momentum}), we write the spinon Hamiltonian,
\bqa
H^f & = & -\frac{J_p}{2} \sum_{k\sigma}^{'} 
 ( f_{k\sigma}^{\dagger}  f_{k+Q,\sigma}^{\dagger}  )
 \left[
 \begin{array}{cc} \chi \gamma _k \cos \theta  & i\chi \varphi_k \sin \theta \\
		  -i\chi \varphi_k \sin \theta  & -\chi \gamma_k \cos \theta
 \end{array}
  \right]
  \left( \begin{array}{c} f_{k\sigma} \\ f_{k+Q\sigma} \end{array} \right)  \nn
&& -\frac{J_p}{2} \Delta_f \sum_{k} [ (\cos \tau^f \gamma_k + i \sin \tau^f \varphi_k )
   ( \sigma f_{k\sigma} f_{-k -\sigma }) + c.c. ] \nn
&& -\mu^f \sum_{k\sigma} f_{k\sigma}^{\dagger}f_{k\sigma} + \mu^f N (1-x),
\label{spinon_momentum}
\eqa
and combining Eqs.(\ref{eq:ke_boson_real_momentum}), (\ref{eq:pairing_boson_real_momentum}) and (\ref{holon_chemical_momentum}), we write the holon Hamiltonian,
\bqa
H^b & = &
 -2t \sum_{k}^{'} 
 ( \begin{array}{cc} b_{k}^{\dagger} & b_{k+Q}^{\dagger} \end{array})
 \left[
 \begin{array}{cc} \chi \gamma _k \cos \theta  & i\chi \varphi_k \sin\theta \\
		  -i\chi \varphi_k \sin\theta  & -\chi \gamma_k \cos\theta
  \end{array}
  \right]
  \left( \begin{array}{c} b_{k} \\ b_{k+Q} \end{array} \right) \nn
&& -J\frac{\Delta_f^2}{2} \Delta_b \sum_{k} [ (\cos \tau^b \gamma_k + i \sin \tau^b \varphi_k )
(  b_{k} b_{-k  }) + c.c. ] \nn
&& -\mu^b \sum_k b_{k}^{\dagger}b_{k} + \mu^b N x.
\label{holon_momentum}
\eqa
\bottom{-2.8cm}
\narrowtext
\noindent

From Eq.(\ref{eq:ke_fermion_real_momentum}) and Eq.(\ref{eq:ke_boson_real_momentum}), the kinetic energy terms of the spinon and holon parts are rewritten,
\begin{eqnarray}
H^f_{K.E.} & = & -\frac{J_p}{2} \sum_{k, \sigma}^{'} 
( \begin{array}{cc} f_{k\sigma}^{\dagger} & f_{k+Q,\sigma}^{\dagger} \end{array} )
        N_k(\chi,\theta)
        \left( \begin{array}{c} f_{k\sigma} \\ f_{k+Q,\sigma} \end{array} \right), \nn
	\label{spinon_ke} \\
H^b_{K.E.} & = & -2t \sum_{k}^{'} 
( \begin{array}{cc} b_{k}^{\dagger} & b_{k+Q}^{\dagger} \end{array} )
        N_k(\chi,\theta)
        \left( \begin{array}{c} b_{k} \\ b_{k+Q} \end{array} \right), \nn
	\label{holon_ke}
\end{eqnarray}
with 
\[
N_k(\chi,\theta) = 
\left[
\begin{array}{cc} \chi \gamma _k \cos\theta  & i\chi \varphi_k \sin\theta \\
                 -i\chi \varphi_k \sin\theta  & -\chi \gamma_k \cos\theta 
\end{array}
\right].
\]

In order to diagonalize the kinetic energy term, we introduce a basis transformation between bare particle (holon $b$ and spinon $f$) basis and quasi-particle ($B$ and $F$) basis,
\begin{eqnarray}
\left( \begin{array}{c} B_{k+} \\ B_{k-} \end{array} \right)
& = &
U_k^\dagger
\left( \begin{array}{c} b_{k} \\ b_{k+Q}  \end{array} \right), \\
\left( \begin{array}{c} F_{k \sigma +} \\ F_{k \sigma -}  \end{array} \right)
& = &
U_k^\dagger
\left( \begin{array}{c} f_{k\sigma} \\ f_{k+Q\sigma} \end{array}  \right)
\label{eq:basis_transformation}
\end{eqnarray}
with $U = \left[ \begin{array}{cc} u_{k+}  & u_{k-} \\
                 v_{k+} &  v_{k-} \end{array} \right]$, 
a unitary transformation matrix with
\begin{eqnarray}
\left( \begin{array}{c} u_{ k\pm} \\ v_{ k\pm}  \end{array} \right) 
& = &
\frac{1}{N_{\pm}} 
\left( \begin{array}{c} \chi \gamma_k \cos\theta \pm \chi \xi_k(\theta) \\ 
                        -i \chi \varphi_k \sin\theta 
  \end{array} \right), 
\label{eigenvector_u1}
\end{eqnarray}
and
\bqa
N_{\pm} & = & 
 \sqrt{ 2 \chi \xi_k(\theta) ( \chi \xi_k(\theta) \pm \chi \gamma_k \cos\theta ) }.
\eqa
In the quasi-particle basis, the kinetic energy term is rewritten,
\bqa
H^{f}_{K.E} & = & \sum_{k,\sigma, s=\pm 1}^{'} \frac{J_p}{2}\chi s \xi_k(\theta)  F_{k\sigma s}^{\dagger} F_{k\sigma s}  \label{sd1} \\
H^{b}_{K.E} & = &  \sum_{k,s=\pm 1}^{'}  2t \chi s \xi_k(\theta) B_{ks}^{\dagger} B_{ks}, \label{hd1}
\eqa
with $\xi_k(\theta) = \sqrt{ \gamma_k^2 \cos^2\theta + \varphi_k^2 \sin^2 \theta }$.
Here, $s=1$ ($-1$) represents the upper (lower) band of the quasiparticle energy spectrum in the reduced Brillouin zone defined by $|k_x| + |k_y| \leq \pi$ and $\sum^{'}$, the summation over the reduced Brillouin zone.

The chemical potential terms in the quasiparticle basis are,
\bqa
 -\mu^f \sum_{k\sigma} f_{k\sigma}^{\dagger}f_{k\sigma} &= & -\mu^f \sum_{k\sigma}^{'} \left( f_{k\sigma}^{\dagger}f_{k\sigma} + f_{k+Q\sigma}^{\dagger}f_{k+Q\sigma} \right) \nn
 & = & -\mu^f \sum_{k\sigma}^{'} \left( F_{k \sigma +}^\dagger  F_{k \sigma +} +  F_{k \sigma -}^\dagger F_{k \sigma -} \right)
 \label{spinon_chem_F}
\eqa
for spinon and
\bqa
 -\mu^b \sum_{k} b_{k}^{\dagger}b_{k} &= & -\mu^b \sum_{k}^{'} \left( b_{k}^{\dagger}b_{k} + b_{k+Q}^{\dagger}b_{k+Q} \right) \nn
 & = & -\mu^b \sum_{k}^{'} \left( B_{k +}^\dagger  B_{k +} +  B_{k -}^\dagger B_{k -} \right)
 \label{holon_chem_B}
\eqa
for holon.

Finally, inserting Eqs.(\ref{sd1}) and (\ref{spinon_chem_F}) into Eq.(\ref{spinon_momentum}) and  Eqs.(\ref{hd1}) and (\ref{holon_chem_B}) into Eq.(\ref{holon_momentum}), we obtain, respectively,
\bqa
H^{f}_{K.E. + chem.} & = & \sum_{k,\sigma,s=\pm 1}^{'} ( \epsilon_{ks}^{f} - \mu^f )  F_{k\sigma s}^{\dagger} F_{k\sigma s} + \mu^fN(1-x) \nn
\label{spinon_kinetic_basis}
\eqa
for spinon and
\bqa
H^{b}_{K.E. + chem.} & = &  \sum_{k,s=\pm 1}^{'} ( \epsilon_{ks}^{b} -\mu^b ) B_{ks}^{\dagger} B_{ks} + \mu^b N x
\label{holon_kinetic_basis}
\eqa
for holon.
Here the spinon quasiparticle (``quasispinon'') energy is $\epsilon_{ks}^{f} = \frac{J_p}{2}\chi s \xi_k(\theta)$ and
the holon quasiparticle (``quasiholon'') energy, $\epsilon_{ks}^{b} = 2t \chi s \xi_k(\theta)$ with
$\xi_k(\theta) = \sqrt{ \gamma_k^2 \cos^2\theta + \varphi_k^2 \sin^2 \theta }$ and $s = \pm 1$.

\widetext
\top{-2.8cm}
The spinon pairing term in Eq.(\ref{eq:pairing_fermion_real_momentum}) is rearranged to yield,
\bqa
H^f_P&=& -\frac{J_p}{2} \Delta_f \sum_{k} [ (\cos \tau^f \gamma_k + i \sin \tau^f \varphi_k )
( \sigma f_{k\sigma} f_{-k -\sigma }) + c.c. ] \nn
&=& -\frac{J_p}{2} \Delta_f \sum_{k}^{'} [ (\cos \tau^f \gamma_k + i \sin \tau^f \varphi_k )
\Bigl(  (f_{k\uparrow} f_{-k \downarrow }  + f_{-k \uparrow } f_{k\downarrow} ) \nn
&& - ( f_{k+Q\uparrow} f_{-k-Q \downarrow } + f_{-k-Q \uparrow } f_{k+Q\downarrow}  )  \Bigr) + c.c. ].
\eqa
Using the symmetry $\gamma_{-k}=\gamma_{k}$ and $\varphi_{-k}=\varphi_k$, we rewrite,
\bqa
H^f_P
&=& -J_p\Delta_f \sum_{k}^{'} [ (\cos \tau^f \gamma_k + i \sin \tau^f \varphi_k )
\left(  f_{k\uparrow} f_{-k \downarrow }   -  f_{k+Q\uparrow} f_{-k-Q \downarrow }  \right) + c.c. ] \nn
&=& -J_p \Delta_f \sum_{k}^{'} [ a_k (\tau^f)
( F_{k\uparrow +}F_{-k\downarrow +} + F_{k\uparrow -}F_{-k\downarrow -} ) + c.c. ],
\label{spinon_pairing_basis}
\eqa
where $a_k (\tau^f)=( \cos \tau^f \gamma_k + i \sin \tau^f \varphi_k) = e^{i \Phi} \xi_k(\tau^f)$ with 
$\xi_k(\tau^f) = \sqrt{ \gamma_k^2 \cos^2\tau^f + \varphi_k^2 \sin^2 \tau^f }$ and  
$\tan \Phi = \frac{ \varphi_k \sin \tau^f }{\gamma_k \cos \tau^f} = \frac{\varphi_k}{\gamma_k} \tan \tau^f$.
Here we used the unitarity of
$\left[
\begin{array}{cc} u_{k+}  & u_{k-} \\
                 v_{k+} &  v_{k-}
\end{array}
\right]$, that is,
\bqa
&& u_{ks}u_{-ks}-v_{ks}v_{-ks} = (u_{ks})^2 - (v_{ks})^2  = |u_{ks}|^2 + |v_{ks}|^2  =1 \nn
&& u_{ks}u_{-k-s}-v_{ks}v_{-k-s} = (u_{ks})^2 - (v_{k-s})^2  = u_{ks}u_{k-s}^* + v_{ks}v_{k-s}^* = 0,
\eqa
and $u_{ks}^* = u_{ks}$ and $v_{ks}^* = -v_{ks}$ for $s=\pm$ from Eq.(\ref{eigenvector_u1}).

Similarly, the holon pairing term in Eq.(\ref{eq:pairing_boson_real_momentum}) is rearranged to yield,
\bqa
H^b_{P}&=& -\frac{J}{2}|\Delta_f|^2 \Delta_b \sum_{k} [ (\cos \tau^b \gamma_k + i \sin \tau^b \varphi_k )
(  b_{k} b_{-k  }) + c.c. ] \nn
&=& -\frac{J}{2}|\Delta_f|^2 \Delta_b \sum_{k}^{'} [ a_k (\tau^b)
( B_{k+}B_{-k+} + B_{k-}B_{-k-} ) + c.c. ].
\label{holon_pairing_basis}
\eqa

From Eqs.(\ref{spinon_kinetic_basis}) and (\ref{spinon_pairing_basis}), we obtain the effective Hamiltonian in the transformed basis,
\bqa
H^{f} & = &  \sum_{k,s,\sigma}^{'} ( \epsilon_{ks}^{f} - \mu^f ) F_{k\sigma s}^{\dagger} F_{k\sigma s} - \sum_{k,s}^{'} [ e^{i \Phi_k(\tau^f)} \Delta^f_0  F_{k\uparrow s}F_{-k\downarrow s} - c.c.] + \mu^f N (1-x) , 
\label{spinon_trans}
\eqa
for spinon and 
from Eqs.(\ref{holon_kinetic_basis}) and (\ref{holon_pairing_basis}), 
\bqa
H^{b} & = & \sum_{k,s=\pm 1}^{'} ( \epsilon_{ks}^{b} - \mu^b) B_{ks}^{\dagger} B_{ks} - \sum_{k,s=\pm 1}^{'} [ e^{i \Phi_k(\tau^b)} \frac{\Delta^b_0}{2} B_{ks}B_{-ks} + c.c.] + \mu^b N x
\label{holon_trans}
\eqa
for holon.
Here $\epsilon_{ks}^{f} = \frac{J_p}{2}s \chi \xi_k(\theta)$, 
$\xi_k(\theta) = \sqrt{ \gamma_k^2 \cos^2\theta + \varphi_k^2 \sin^2 \theta }$,
$\Delta^f_0 = J_p \Delta_{f} \xi_k(\tau^f)$, 
$\epsilon_{ks}^{b} = 2ts \chi \xi_k(\theta)$ and 
$\Delta^b_0 = J \Delta_f^2 \Delta^{b} \xi_k(\tau^b)$.

Taking $F_{k\sigma s} = e^{-i \Phi_k(\tau^f)/2} F^{'}_{ks\sigma}$ for spinon 
and $B_{ks} = e^{-i \Phi_k(\tau^b)/2} B^{'}_{ks}$ for holon, we rewrite Eqs.(\ref{spinon_trans}) and (\ref{holon_trans}) respectively
\bqa
H^{f} & = &  \sum_{k,s,\sigma}^{'} (\epsilon_{ks}^{f}-\mu^f) F_{k\sigma s}^{' \dagger} F_{k\sigma s}^{'} - \sum_{k,s}^{'} \Delta^f_0 ( F^{'}_{ks\uparrow}F^{'}_{-ks\downarrow} - F_{k\uparrow s}^{' \dagger}F_{-k\downarrow s}^{' \dagger}) + \mu^f N (1-x), \label{spinon_dia} \\
H^{b} & = & \sum_{k,s=\pm 1}^{'} (\epsilon_{ks}^{b} - \mu^b) B_{ks}^{' \dagger} B^{'}_{ks} - \sum_{k,s=\pm 1}^{'} \frac{\Delta^b_0}{2} ( B^{'}_{ks}B^{'}_{-ks} + B_{ks}^{' \dagger}B_{-ks}^{' \dagger}) + \mu^b N x.
\label{holon_dia}
\eqa
For simplicity, we will omit the prime symbol $'$ in $F^{'}_{ks\sigma}$  and $B^{'}_{ks}$.

Introducing the Bogoliubov-Valatin transformation for spinon quasiparticle operators 
$\alpha_{ks\sigma}$ and $\alpha_{ks\sigma}^\dagger$,
\bqa
 F_{k\uparrow s} &=& \cos \zeta_{ks} \alpha_{ks\uparrow} +  \sin \zeta_{ks} \alpha_{-ks\downarrow}^{\dagger}, \nn
 F_{-k\downarrow s} &=& -\sin \zeta_{ks} \alpha_{ks\uparrow}^{\dagger} +  \cos \zeta_{ks} \alpha_{-ks\downarrow},
\label{eq:fermion_bogoliubov}
\eqa
we rewrite Eq.(\ref{spinon_dia}),
\bqa
H^f & = & \sum_{k,s}^{'} E_{ks}^f ( \alpha_{ks\uparrow}^{\dagger}  \alpha_{ks\uparrow} + \alpha_{ks\downarrow}^{\dagger} \alpha_{ks\downarrow} ) - \sum_{k,s}^{'} E_{ks}^f  - N \mu^f  + \mu^f N (1-x)\nn
 & = & \sum_{k,s}^{'} E_{ks}^f ( \alpha_{ks\uparrow}^{\dagger}  \alpha_{ks\uparrow} - \alpha_{ks\downarrow}^{\dagger} \alpha_{ks\downarrow} ) - N x \mu^f \nn,
\label{eq:fermion_diagonalized_hamiltonian}
\eqa
where $E_{ks}^f$ is the spinon quasiparticle (quasi-spinon) energy, 
$E_{ks}^f = \sqrt{ (\epsilon_{ks}^{f}-\mu^f)^2 +(\Delta^f_0)^2 }$
and
$\Delta^f_0$ is the spinon pairing energy (gap),
$\Delta^f_0 = J_p \Delta_{f} \xi_k(\tau^f)$. 

Introducing the Bogoliubov-Valatin transformation concerned with holon quasiparticle operators $\beta_{ks}$ and $\beta_{ks}^\dagger$,
\bqa
 B_{ks} &=& \cosh \eta_{ks} \beta_{ks} +  \sinh \eta_{ks} \beta_{-ks}^{\dagger}, \nn
 B_{-ks} & =&  \sinh \eta_{ks} \beta_{ks}^{\dagger} +  \cosh \eta_{ks} \beta_{-ks},
\label{eq:boson_bogoliubov}
\eqa
we rewrite the holon Hamiltonian,
\bqa
H^b& = & \sum_{k,s=\pm 1}^{'}  E_{ks}^b  \beta_{ks}^{\dagger}  \beta_{ks} + \sum_{k,s=\pm 1}^{'} \frac{1}{2}(E_{ks}^b + \mu^b) + \mu^b N x, 
\label{eq:boson_diagonalized_hamiltonian}
\eqa
where $E_{ks}^b$ is the holon quasiparticle (quasi-holon) energy, 
$E_{ks}^b = \sqrt{(\epsilon_{ks}^{b} - \mu^b)^2 - (\Delta^b_0)^2}$
and
$\Delta^b_0$ is the holon pairing energy,
$\Delta^b_0 = J \Delta_f^2 \Delta_{b} \xi_k(\tau^b)$.
In the last term we used the identity $\sum_{k,s=\pm 1}^{'} (-\epsilon_{ks}^{b} + \mu^b) = -\sum_{k,s=\pm 1}^{'} ( 2ts \chi \xi_k(\theta) - \mu^b ) = \sum_{k,s}^{'} \mu^b$.

Combining Eqs.(\ref{u1_saddle_point}), (\ref{eq:fermion_diagonalized_hamiltonian}) and (\ref{eq:boson_diagonalized_hamiltonian}), we obtain the total quasiparticle Hamiltonian,
\bqa
H^{MF}_{U(1)} & = &
N J \Bigl[ \Delta_f^2 \Delta_b^{2}  + \Delta_f^2 x^2 \Bigr]
+ N J_p \Bigl[ \Delta_{f}^{2} + \frac{1}{2} \chi^{2} +\frac{1}{4} \Bigr]  \nn
& & + \sum_{k,s}^{'} E_{ks}^f ( \alpha_{ks\uparrow}^{\dagger}  \alpha_{ks\uparrow} - \alpha_{ks\downarrow}^{\dagger} \alpha_{ks\downarrow} ) - N x \mu^f \nn
 & & + \sum_{k,s=\pm 1}^{'}  E_{ks}^b  \beta_{ks}^{\dagger}  \beta_{ks} + \sum_{k,s=\pm 1}^{'} \frac{1}{2}(E_{ks}^b + \mu^b) + \mu^b N x.
 \label{u1_diagonalized_hamiltonian_app}
 \eqa

\renewcommand{\theequation}{B\arabic{equation}}
\setcounter{equation}{0}
\section*{APPENDIX B: diagonalization of the SU(2) Hamiltonian}
Here we diagonalize the SU(2) Hamiltonians for both the spinon and holon in Eqs.(\ref{su2_spinon_sector}) and (\ref{su2_holon_sector}).
For the saddle point energy (the second term) in Eq.(\ref{su2_mean}), we write
\bqa
H^{\Delta,\chi}_{SU(2)} 
& = & 
\frac{J}{2} \sum_{<i,j>} |\Delta^f_{ij}|^2 \Bigl[ \sum_{\alpha,\beta} |\Delta_{ij; \alpha \beta}^{b}|^{2} + x^2 \Bigr] 
+ \frac{J_p}{2} \sum_{<i,j>} \Bigl[ |\Delta_{ij}^{f}|^{2} + \frac{1}{2} |\chi_{ij}|^{2} + \frac{1}{4} \Bigr]   \nn
& = & 
N J \Delta_f^2 ( 2  \Delta_{b}^{2} + x^2 )
+ N J_p ( \Delta_{f}^{2} + \frac{1}{2} \chi^{2} + \frac{1}{4} ),
 \label{su2_saddle_energy}
\eqa
where we used $\Delta_{ij; \alpha \beta}^{b} = \Delta_b ( \delta_{\alpha,1}\delta_{\beta,1} \pm \delta_{\alpha,2} \delta_{\beta,2} )$.
The spinon Hamiltonian in Eq.(\ref{su2_spinon_sector}) is rewritten,
\bq
H^f_{SU(2)}  =  -\frac{J_p}{4} \sum_{<i,j>} [ \psi_i^{\dagger} U_{ij} \psi_j + c.c. ] - \vec a \cdot \sum_i \psi_i^\dagger \vec \tau \psi_i.
\eq
where 
$\psi_{i} = \left( \begin{array}{c}  f_{i1} \\ f_{i2}^{\dagger} \end{array} \right)$
and
$
U_{i,j}  =  
\left( \begin{array}{cc} \chi_{ij}^* & -2 \Delta^f_{ij} \\
			   -2 \Delta^{f*}_{ij} & -\chi_{ij} \end{array} \right)
$
with
$a^{(1)} = i\frac{ \lambda^{1} + \lambda^{2} }{2}$,
$a^{(2)} = \frac{ -\lambda^{1} + \lambda^{2} }{2}$,
$a^{(3)} = i \lambda^{3}$ and
$\vec \tau$, the Pauli spin matrices.

For the uniform phase of the hopping order parameter and the d-wave phase of the spinon pairing order parameter ( $\chi_{ij}=\chi$, $\Delta_{i,i+x}^f = -\Delta_{i,i+y}^f = \Delta_f$ ), the order parameter matrix $U_{ij}$ is given by,
\bqa
U_{i,i+x} & = & 
\left( \begin{array}{cc} \chi & -2 \Delta_f \\
			   -2 \Delta_f & -\chi \end{array} \right) \nn
U_{i,i+y} & = & 
\left( \begin{array}{cc} \chi & 2 \Delta_f \\
			   2 \Delta_f & -\chi \end{array} \right) .
\label{su2_spinon_5}
\eqa
Using Eq.(\ref{su2_spinon_5}), we rewrite the spinon Hamiltonian in momentum space, 
\bqa
H^f_{SU(2)} & = & -\frac{J_p}{4} \sum_{<i,j>} [ \psi_i^{\dagger} U_{ij} \psi_j + c.c. ] - \vec a \cdot \sum_i \psi_i^\dagger \vec \tau \psi_i \nn
& = & \sum_{k} \psi_{k}^{\dagger} 
        \Bigl(   \begin{array}{cc} -\frac{J_p}{2} \chi \gamma_k - a^{(3)} & J_p \Delta_f \varphi_k - a^{(1)} + i a^{(2)} \\
                                J_p \Delta_f \varphi_k - a^{(1)} - i a^{(2)} & \frac{J_p}{2} \chi \gamma_k + a^{(3)}
                 \end{array} \Bigr) \psi_{k}, 
\label{su2_spinon_momentum}
\eqa
where
$\psi_{k} = \left( \begin{array}{c}  f_{k1} \\ f_{-k2}^{\dagger} \end{array} \right)$ 
and
$\gamma_k=( \cos k_x + \cos k_y)$ and $\varphi_k=( \cos k_x - \cos k_y)$.
\bottom{-2.8cm}
\narrowtext
\noindent

The spinon Hamiltonian in Eq.(\ref{su2_spinon_momentum}) is rewritten,
\bqa
H^{f}_{SU(2)} & = &  \sum_{k} \psi_{k}^{\dagger}
        \Bigl(   \begin{array}{cc} \epsilon^f_k - a^{(3)}   & \Delta^f_k \\
                                \Delta^{f*}_k  &  -\epsilon^f_k + a^{(3)}  
                 \end{array} \Bigr) \psi_{k},
\label{eq:eff_su2_spinon}
\eqa
where $\epsilon^f_k = -\frac{J_p}{2} \chi \gamma_k$ and $ \Delta^f_k = J_p \Delta_f \varphi_k - a^{(1)} + i a^{(2)}$.
To diagonalize the Hamiltonian, the unitary (Bogoliubov-Valatin) transformation is introduced,
\bqa
\left(   \begin{array}{c} f_{k1} \\  f_{-k2}^{\dagger} \end{array} \right) & = & 
\left(   \begin{array}{cc} u_k^f & -v_k^{f*} \\  v_k^f & u_k^f  \end{array} \right)
\left( \begin{array}{c} \alpha_{k1} \\  \alpha_{-k2}^{\dagger} \end{array} \right),
\eqa
with $u_k^{f} = \frac{1}{\sqrt{2}}\sqrt{ 1 + \frac{ \epsilon_k^f - a^{(3)} }{E^{f}_k } }$ and $v_k^{f} = \frac{e^{-i\phi}}{\sqrt{2}}\sqrt{ 1- \frac{ \epsilon_k^{f} - a^{(3)} }{E^{f}_k} }$  with $e^\phi = \frac{ \Delta^f_k}{|\Delta^f_k|}$.
In the quasiparticle basis, the spinon Hamiltonian (\ref{eq:eff_su2_spinon}) is diagonalized, 
\bqa
H^{f}_{SU(2)} & = & \sum_k \psi_k^{' \dagger} 
\Bigl(   \begin{array}{cc} 
E_{k}^f & 0 \\
0 & -E_{k}^f
\end{array} \Bigr)
\psi_k^{'} \nn
& = & \sum_k E^f_k ( \alpha_{k1}^{\dagger}\alpha_{k1} - \alpha_{k2}\alpha_{k2}^{\dagger} ),
\label{diagonalized_spinon_hamiltonian}
\eqa
with $ E_k^f = \sqrt{ (\epsilon_k^f - a^{(3)} )^2 + (\Delta^f_0)^2 }$, $\epsilon^f_k = -J_p  \chi \gamma_k$ and 
$\Delta^f_0 = | \Delta^f_k | = \sqrt{ (J_p \Delta_f \varphi_k -a^{(1)} )^2  + (a^{(2)})^2 }$.


\widetext
\top{-2.8cm}
The holon Hamiltonian in Eq.(\ref{su2_holon_sector}) is rewritten,
\bqa
H^b_{SU(2)} & = & -\frac{t}{2} \sum_{<i,j>} [ h_i^{\dagger} U_{ij} h_j + c.c. ] \nn
 &&  - \frac{J}{2} \sum_{<i,j>} |\Delta^f_{ij}|^2 \Bigl[ h_i^\dagger \Delta^B_{ij} (h_j^\dagger)^T + c.c. \Bigr] \nn
&& - \sum_i h_i^\dagger ( \vec a_i \cdot \tau + \mu^b ) h_i \nn
&& + \mu^b N x,
\label{h1}
\eqa
where
$h_{i} = \left( \begin{array}{c}  b_{i1} \\ b_{i2} \end{array} \right)$ and
$
U_{i,j}  =  
\left( \begin{array}{cc} \chi_{ij}^* & - \Delta^f_{ij} \\
			   - \Delta^{f*}_{ij} & -\chi_{ij} \end{array} \right)
$
with
$a^{(1)} = i\frac{ \lambda_{i}^{1} + \lambda_{i}^{2} }{2}$,
$a^{(2)} = \frac{ -\lambda_{i}^{1} + \lambda_{i}^{2} }{2}$,
$a^{(3)} = i \lambda_{i}^{3}$ and
$\vec \tau$, the Pauli spin matrices.

Taking the uniform phase of the hopping order parameter ($\chi_{ij}=\chi$) and the d-wave symmetry of the spinon pairing order parameter ($\Delta_{i,i+x}^f = -\Delta_{i,i+y}^f = \Delta_f$), we write,
\bqa
U_{i,i+x} & = & 
\left( \begin{array}{cc} \chi & - \Delta_f \\
			   - \Delta_f & -\chi \end{array} \right) \nn
U_{i,i+y} & = & 
\left( \begin{array}{cc} \chi &  \Delta_f \\
			    \Delta_f & -\chi \end{array} \right) .
\label{U_su2}
\eqa
Using the above expression in Eq.(\ref{U_su2}), we rewrite the first and third terms of Eq.(\ref{h1}) in momentum space,
\bqa
H^b_{KE} & = & -\frac{t}{2} \sum_{<i,j>} [ h_i^{\dagger} U_{ij} h_j + c.c. ] -  \sum_i h_i^\dagger ( \vec a \cdot \tau + \mu^b ) h_i \nn
& = & \sum_{k} h_{k}^{\dagger} 
        \Bigl(   \begin{array}{cc} -t \chi \gamma_k - a^{(3)} -\mu^b & t \Delta_f \varphi_k - a^{(1)} + i a^{(2)} \\
                                 t \Delta_f \varphi_k - a^{(1)} - i a^{(2)} & t \chi \gamma_k + a^{(3)} - \mu^b 
                 \end{array} \Bigr) h_{k}, 
\label{h3}
\eqa
where
$h_{k} = \left( \begin{array}{c}  b_{k1} \\ b_{k2} \end{array} \right)$
and
$\gamma_k=( \cos k_x + \cos k_y)$ and $\varphi_k=( \cos k_x - \cos k_y)$.

For the s-wave symmetry of the holon-pair order parameter ($\Delta^B_{ii+x} = \Delta^B_{ii+y} = \Delta^B$) with the same amplitude for the $b_1-b_1$ and $b_2-b_2$ holon pairs, the holon-pair order parameter matrix is written,
\bqa
\Delta^B & = & 
\left( \begin{array}{cc} \Delta_b & 0 \\
			0 & e^{i\phi}  \Delta_b \end{array} \right),
\eqa
where $e^{i\phi}$ is the phase difference between the $b_1-b_1$ and $b_2-b_2$ order parameter. 
Here we consider the two cases of $e^{i\phi} = \pm 1$.
Transforming the holon pairing term (the second term of Eq.(\ref{h1})) into momentum space, we obtain
\bqa
H^b_{P} &=& - \frac{J}{2} \sum_{<i,j>} |\Delta^f_{ij}|^2 \Bigl[ h_i^\dagger \Delta^B_{ij} (h_j^\dagger)^T + c.c. \Bigr] \nn
&=& - \frac{J}{2} |\Delta_f|^2 \sum_k  \Bigl[ \gamma_k  h_k^\dagger \Delta^B (h_{-k}^\dagger)^T + c.c. \Bigr]
\label{h4}
\eqa
with $\gamma_k=( \cos k_x + \cos k_y)$.
Here we used the identity
$h_{-k}^\dagger \Delta^B (h_{k}^\dagger)^T = 
b_{-k \alpha}^\dagger \Delta^B_{\alpha \beta} b_{k \beta}^\dagger = 
b_{k \beta}^\dagger \Delta^B_{\alpha \beta} b_{-k \alpha}^\dagger = 
h_{k}^\dagger ( \Delta^B )^T (h_{-k}^\dagger)^T = 
h_{k}^\dagger  \Delta^B  (h_{-k}^\dagger)^T$ and
$(\Delta^B)^T = \Delta^B$,
where $(\Delta^B)^T$ is the transpose of $\Delta^B$ 

Inserting Eq.(\ref{h3}) and Eq.(\ref{h4}) into Eq.(\ref{h1}), we obtain the holon Hamiltonian in the momentum space
\bqa
H^b_{SU(2)} & = & \sum_{k} h_{k}^{\dagger} 
        \Bigl(   \begin{array}{cc} -t \chi \gamma_k - a^{(3)} -\mu^b & t \Delta_f \varphi_k - a^{(1)} + i a^{(2)} \\
                                 t \Delta_f \varphi_k - a^{(1)} - i a^{(2)} & t \chi \gamma_k + a^{(3)} - \mu^b 
                 \end{array} \Bigr) h_{k} \nn
&& - \frac{J}{2} |\Delta_f|^2 \sum_k  \Bigl[ \gamma_k  h_k^\dagger \Delta^B (h_{-k}^\dagger)^T + c.c. \Bigr] \nn
&& + \mu^b N x.
\eqa
The above holon Hamiltonian is rewritten,
\bqa
H^b_{SU(2)} &=& \sum_k  \Bigl[ 2 h_{k}^{\dagger} M_k  h_{k} 
+  h_k^\dagger  N_k (h_{-k}^\dagger)^T + h_{-k}^T N_k^\dagger h_k \Bigr] + \mu^b N x \nn
&= & \sum_k  \Bigl[  h_{k}^{\dagger} M_k  h_{k} + h_{-k}^{\dagger} M_{-k}  h_{-k} \Bigr] \nn
&& + \sum_k  \Bigl[  h_k^\dagger  N_k (h_{-k}^\dagger)^T + h_{-k}^T N_k^\dagger h_k \Bigr] + \mu^b N x,
\label{h6}
\eqa
where 
\bqa
M_k & \equiv &  \frac{1}{2} 
 \Bigl(   \begin{array}{cc} -t \chi \gamma_k - a^{(3)} -\mu^b & t \Delta_f \varphi_k - a^{(1)} + i a^{(2)} \\
				   t \Delta_f \varphi_k - a^{(1)} - i a^{(2)} & t \chi \gamma_k + a^{(3)} - \mu^b
     \end{array} \Bigr), \nn
N_k & \equiv & 
- \frac{J}{2} |\Delta_f|^2  \gamma_k  \Delta^B.
\eqa

In order to represent the above holon Hamiltonian in terms of the $4$-component holon field operator, we rearrange the second term of Eq.(\ref{h6}),
\bqa
h_{-k}^{\dagger} M_{-k}  h_{-k} & = &
( b_{-k1}^\dagger,  b_{-k2}^\dagger ) 
 \Bigl(   \begin{array}{cc} M_{-k}^{11} &  M_{-k}^{12} \\
			 M_{-k}^{21} &  M_{-k}^{22}
     \end{array} \Bigr)
\left( \begin{array}{c}  b_{-k1} \\ b_{-k2} \end{array} \right) \nn
& = & M_{-k}^{11} b_{-k1}^\dagger b_{-k1}
 + M_{-k}^{12} b_{-k1}^\dagger b_{-k2}
 + M_{-k}^{21} b_{-k2}^\dagger b_{-k1}
 + M_{-k}^{22} b_{-k2}^\dagger b_{-k2} \nn
&=& h_{-k}^T M_{-k}^T (h_{-k}^\dagger)^T + \mu^b.
\label{h7}
\eqa
Inserting Eq.(\ref{h7}) into Eq.(\ref{h6}), the holon Hamiltonian leads to
\bqa
H^b_{SU(2)}  &= & \sum_k  \Bigl[  h_{k}^{\dagger} M_k  h_{k} + h_{-k}^T (M_{-k})^T  (h_{-k}^\dagger)^T \Bigr] \nn
&& + \sum_k  \Bigl[  h_k^\dagger  N_k (h_{-k}^\dagger)^T + h_{-k}^T N_k^\dagger h_k \Bigr] + \mu^b N (1+x) \nn
& = & \frac{1}{2} \sum_k H_k^\dagger M^b_k H_k + \mu^b N (1+x),
\eqa
where $H_k = \left( \begin{array}{c}  b_{k1} \\ b_{k2}  \\ b_{-k1}^\dagger \\ b_{-k2}^\dagger \end{array} \right)$ and
\bq
M^b_{\bf k}  =  
 \left[ \begin{array}{cccc}
-t\chi \gamma_{\bf k} - \mu^b - a^{(3)} & t\Delta_f \varphi_{\bf k} - a^{(1)} + i a^{(2)} & 
-J \Delta_f^2 \Delta^b_{11} \gamma_{\bf k} & -J \Delta_f^2 \Delta^b_{12} \gamma_{\bf k} \nn

t\Delta_f \varphi_{\bf k} - a^{(1)} - i a^{(2)} & t\chi \gamma_{\bf k} - \mu^b + a^{(3)} & 
-J \Delta_f^2 \Delta^b_{21} \gamma_{\bf k} & -J \Delta_f^2 \Delta^b_{22} \gamma_{\bf k} \nn 

-J \Delta_f^2 \Delta^{b*}_{11} \gamma_{\bf k} & -J \Delta_f^2 \Delta^{b*}_{21} \gamma_{\bf k} &
-t\chi \gamma_{\bf k} - \mu^b - a^{(3)} & t\Delta_f \varphi_{\bf k} - a^{(1)} - i a^{(2)}  \nn

-J \Delta_f^2 \Delta^{b*}_{12} \gamma_{\bf k} & -J \Delta_f^2 \Delta^{b*}_{22} \gamma_{\bf k} &
t\Delta_f \varphi_{\bf k} - a^{(1)} + i a^{(2)} & t\chi \gamma_{\bf k} - \mu^b + a^{(3)} 
  \end{array} \right]
  \label{h10}
\eq
with $\gamma_{\bf k} = (\cos k_x + \cos k_y)$ and $\varphi_{\bf k} = (\cos k_x - \cos k_y)$.
\bottom{-2.8cm}
\narrowtext
\noindent

To diagonalize the above Hamiltonian, we consider the Bogoliubov-Valatin transformation for the holon field operator $H_{\bf k}$,
\bqa
H_{\bf k} = W({\bf k}) H^{'}_{\bf k},
\label{eq_bogoliubov:bogoliubov}
\eqa
where $W({\bf k})$ is a $4 \times 4$ matrix and $H^{'}_{\bf k}$, holon quasiparticle (``quasiholon'') operator.
For the quasiholon operator $H^{'}_{\bf k}$ to satisfy the boson commutation relation, the transformation matrix $W({\bf k})$ should satisfy
\bqa
W({\bf k}) {\bf J} W({\bf k})^\dagger &=& {\bf J},
\eqa
where  
$J =
\left( \begin{array}{cccc}
1 & 0 & 0 & 0 \\
0 & 1 & 0 & 0 \\
0 & 0 & -1 & 0 \\
0 & 0 & 0 & -1
\end{array} \right)$.
It is noted that $W({\bf k})$ is not a unitary matrix unlike the Bogoliubov-Valatin transformation for fermion operators.
$W({\bf k})$ is chosen so that the transformed Hamiltonian matrix $W({\bf k})^\dagger M^b_{\bf k} W({\bf k})$ is diagonalized, that is, 
\bqa
W({\bf k})^\dagger M^b_{\bf k} W({\bf k}) & = &  
\left( \begin{array}{cccc}
E_1 & 0 & 0 & 0 \\
0 & E_2 & 0 & 0 \\
0 & 0 & -E_3 & 0 \\
0 & 0 & 0 & -E_4 
\end{array} \right),
\label{eq_app_su2:holon_diagonalize}
\eqa
where $E_\alpha$ is the quasiparticle energy.
The Eq.(\ref{eq_app_su2:holon_diagonalize}) can be rewritten,
\bqa
\left[ M^b_{\bf k} - E_\alpha J \right] \cdot {\bf w}_{\bf k}^\alpha  & = & 0,
\eqa
where ${\bf w}_{\bf k}^\alpha$ is the $\alpha$-th column vector of $W({\bf k})$ with $\alpha = 1,2,3,4$.
The quasiholon energies and the corresponding eigenvectors are obtained from the above equation.
This equation is solved by the symbolic calculation of {\it Mathematica}\cite{MATHEMATICA}.

Only pairing between the same kind of holons allows the pairing state of the zero center of mass momentum.
At low temperature $b_1$ and $b_2$ holons are equally populated at ${\bf k} = (0,0)$ and $(\pi,\pi)$ respectively.
Thus, we consider the same amplitude for $b_1$-$b_1$ boson pairing and $b_2$-$b_2$ boson pairing, but not for $b_1$-$b_2$ boson pairing.
To allow the phase difference between the pairing order parameters of these two channels, we consider the two cases of $\Delta^b_{ \alpha \beta} = \Delta_b ( \delta_{\alpha,1}\delta_{\beta,1} \pm \delta_{\alpha,2} \delta_{\beta,2} )$ with $+$ sign for no phase difference and $-$ sign for the phase difference $\pi$ between the $b_1$ and $b_2$ boson pairing. 
For the case  of
$
\left( \begin{array}{cc}
\Delta^b_{11} & \Delta^b_{12} \\
\Delta^b_{21} & \Delta^b_{22} 
\end{array} \right)
= \Delta_b
\left( \begin{array}{cc}
1 & 0 \\
0 & -1 
\end{array} \right)$, 
the quasiparticle energy eigenvalues are given by
\widetext
\top{-2.8cm}
\bqa
E_{k1}^{b}  & =&   \sqrt{ (E_k^b)^2 + (\mu^b)^2 - ( \Delta_b^{'} )^{2} + 2 \sqrt{ (\mu^b E_{k}^{b})^2 - ( \Delta_b^{'} )^{2} ( \Delta_f^{''} - a^{(1)})^2} }, \nn
E_{k2}^{b}  & =&   \sqrt{ (E_k^b)^2 + (\mu^b)^2 - ( \Delta_b^{'} )^{2} - 2 \sqrt{ (\mu^b E_{k}^{b})^2 - ( \Delta_b^{'} )^{2} ( \Delta_f^{''}  - a^{(1)})^2} }, \nn
E_{k3}^{b}  & =&   -\sqrt{ (E_k^b)^2 + (\mu^b)^2 - ( \Delta_b^{'} )^{2} + 2 \sqrt{ (\mu^b E_{k}^{b})^2 - ( \Delta_b^{'} )^{2} ( \Delta_f^{''} - a^{(1)})^2} }, \nn
E_{k4}^{b}  & =&   -\sqrt{ (E_k^b)^2 + (\mu^b)^2 - ( \Delta_b^{'} )^{2} - 2 \sqrt{ (\mu^b E_{k}^{b})^2 - ( \Delta_b^{'} )^{2} ( \Delta_f^{''}  - a^{(1)})^2} }, 
\label{holon_quasiparticle_energy_m}
\eqa
and for the case  of
$
\left( \begin{array}{cc}
\Delta^b_{11} & \Delta^b_{12} \\
\Delta^b_{21} & \Delta^b_{22} 
\end{array} \right)
= \Delta_b
\left( \begin{array}{cc}
1 & 0 \\
0 & 1 
\end{array} \right)$, 
\bqa
E_{k1}^{b}  & =&   \sqrt{ (E_k^b)^2 + (\mu^b)^2 - ( \Delta_b^{'} )^{2} + 2 \sqrt{ (\mu^b E_{k}^{b})^2 - ( \Delta_b^{'} )^{2} ( a^{(2)})^2} }, \nn
E_{k2}^{b}  & =&   \sqrt{ (E_k^b)^2 + (\mu^b)^2 - ( \Delta_b^{'} )^{2} - 2 \sqrt{ (\mu^b E_{k}^{b})^2 - ( \Delta_b^{'} )^{2} ( a^{(2)})^2} }, \nn
E_{k3}^{b}  & =&   -\sqrt{ (E_k^b)^2 + (\mu^b)^2 - ( \Delta_b^{'} )^{2} + 2 \sqrt{ (\mu^b E_{k}^{b})^2 - ( \Delta_b^{'} )^{2} ( a^{(2)})^2} }, \nn
E_{k4}^{b}  & =&   -\sqrt{ (E_k^b)^2 + (\mu^b)^2 - ( \Delta_b^{'} )^{2} - 2 \sqrt{ (\mu^b E_{k}^{b})^2 - ( \Delta_b^{'} )^{2} ( a^{(2)})^2} }, \nn
\label{holon_quasiparticle_energy_p}
\eqa
where $E^b_k = \sqrt{(-t\chi\gamma_{\bf k})^2 + ( \Delta_f^{''}({\bf k}))^2 }$, $\Delta_f^{''} = t\Delta_f \varphi_{\bf k}$, and $\Delta_b^{'} = J\Delta_f^2 \Delta_b \gamma_{\bf k}$.
\bottom{-2.8cm}
\narrowtext
\noindent

Inserting Eq.(\ref{holon_quasiparticle_energy_m}) or Eq.(\ref{holon_quasiparticle_energy_p}) into Eq.(\ref{eq_app_su2:holon_diagonalize}), we obtain 
\bqa
W({\bf k})^\dagger M^b_{\bf k} W({\bf k}) & = &
\left( \begin{array}{cccc}
E_{k1}^b & 0 & 0 & 0 \\
0 & E_{k2}^b & 0 & 0 \\
0 & 0 & E_{k1}^b & 0 \\
0 & 0 & 0 & E_{k2}^b
\end{array} \right).
\label{eq_app_su2:holon_diagonalize2}
\eqa
and the diagonalized holon Hamiltonian,
\bqa
H^b_{SU(2)}  & = & \frac{1}{2} \sum_{\bf k} H^\dagger_{\bf k} M^b_{\bf k} H_{\bf k} + \mu^b N (1+x)\nn
& = & \frac{1}{2} \sum_{\bf k} H^{' \dagger}_{\bf k} W({\bf k})^\dagger M^b_{\bf k} W({\bf k}) H^{'}_{\bf k} + \mu^b N (1+x) \nn
& = & \frac{1}{2} \sum_{\bf k} \Bigl[
E_{k1}^b ( \beta_{k1}^{\dagger} \beta_{k1} + \beta_{k1} \beta_{k1}^{\dagger} ) \nn
&& + E_{k2}^b ( \beta_{k2}^{\dagger} \beta_{k2} + \beta_{k2}  \beta_{k2}^{\dagger} ) \Bigr] + \mu^b N (1+x),
\eqa
where we used the fact that $E_{k\alpha}^b$ is the even function of $k$.
After normal ordering, we obtain the diagonalized Hamiltonian, 
\bqa
H^b_{SU(2)}  & = & \sum_{{\bf k},\alpha=1,2} \Bigl[ E_{k\alpha}^b \beta_{k\alpha}^{\dagger} \beta_{k\alpha} + \frac{E_{k\alpha}^b }{2} \Bigr] + \mu^b N (1+x),
\label{diagonalized_holon_hamiltonian}
\eqa
where
\bqa
H_{\bf k} = W({\bf k}) H^{'}_{\bf k}.
\eqa

Combining Eqs.(\ref{su2_saddle_energy}), (\ref{diagonalized_spinon_hamiltonian}) and (\ref{diagonalized_holon_hamiltonian})
we obtain the diagonalized total Hamiltonian,
\begin{eqnarray}
H^{MF}_{SU(2)}  & = &
NJ\Delta_f^2 ( 2\Delta_b^2 + x^2 )
+ N J_p \Bigl( \frac{1}{2}\chi^{2} + \Delta_f^{2} + \frac{1}{4} \Bigr) \nn
& + & \sum_{k} E_{k}^{f} (\alpha_{k1}^{\dagger}\alpha_{k1} - \alpha_{k2}\alpha_{k2}^{\dagger}) \nn
& + &  \sum_{k,\alpha=1,2} \Bigl[ E_{k\alpha}^{b} \beta_{k\alpha}^{\dagger}  \beta_{k\alpha}  +  \frac{1}{2}( E_{k\alpha}^b + \mu^b ) \Bigr] + \mu^b N x. \nn
\label{su2_diagonalized_hamiltonian_app}
\end{eqnarray}


\newpage
\begin{figure}
	\epsfxsize=7.9cm
	\epsfysize=5.7cm
	\epsffile{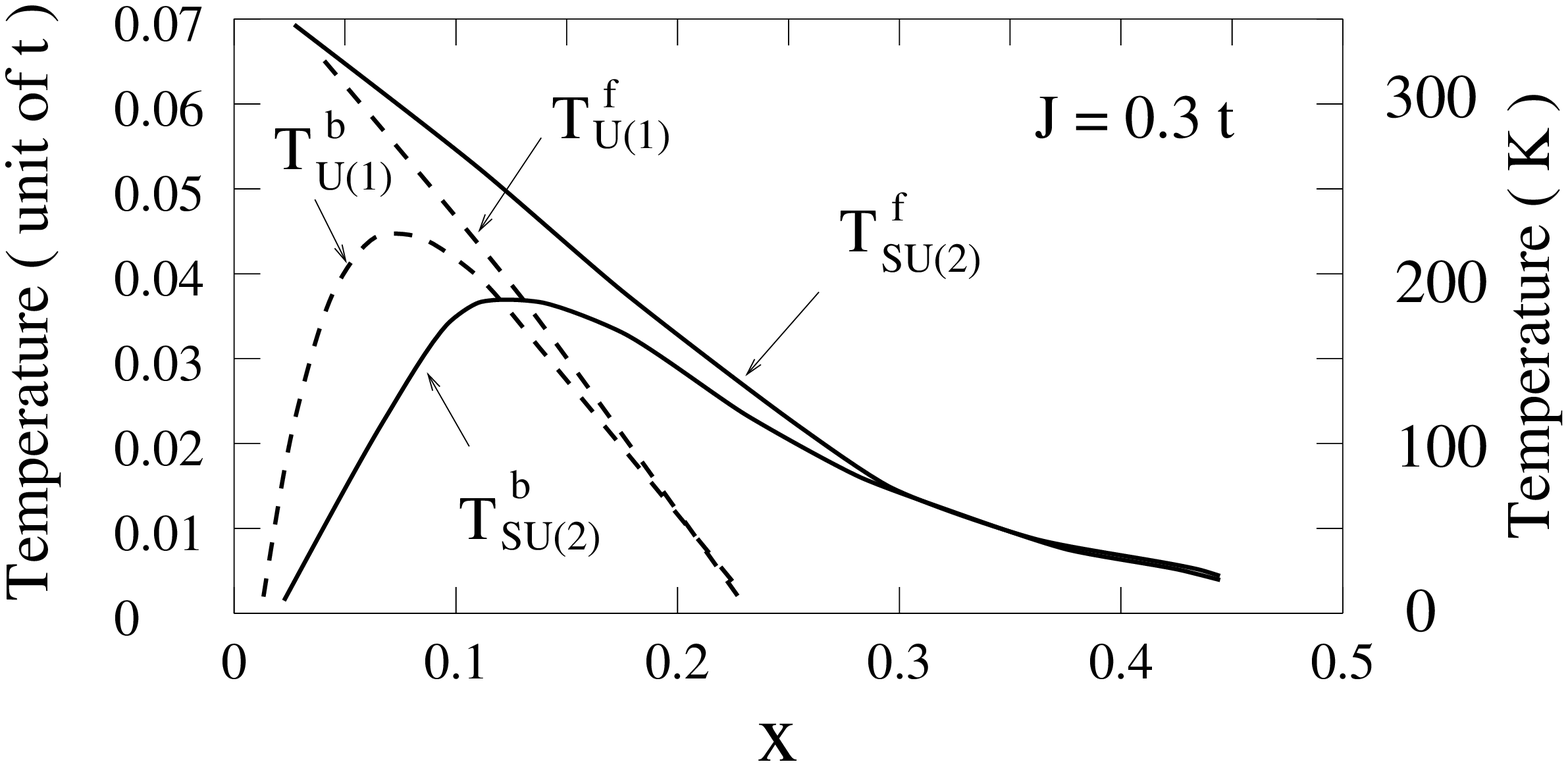}
\label{fig:1}
\caption{
Computed phase diagrams with $J/t=0.3$.  $T^*_{SU(2)}$ ($T^*_{U(1)}$) denotes the pseudogap temperature and $T_c^{SU(2)}$ ($T_c^{U(1)}$), the holon pair bose condensation temperature predicted from the SU(2) (solid lines) and (U(1)) (dotted lines) slave-boson theories respectively.  The scale of temperature is based on $t=0.44eV$[43].
}
\end{figure}

\begin{figure}
	\epsfxsize=8.1cm
	\epsfysize=7.1cm
	\epsffile{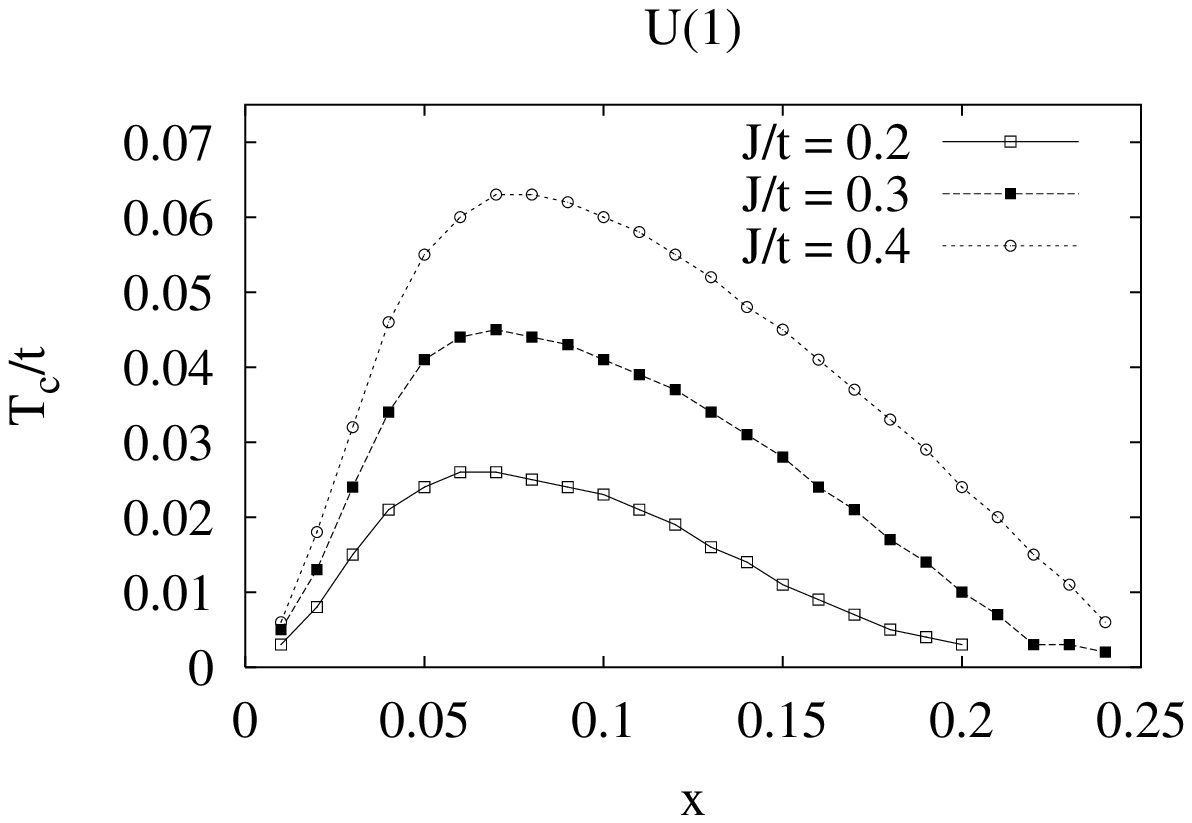}
	\epsfxsize=8.1cm
	\epsfysize=7.1cm
	\epsffile{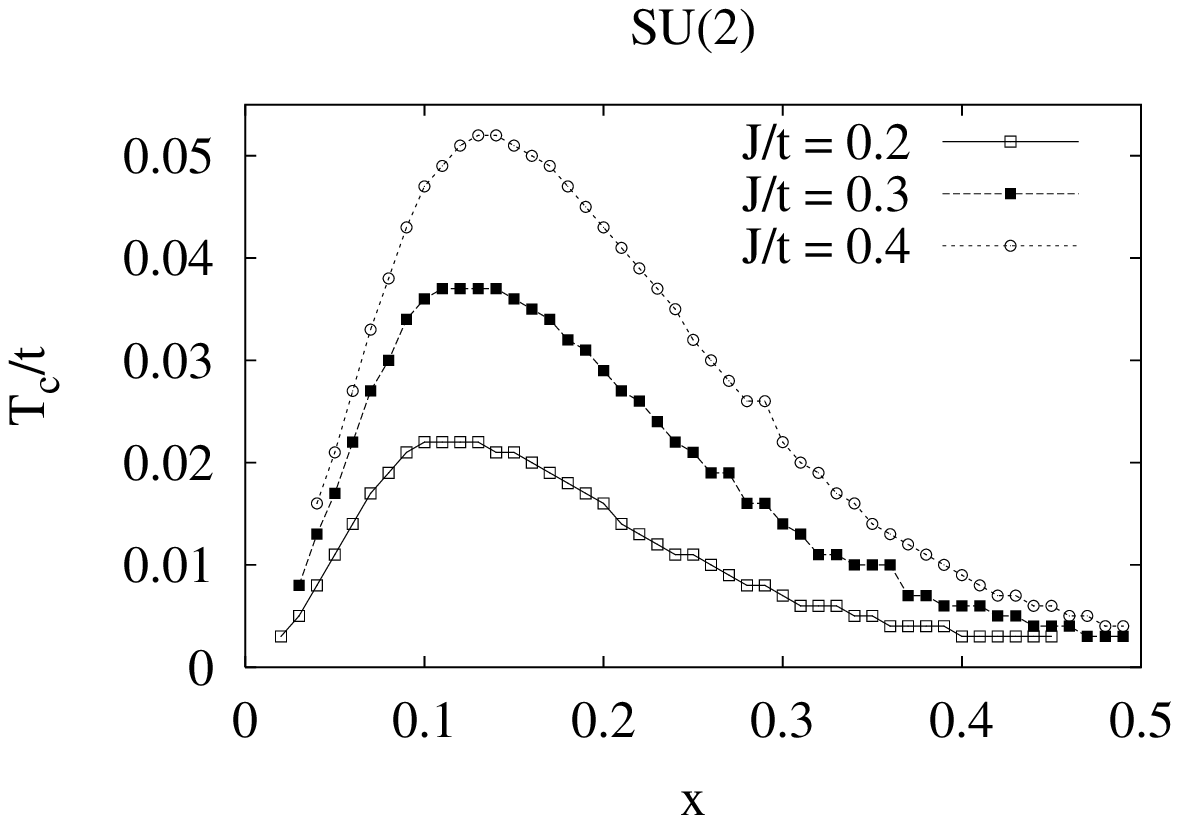}
\label{fig:2}
\caption{
Doping dependence of the holon pair bose condensation (superconducting transition) temperature $T_c$ for $J/t=0.2$, $0.3$ and $0.4$ based on the U(1) and SU(2) holon-pair boson theories. 
}
\end{figure}

\begin{figure}
	\epsfxsize=8.1cm
	\epsfysize=7.1cm
	\epsffile{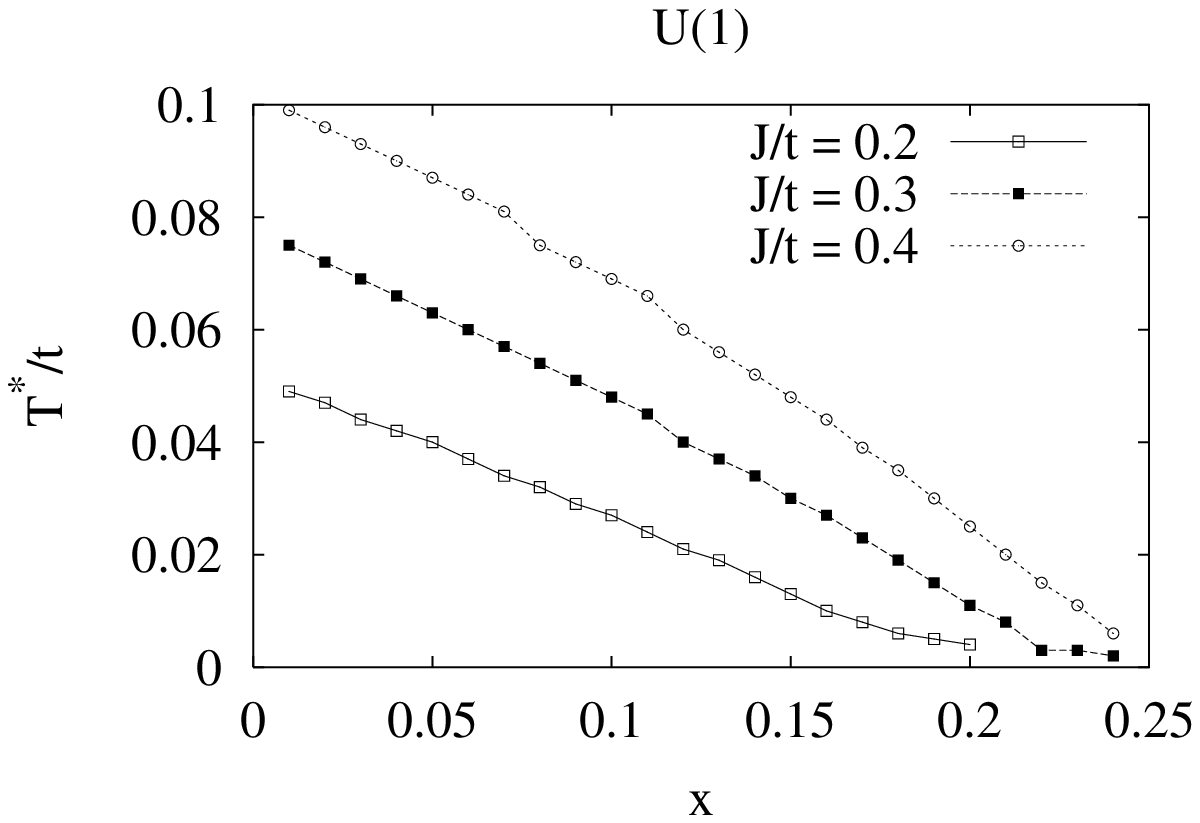}
	\epsfxsize=8.1cm
	\epsfysize=7.1cm
	\epsffile{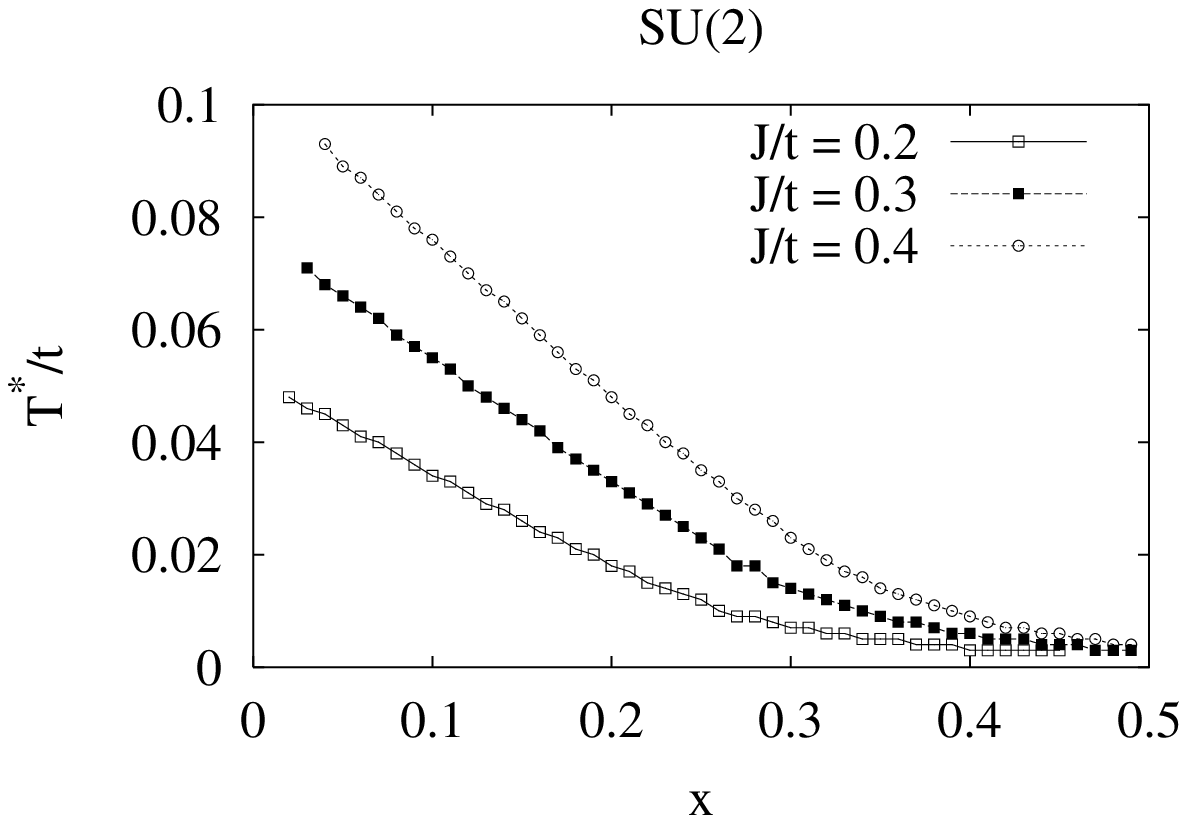}
\label{fig:3}
\caption{
Doping dependence of the spin gap (pseudogap) temperature $T^*$ for $J/t=0.2$, $0.3$ and $0.4$ based on the U(1) and SU(2) holon-pair boson theories. 
}
\end{figure}

\begin{figure}
	\epsfxsize=8.1cm
	\epsfysize=7.1cm
	\epsffile{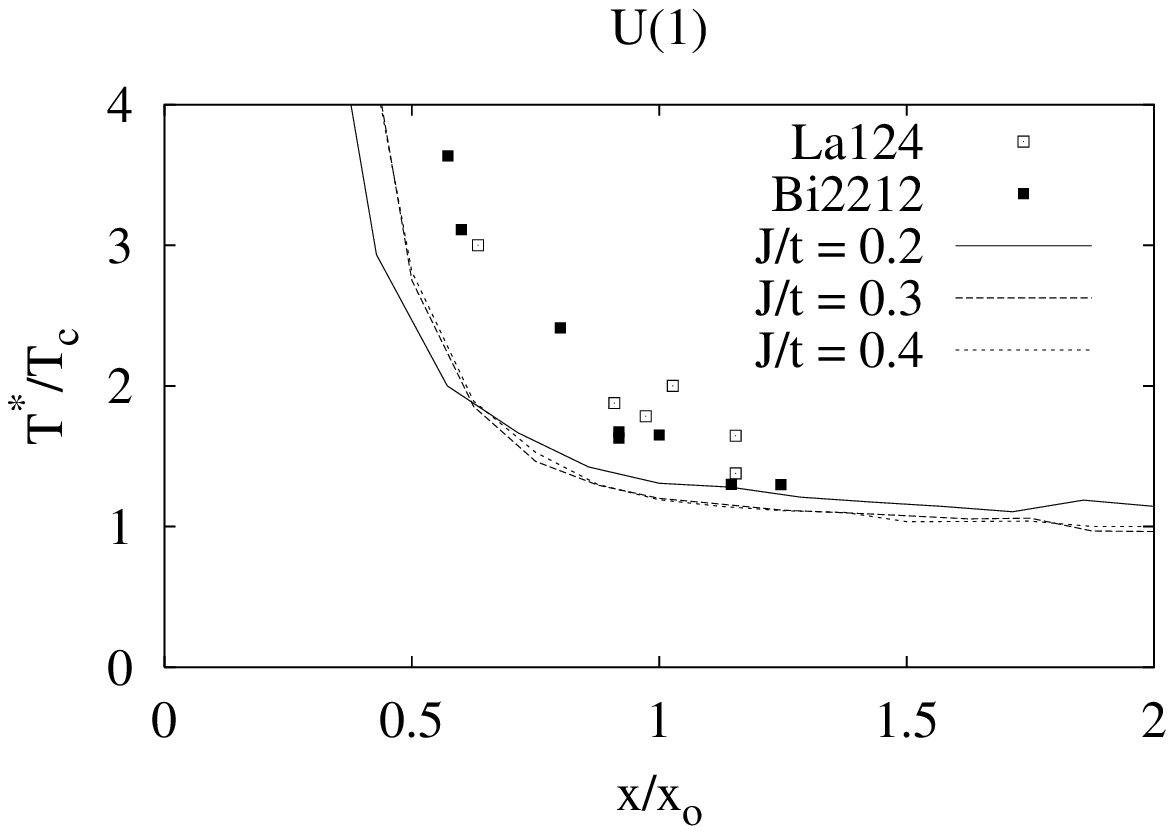}
	\epsfxsize=8.1cm
	\epsfysize=7.1cm
	\epsffile{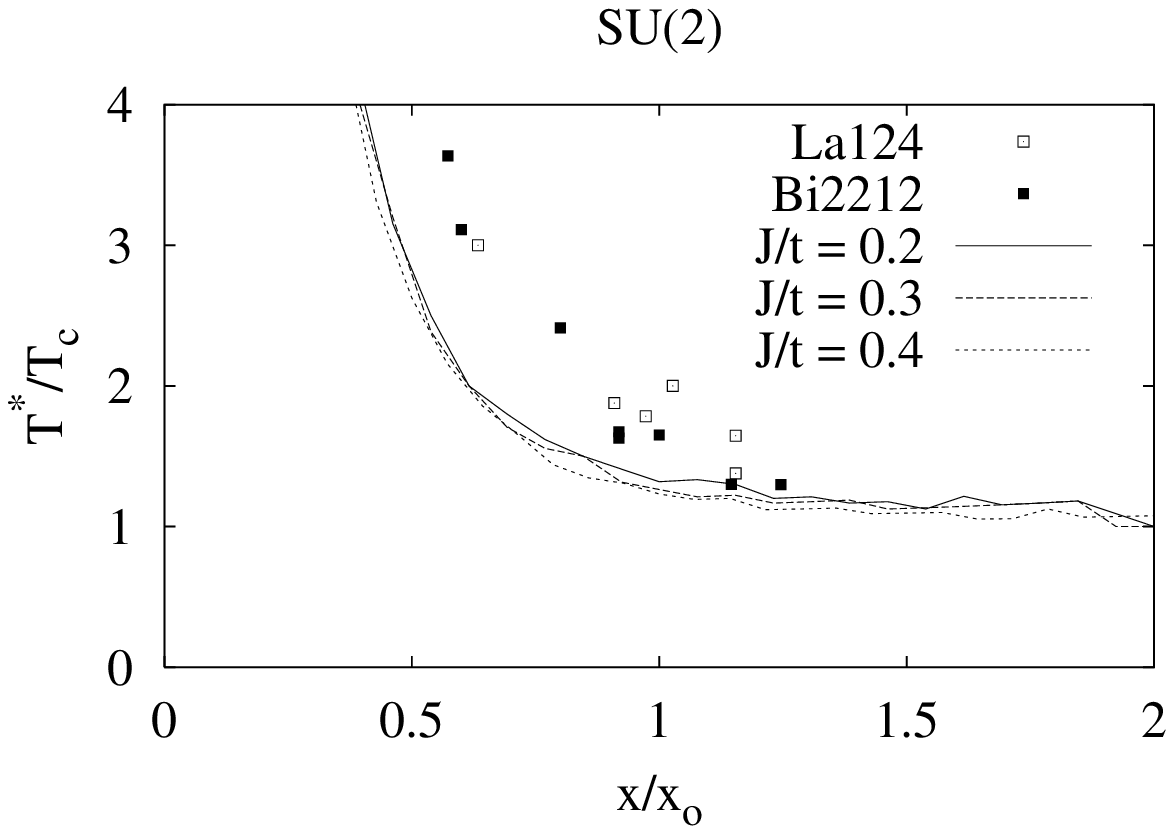}
\label{fig:4}
\caption{
The ratio of the pseudogap temperature to the superconducting temperature, $T^*/T_c$ as a function of scaled hole concentration.  Experimental ratios (squares) are for $Bi_2 Sr_2 Ca Cu_2 O_{8+\delta}$ and $La_{2-x} Sr_x Cu O_4$[8,21].  $x_o$ is the optimal doping rate.
}
\end{figure}

\begin{figure}
	\epsfxsize=8.1cm
	\epsfysize=7.1cm
	\epsffile{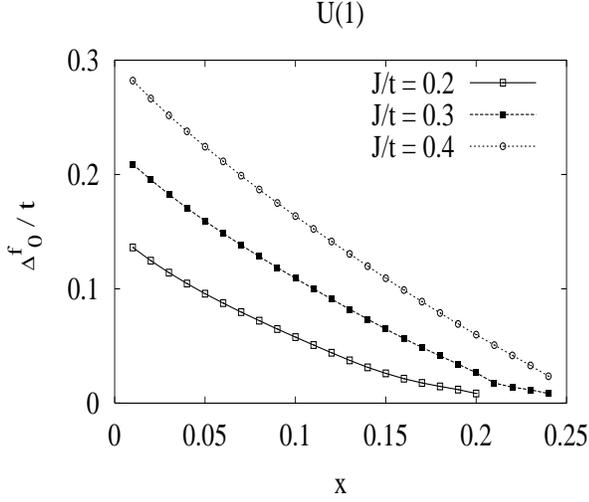}
	\epsfxsize=8.1cm
	\epsfysize=7.1cm
	\epsffile{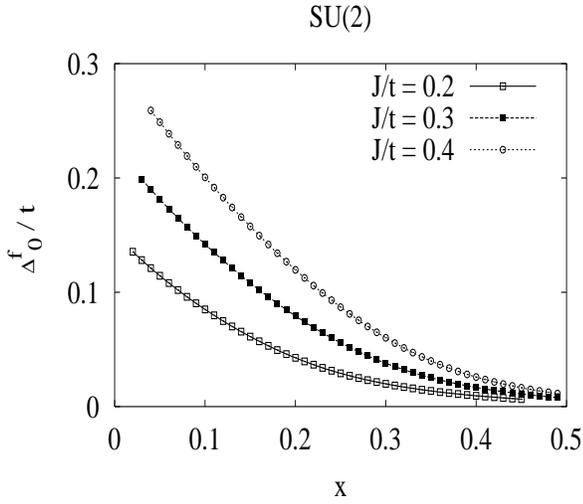}
\label{fig:5}
\caption{
The doping dependence of the superconducting gap $\Delta^f_0 = 2 J_p \Delta_f$ at $T=0$ for $J/t=0.2$, $0.3$ and $0.4$ based on the U(1) and SU(2) theories.
}
\end{figure}

\begin{figure}
	\epsfxsize=8.1cm
	\epsfysize=7.1cm
	\epsffile{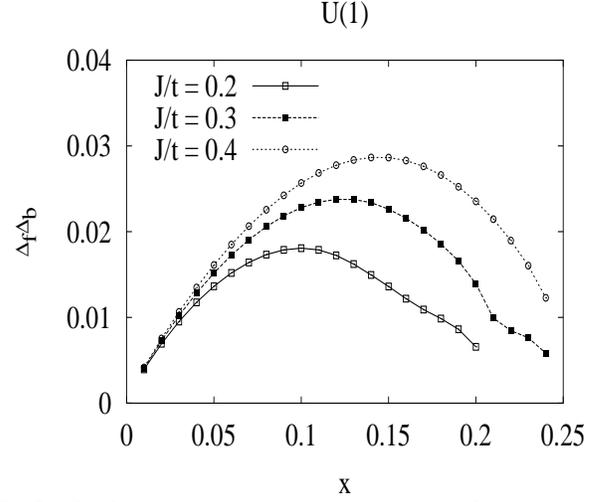}
\caption{
Doping dependence of the Cooper pair order parameter, $(\Delta_f \Delta_{b})$ at $T=0$ for $J/t=0.2$, $0.3$ and $0.4$ based on the U(1) theory. 
}
\label{fig:6}
\end{figure}

\begin{figure}
	\epsfxsize=8.1cm
	\epsfysize=7.1cm
	\epsffile{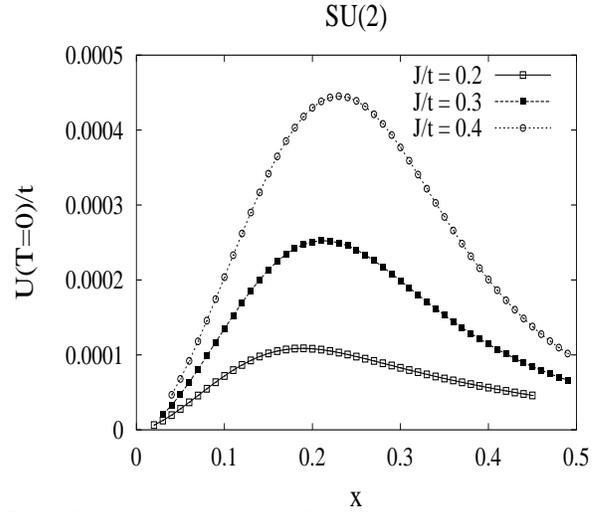}
\caption{
Doping dependence of the condensation energy $U$ at $T=0$ for $J/t=0.2$, $0.3$ and $0.4$ based on the SU(2) theory.
}
\label{fig:7}
\end{figure}

\begin{figure}
	\epsfxsize=8.1cm
	\epsfysize=7.1cm
	\epsffile{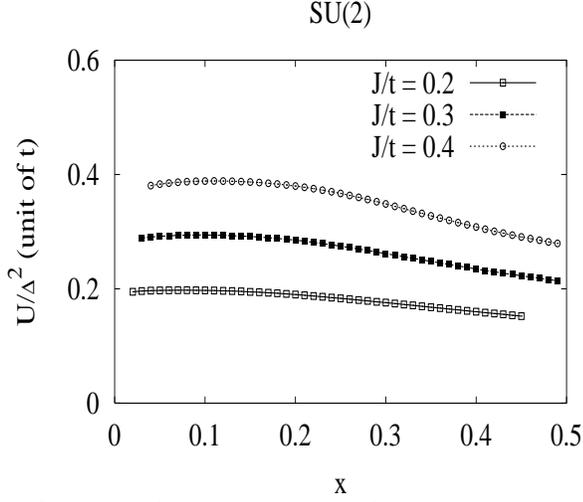}
\caption{
The ratio of the condensation energy, $U$ to the square of the Cooper pair order parameter $\Delta^2$ as a function of hole concentration at $T=0$ for $J/t=0.2$, $0.3$ and $0.4$ based on the SU(2) theory.
}
\label{fig:8}
\end{figure}

\begin{figure}
	\epsfxsize=8cm
	\epsfysize=11cm
	\epsffile{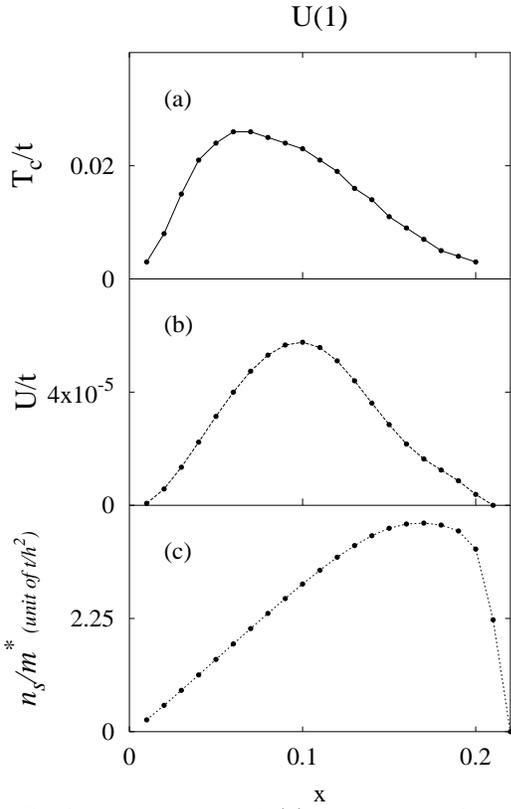}
\caption{
Doping dependence of (a) superconducting temperature, (b) the condensation energy and (c) the superfluid weight with $J/t=0.2$ for the U(1) theory.  The superfluid weight and the condensation energy was obtained at $T/t=0.001$ (equivalent to $5K$ with the use of $t=0.44eV$[43]).
}
\label{fig:9}
\end{figure}

\begin{figure}
	\epsfxsize=8.1cm
	\epsfysize=7.1cm
	\epsffile{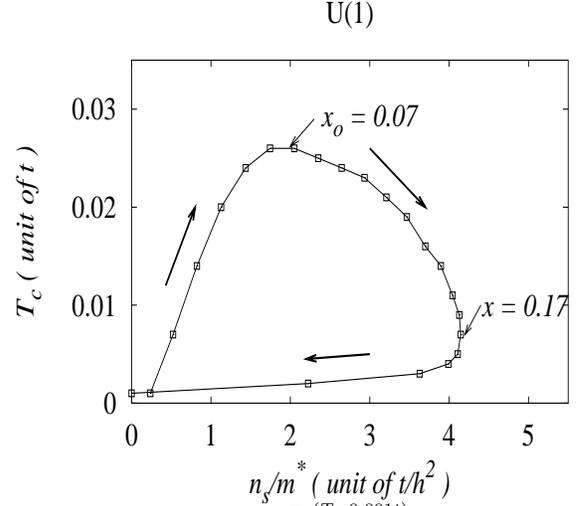}
\caption{
Superfluid weight, $\frac{n_s(T = 0.001t )}{m^*}$ vs superconducting transition temperature, $T_c$ in the U(1) theory with $J/t=0.2$.  The directions of thick arrows denote increasing hole doping concentration.  $x_o=0.07$ is the predicted optimal doping rate. 
}
\label{fig:10}
\end{figure}

\begin{figure}
	\epsfxsize=8.1cm
	\epsfysize=7.1cm
	\epsffile{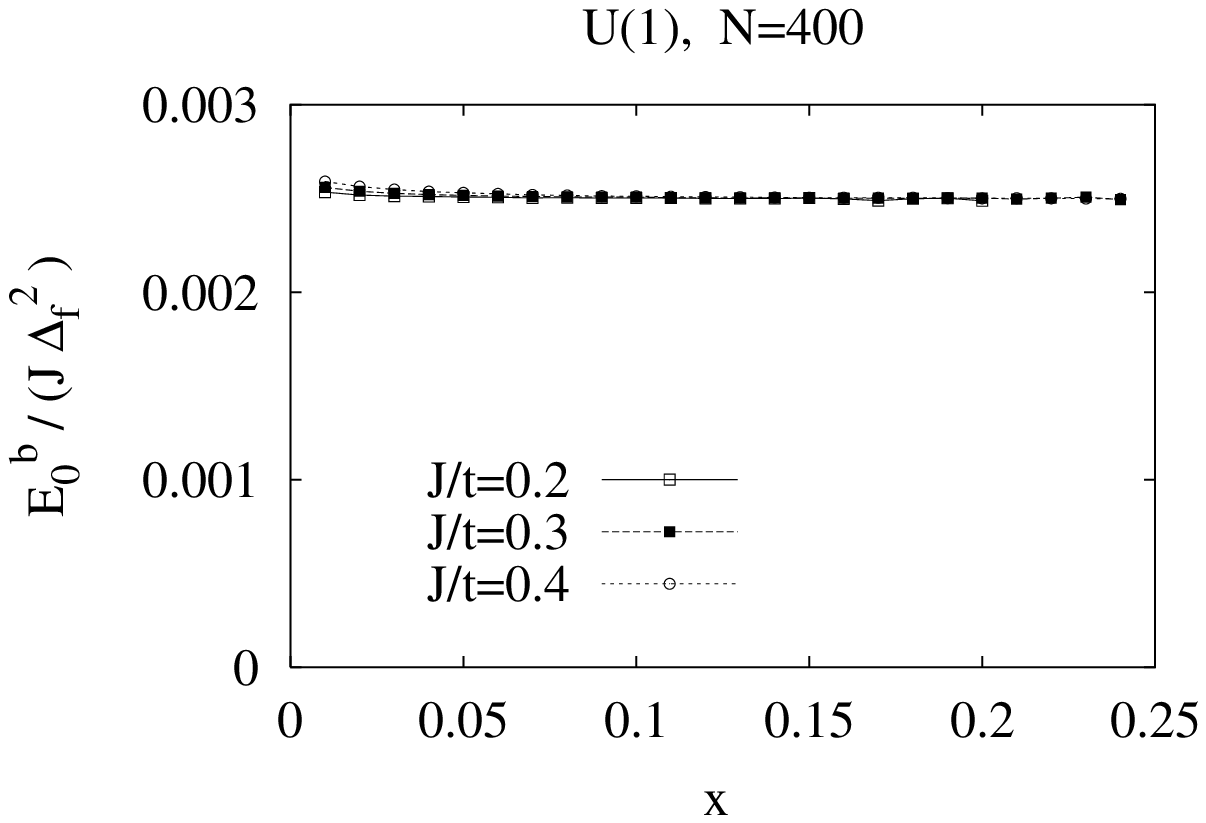}
	\epsfxsize=8.1cm
	\epsfysize=7.1cm
	\epsffile{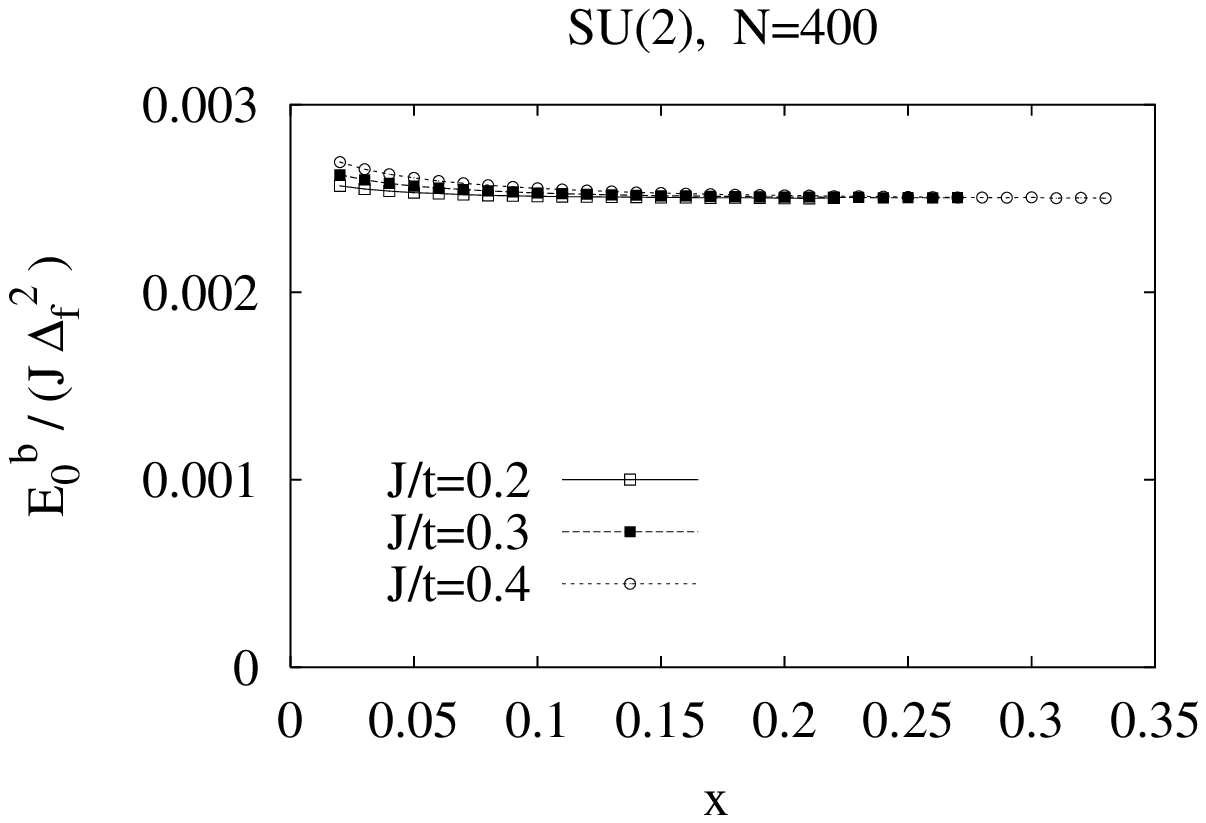}
\caption{
Doping dependence of $\frac{E_0^b}{J\Delta_f^2}$, the ratio of the holon quasiparticle excitation energy gap $E_0^b$ to the effective holon pairing interaction strength $J \Delta_f^2$ at $T=0$ for $J/t=0.2$, $0.3$ and $0.4$ using the U(1) and SU(2) theories respectively. $N = 20 \times 20$ lattice is used.
}
\label{fig:11}
\end{figure}

\begin{figure}
	\epsfxsize=8.1cm
	\epsfysize=7.1cm
	\epsffile{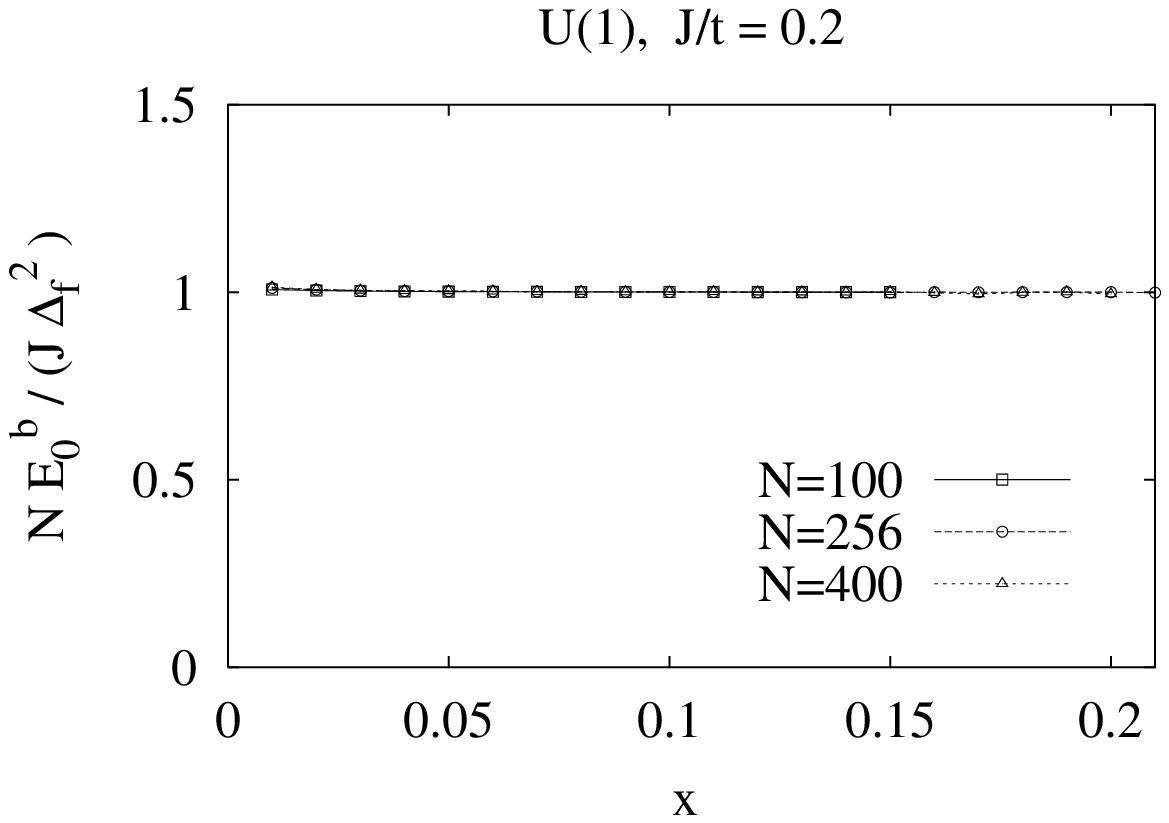}
	\epsfxsize=8.1cm
	\epsfysize=7.1cm
	\epsffile{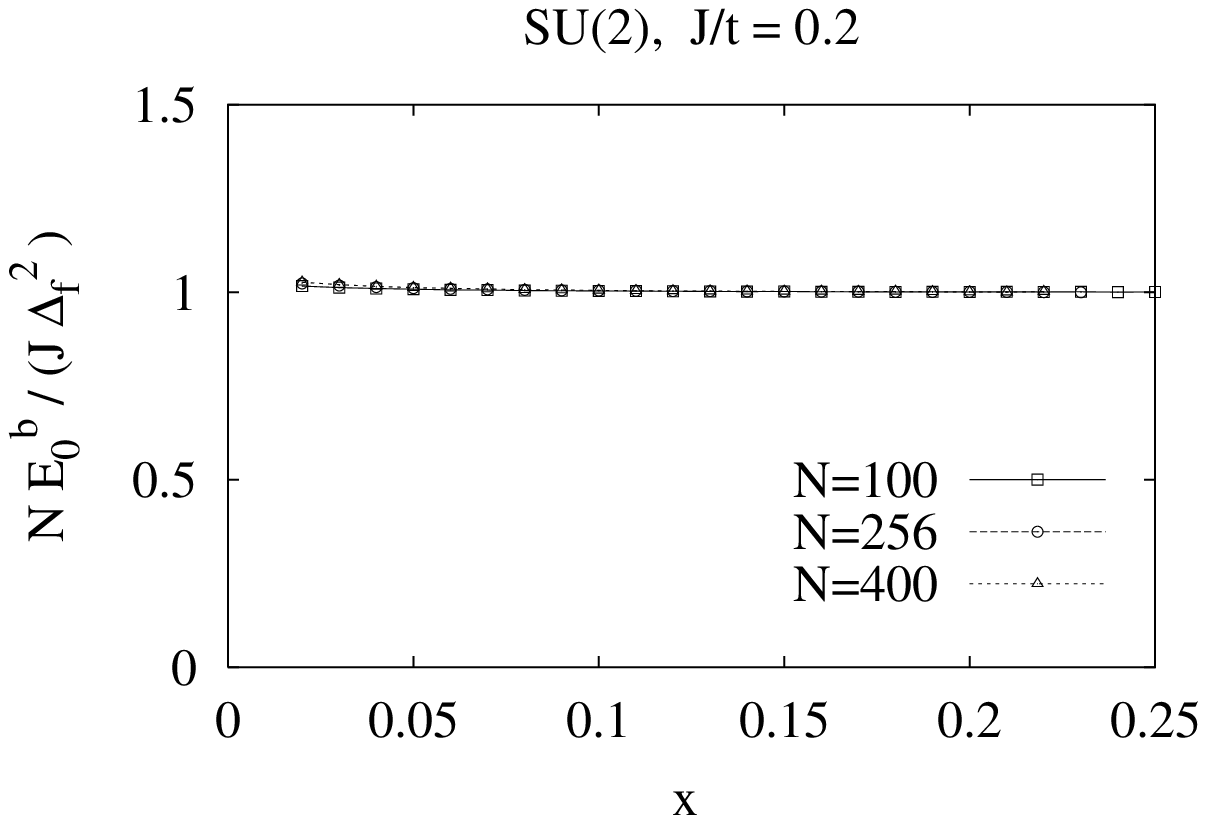}
\caption{
Doping dependence of $N \frac{E_0^b}{J\Delta_f^2}$, the holon quasiparticle excitation energy gap $E_0^b$ to the effective holon pairing interaction strength $J\Delta_f^2$ multiplied by the lattice size $N$ at $T=0$ using the U(1) and SU(2) theories respectively. Three different lattice sizes of $N = 10 \times 10$, $16 \times 16$ and $20 \times 20$ are used with the choice of $J/t=0.2$.
}
\label{fig:12}
\end{figure}

\begin{figure}
	\epsfxsize=8.1cm
	\epsfysize=7.1cm
	\epsffile{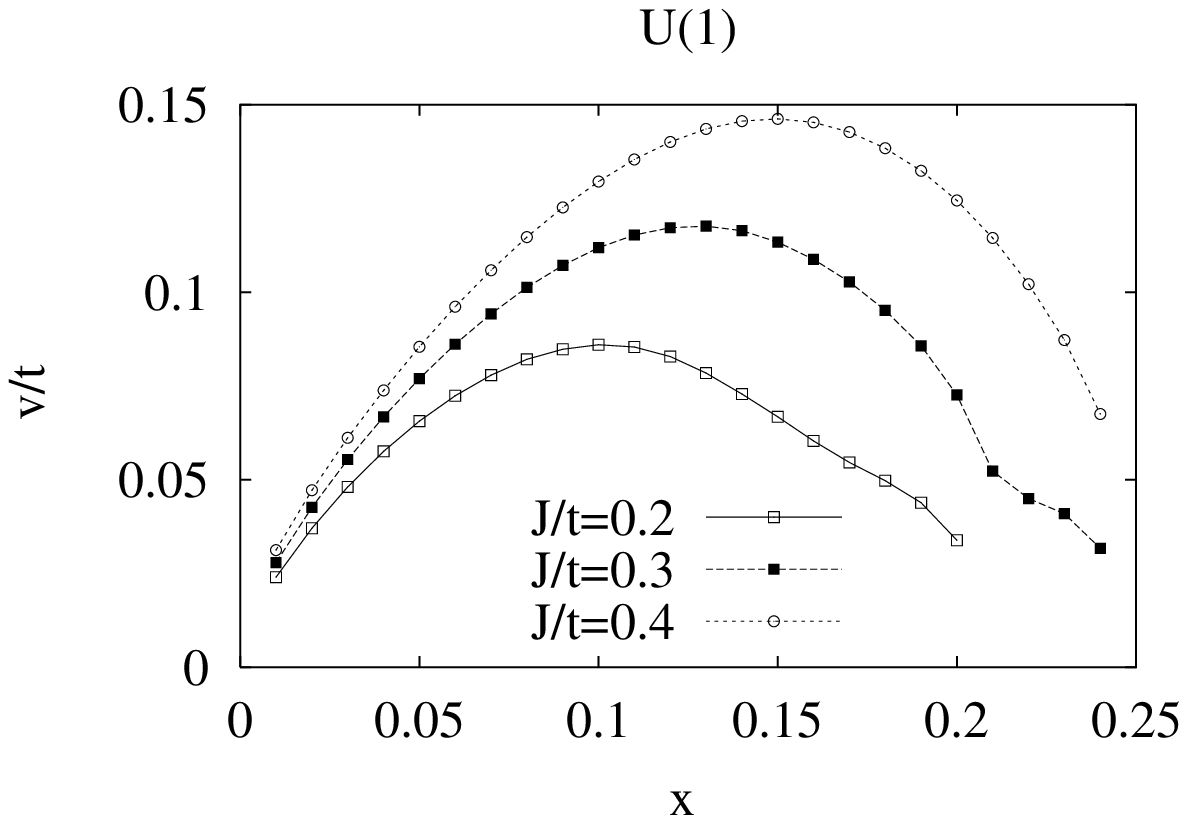}
	\epsfxsize=8.1cm
	\epsfysize=7.1cm
	\epsffile{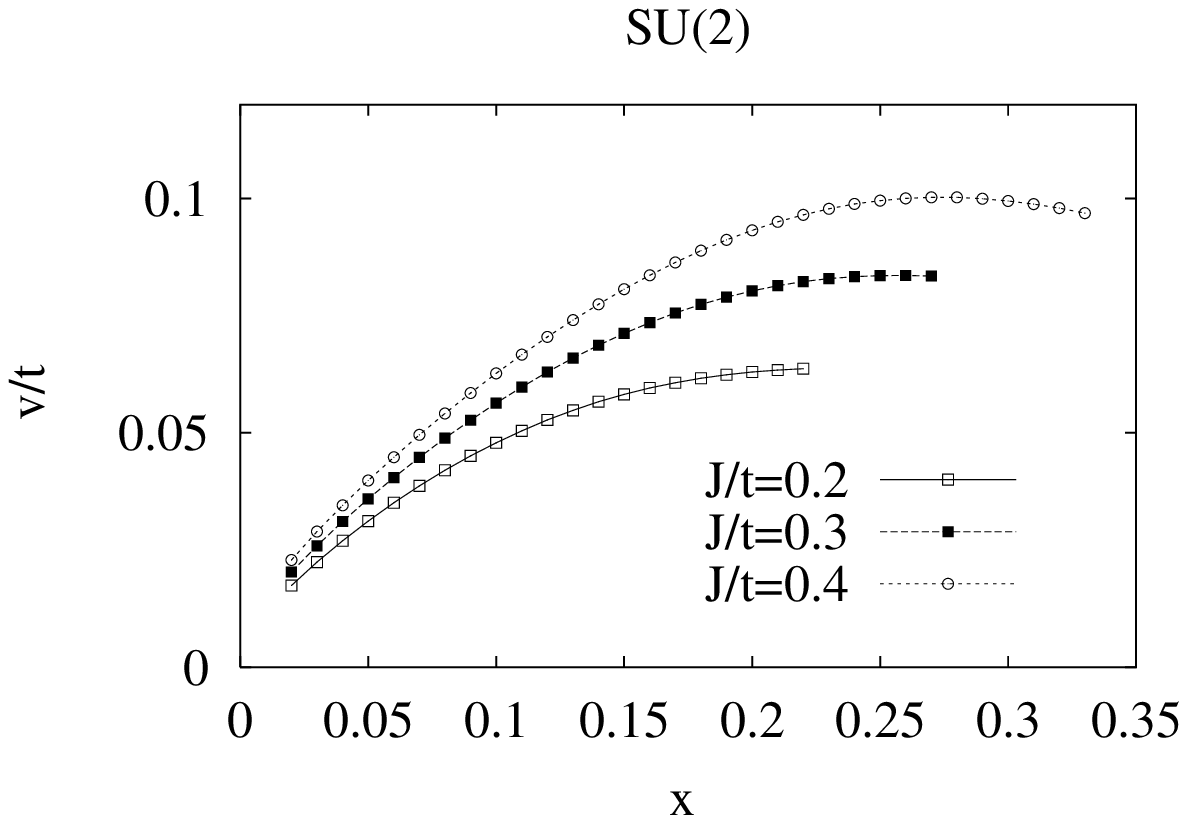}
\caption{
Doping dependence of quasiholon velocity at $T=0$ based on the U(1) and SU(2) theories respectively. $N = 20 \times 20$ is used.
}
\label{fig:13}
\end{figure}

\end{multicols}

\end{document}